\documentclass[preprint,10pt]{elsarticle}
\usepackage[utf8]{inputenc}
\usepackage{graphicx}
\usepackage{bm}%
\usepackage{siunitx}
\usepackage{textcomp}
\usepackage{lineno}
\usepackage{xcolor}
\usepackage{subfigure}
\usepackage{amsmath}
\usepackage{textcomp}
\usepackage{rotating}

\journal{XXXX}

\begin{document}

\begin{frontmatter}


\title{In-Silico Optimisation of Tileable Philips Digital SiPM Based Thin Monolithic Scintillator Detectors for SPECT Applications}



\author{Jeremy~M.~C.~Brown}
\address{Department of Radiation Science and Technology, Delft University of Technology, The Netherlands}

\begin{abstract}
Over the last decade one of the most significant technological advances made in the field of radiation detectors for nuclear medicine was the development of Silicon Photomultipler (SiPM) sensors. At present a only small number of SiPM based radiation detectors for Single Photon Emission Computed Tomography (SPECT) applications have been explored, and even fewer experimental prototypes developed. An in-silico investigation into the optimal design of a Philips DPC3200 SiPM photosensor-based thin monolithic scintillator detector for SPECT applications was undertaken using the Monte Carlo radiation transport modelling toolkit Geant4 version 10.5. The performance of the 20 different SPECT radiation detector configurations, 4 scintillator materials (NaI(Tl), GAGG(Ce), CsI(Tl) and LaBr$_{3}$(Ce)) and 5 thicknesses (1 to 5 mm), were determined through the use of seven figures of merit. It was found that a crystal thickness range of 4 to 5 mm was required for all four materials to ensure acceptable energy resolution, sensitivity and spatial resolution performance with the Philips DPC3200 SiPM. Any thinner than this and the performance of all four materials was found to degrade rapidly due to a high probability of material specific fluorescence x-ray escape after incident gamma/x-ray photoelectric absorption. When factoring in each material's magnetic resonance imaging compatibility, hygroscopy, and cost, it was found that CsI(Tl) represents the most promising material to construct tileable Philips digital SiPM based thin monolithic scintillator detectors for SPECT applications. 

\end{abstract}


\end{frontmatter}

\section{Introduction}

Single Photon Emission Computed Tomography (SPECT) is one of the primary emission imaging modalities utilised in nuclear medicine. This imaging modality is based on the use of a mechanical collimator to restrict the solid angle of gamma/x-rays incident upon the surface of a position-and-energy-resolving radiation detector \cite{Kupinski2005,Brown2013}. These two key system elements, the mechanical collimator and radiation detector, define the fundamental limit of any SPECT imaging system's performance \cite{Cherry2003,Bushberg2011}. The restriction of the solid angle via the mechanical collimator enables the incident trajectory of a detected gamma/x-ray to be estimated. The accuracy of this estimated incident trajectory and the fraction of emitted radiation allowed to pass through an aperture are inversely proportional, resulting in a trade-off between spatial resolution and sensitivity \cite{Kupinski2005,Cherry2003}. Conversely in the case of the radiation detector, a three way trade-off exists between spatial resolution, sensitivity and energy resolution depending on its gamma/x-ray detection mechanism (direct or indirect), detection material type/geometry, and signal processing electronics/optical photosensor combination \cite{Kupinski2005,Cherry2003}. 

One of the most recent significant technological advances made in the field of radiation detectors for nuclear medicine was the development of Silicon Photomultipler (SiPM) sensors. These compact novel optical photosensors have enabled significant gains in radiation detector performance for Positron Emission Tomography (PET) applications \cite{Lewellen2008,Schaart2016,Bisogni2018}, and at the same time enabled full integration of PET with Magnetic Resonance Imaging (MRI) \cite{Judenhofer2007,Schlemmer2008,Wehrl2009,Delso2011,Weissler2015,Gonzalez2016,Benlloch2018,Gonzalez2018,Hypmed2019}. Furthermore a number of these SiPM units, such as the Philips DPC3200 SiPM \cite{Frach2009,Frach2010}, are four-side buttable enabling their tiling to create large surface area MRI compatible radiation detectors ideal for clinical SPECT applications. However at present only a small number of SPECT SiPM radiation detectors has been developed \cite{Heemskerk2010,Georgiou2014,Busca2015,David2015}, and a single simultaneous acquisition SPECT/MR clinical prototype constructed as part of the INSERT program \cite{Busca2014,Hutton2018,Carminati2019}.

This work presents an in-silico investigation into the optimal design of a Philips DPC3200 SiPM photosensor-based thin monolithic scintillator detector for SPECT applications. The Philips DPC3200 SiPM photosensor was selected as the subject SiPM of this study due to its ability to assess and suppress background noise (dark count-rate) effects on an individual photon detection element basis (increasing energy resolution and decreasing detector dead-time), moderate SiPM pixel size (3.2 $\times$ 3.8 mm$^2$), and proven effectiveness in a clinical setting (i.e. it is the photosensor used in Philips' Vereos PET/CT) \cite{Frach2009,Frach2010}. Four different monolithic scintillator crystal material types, NaI(Tl), GAGG(Ce), CsI(Tl) and LaBr$_{3}$(Ce) \cite{Lecoq2016}, directly bonded to the SiPM photosensor were explored for the primary gamma/x-ray emissions from $^{99m}$Tc, $^{123}$I, $^{131}$I and $^{201}$Tl as a function of crystal thickness over the range of 1 to 5 mm. Section \ref{sec:M} describes the developed simulation platform, detector response/readout modelling, and detection performance assessment/optimisation methodology. The results from this in-silico investigation, their discussion and an overall conclusion then follows in Section \ref{sec:R}, \ref{sec:D} and \ref{sec:C} respectively. 

\section{Method}
\label{sec:M}

A simulation platform was constructed using the Monte Carlo radiation transport modelling toolkit Geant4 version 10.5 \cite{G42003,G42006,G42016} to determine the optimal design of a Philips DPC3200 SiPM photosensor-based thin monolithic scintillator detector for SPECT applications. The methodology of the investigation may be separated into four primary areas: 1) simulated detector geometry and materials, 2) physics and optical surface modelling, 3) photosensor response and radiation detector readout modelling, and 4) radiation detector performance assessment/optimisation.

\subsection{Simulated Detector Geometry and Materials}

\begin{figure}[tbh]   
   \centering
   \includegraphics[width=0.8\textwidth]{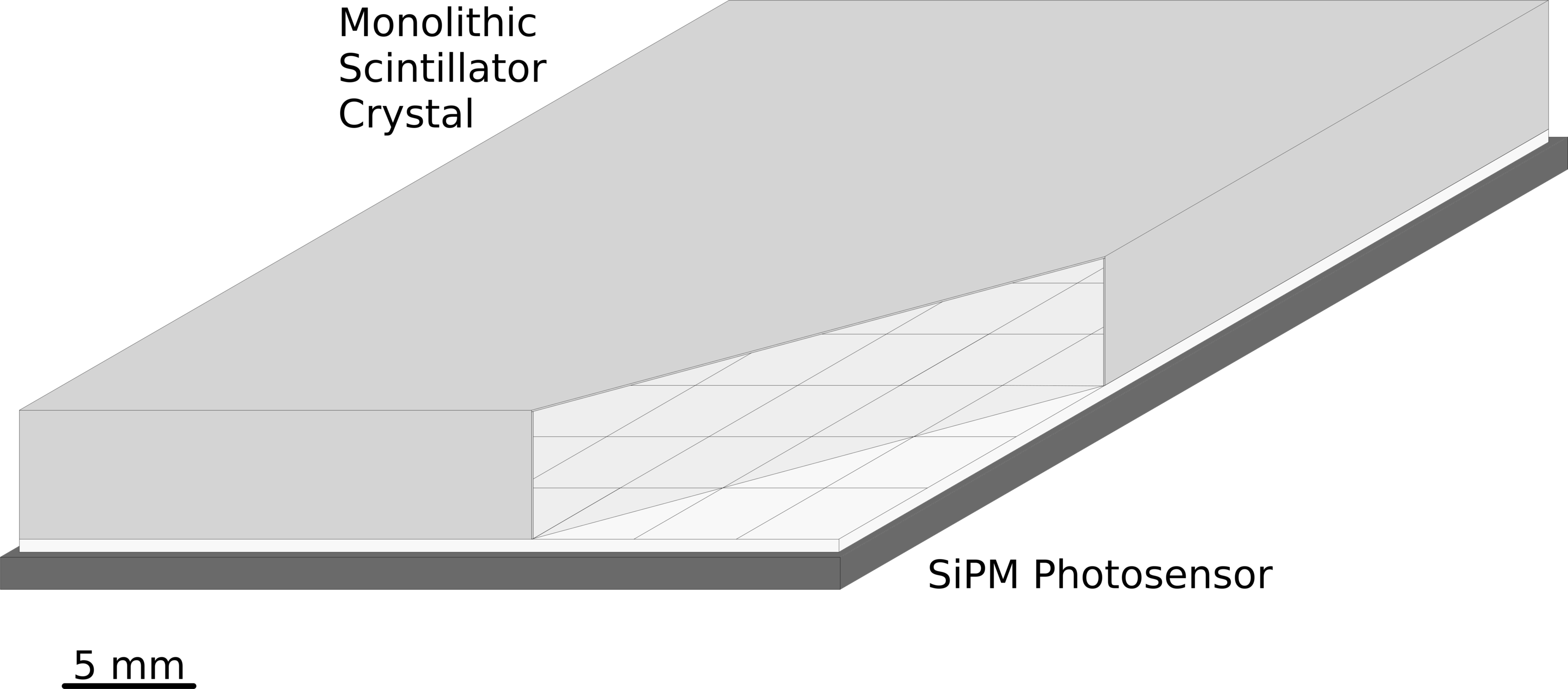}
\caption{A schematic of the SPECT radiation detector geometry constructed within the Geant4 simulation platform. Here a section of the monolithic crystal is removed to illustrate the implemented 8 $\times$ 8 Si pixel footprint of the Philips DPC3200 SiPM photosensor.}
\label{fig:1}
\end{figure}

A schematic of the simulated SPECT radiation detector geometry composed of a thin monolithic scintillator crystal coupled to a Philips DPC3200 Silicon Photomultiplier (SiPM) \cite{Frach2009,Frach2010} with a 100 {\textmu}m layer of DELO photobond 4436 glue is shown in Fig. \ref{fig:1}. The SiPM photosensor consists of a four-side buttable, 4 $\times$ 4 array of independent die sensors distributed in an $\sim$ 8 mm pitch over 32.6 $\times$ 32.6 mm$^2$ unit footprint. Each independent die sensor possesses a 2 $\times$ 2 array of 3.2$\times$3.8 mm$^2$ SiPM pixels, that are each composed of 3200 59.4 {\textmu}m $\times$ 64 {\textmu}m Single Photon Avalanche Diodes (SPADs), and share a common controller and Time-to-Digital-Converter (TDC). Through this common controller it is possible to turn on or off individual SPADs, and select trigger conditions  \cite{DPCManual2016}. The cross-sectional area of the coupled scintillator crystal surface was set to match the approximate active Philips DPC3200 SiPM photosensor region (32$\times$32 mm), with the other five crystal surfaces made light-tight through the mounting of a layer of Vikuiti ESR foil via a 35 {\textmu}m thick layer of DELO photobond 4436 glue. All 6 surfaces of the monolithic crystal were assumed to be polished, with four different scintillator crystal types, and its thickness varied as part of the optimisation process (see Section \ref{sec:M:DPAO} for more details). Implementation of the Philips DPC3200 SiPM photosensor followed the same approach as outlined in \cite{Brown2019}. Here, the photosensor layer structure, dimensions and locations of the quartz light guide, glue layers, 8 $\times$ 8 array of SiPM pixels, and printed circuit board was based on version 1.02 of the unit manual \cite{DPCManual2016}. Finally, the density, elemental composition, and optical/scintillation properties of all materials can be found in \ref{appendix1}.

\subsection{Physics and Optical Surface Modelling}

X-ray, gamma-ray and electron transport was simulated using the Geant4 Option4 EM physics list (G4EmStandardPhysics\_option4 \cite{G42016}) with atomic deexcitation enabled, a maximum particle step length of 10 {\textmu}m, and a low-energy cut off of 250 eV. Optical photon generation and transport was included for the processes of scintillation, absorption, refraction and reflection via the Geant4 implementation of the ``Unified" model \cite{Levin1996,G4Phys2019}. With the exception of the ESR foil-to-DELO-glue material interfaces (modelled as a dielectric-to-metal with reflectivity outlined in \ref{appendix1}), all other material optical interfaces were modelled as dielectric-to-dielectric. Finally, every surface interface between two materials was described as a ground surface with surface roughness of 0.1 degrees because it is not possible for surfaces to be ``perfectly smooth"\cite{VanderLaan2010,Nilsson2015}. 

\subsection{Photosensor Response and Detector Readout Modelling}

The implemented photosensor response model was further developed from that outlined in Brown et al. \cite{Brown2019}. Here, the photosensor response was realised in two steps: 1) physical geometry, and 2) electronic response. The physical geometry of the SiPM was implemented through the definition of scoring boundaries that mimicked the shape and location of all 3200 59.4 {\textmu}m $\times$ 64 {\textmu}m Single Photon Avalanche Diodes (SPADs) \cite{Frach2009,Frach2010} in each of the 64 SiPM pixels. The electronic behaviour of the SiPM photosensor was modelled based on five assumptions: 

\begin{enumerate}
    \item the probability of a photoelectrically absorbed optical photon triggering a SPAD is proportional to the energy/wavelength dependent Photon Detection Efficiency (PDE) outlined in Frach et al. \cite{Frach2009}\footnote{For the scintillation photon emission spectra presented in Fig. \ref{fig:a2} of \ref{appendix1}, the Philips DPC3200 SiPM’s effective PDE for NaI(Tl), GAGG(Ce), CsI(Tl), and LaBr$_{3}$(Ce) are 20.1\%, 18.4\%, 19.5\%, and 9.4\% respectively.},
    \item a given SPAD can only trigger once per simulated primary gamma/x-ray,
    \item each SiPM pixel has a 913.08 thousand counts per second dark count-rate with 90\% SPAD activation operating at \ang{20}C \cite{DPCManual2016},    
    \item the SiPM's trigger logic was set to ``scheme 3" with each pixel's triggering probability model based on Table 5 in version 1.02 of the unit manual \cite{DPCManual2016}, and
    \item for the event acquisition sequence of the DPC3200 the integration time was set to 5125 ns.
\end{enumerate}

\noindent Finally, to enable a more in-depth crystal material dependent ``dead-time" analysis of the radiation detector design under investigation, each 8 $\times$ 8 SiPM pixel SPAD trigger count was also accompanied by a full list of their respective timestamps relative to the first incoherent interaction time of the gamma/x-ray within the monolithic scintillator crystals.

The interaction position $(x,y)$ of each simulated gamma/x-ray within the monolithic scintillator crystals was determined through the use of a truncated Centre of Gravity (CoG) algorithm \cite{Georgiou2014,Steinbach1996}. Each estimated interaction position $(X,Y)$ was determined using the photosensor response model output by:

\begin{equation}
    \displaystyle X =  \frac{\sum^{8}_{i=1}x_i\sum^{8}_{j=1}n_{i,j,\alpha}}{\sum^{8}_{i=1}\sum^{8}_{j=1}n_{i,j,\alpha}}
\end{equation}

\begin{equation}
    \displaystyle Y =  \frac{\sum^{8}_{j=1}y_i\sum^{8}_{i=1}n_{i,j,\alpha}}{\sum^{8}_{j=1}\sum^{8}_{i=1}n_{i,j,\alpha}}
\end{equation}

\noindent where $n_{i,j,\alpha}$ is the truncated SPAD trigger counts of each SiPM pixel $(i,j)$ at location ($x_i$,$y_j$), and:

\begin{equation}
\begin{aligned}
    \displaystyle n_{i,j} - \alpha \sum^{8}_{i=1}\sum^{8}_{j=1} n_{i,j} > 0 & \Rightarrow n_{i,j,\alpha} = n_{i,j} - \alpha \sum^{8}_{i=1}\sum^{8}_{j=1} n_{i,j} \\
    \displaystyle n_{i,j} - \alpha \sum^{8}_{i=1}\sum^{8}_{j=1} n_{i,j} \leq 0 & \Rightarrow n_{i,j,\alpha} = 0 \\
\end{aligned}
\end{equation}

\noindent for the initial number of SPAD trigger counts of each SiPM pixel ($n_{i,j}$) and a given truncation factor $\alpha$. The truncation factor $\alpha$ serves as a scalable mechanism to suppress the impact of SiPM photosensor dark-count rates and inherent statistical fluctuations associated with photon counting measurements, and has been shown to improve position-of-interaction identification and spatial linearity within scintillator gamma/x-ray detectors \cite{Georgiou2014,Wojcik2001}. In this work the impact of the two values for $\alpha$ were explored on detector performance: 1) $\alpha=0$ corresponding to no truncation, and 2) $\alpha = 0.02$ corresponding to the fraction uncertainty of when all active pixel SPADs are triggered.

\subsection{Detector Performance Assessment/Optimisation}
\label{sec:M:DPAO}

Two physical properties were optimised to maximise the performance of the proposed SPECT radiation detector: monolithic scintillator crystal material, and monolithic scintillator crystal thickness. Four different scintillator types, NaI(Tl), GAGG(Ce), CsI(Tl) and LaBr$_{3}$(Ce), over a thickness range of 1 to 5 mm were explored as they represent four of the most commonly used scintillator materials in radiation detectors for SPECT applications. An upper crystal thickness limit of 5 mm was selected to maximise the effective three dimensional radiation detector spatial resolution, through reducing the impact of oblique gamma/x-ray spatial resolution degradation and Depth of Interaction (DoI) effects, at a minimal cost to detection efficiency at 140 keV \cite{Kupinski2005,Cherry2003,Bushberg2011,Garcia2011}. For each configuration 16 different infinitely thin pencil beam irradiation positions were simulated for five different gamma/x-ray energies (28, 72, 140, 159 and 365 keV) that represent the primary emissions from $^{125}$I, $^{201}$Tl, $^{99m}$Tc, $^{123}$I, and $^{131}$I. These 16 irradiation positions were composed of a horizontal $x$-axis and diagonal $x$-$y$ axis sweep from the centre of the SPECT radiation detector, starting at 1 mm and continuing in 2 mm steps to its edge (i.e. 1mm, 3mm, 5mm, ... 15 mm). At each location 50,000 gamma/x-rays were simulated for the infinitely thin pencil beam originating at and orientated perpendicular to the monolithic radiation detector’s surface.

The performance of the 20 different SPECT radiation detector configurations was determined through the use of seven Figures of Merit (FoM): gamma/x-ray total absorption fraction, gamma/x-ray photoelectric absorption fraction on the first interaction, photopeak Full Width at Half Maximum (FWHM) energy resolution, energy spectrum linearity, relative final SPAD trigger time per gamma/x-ray, FWHM of estimated gamma/x-ray irradiation locations, and linearity of estimated gamma/x-ray irradiation locations. All FWHM values were calculated assuming a Gaussian distribution, and the energy/spatial linearity assessed through the use the correlation coefficient ($R^2$) from linear regression with respect to known incident gamma/x-ray energies and irradiation locations. Furthermore, the FWHM and linearity of estimated irradiation locations was applied to gamma/x-rays that underwent photoelectric absorption on their first interaction within the scintillator crystal. Filtering the data in this manner, rather than using an energy window approach which typically also includes gamma/x-rays that deposit their total/near total energy in the scintillator crystals through multiple interactions, enables quantification of the ``true" number of detector events/spatial resolution of the SPECT radiation detector (relevant in calculation of SPECT system sensitivity).

\section{Results}
\label{sec:R}

\begin{figure}[tbh]    
    \centering
    \begin{subfigure}
    \centering 
        \includegraphics[width=0.425\textwidth, trim = {0 0 25 10},clip]{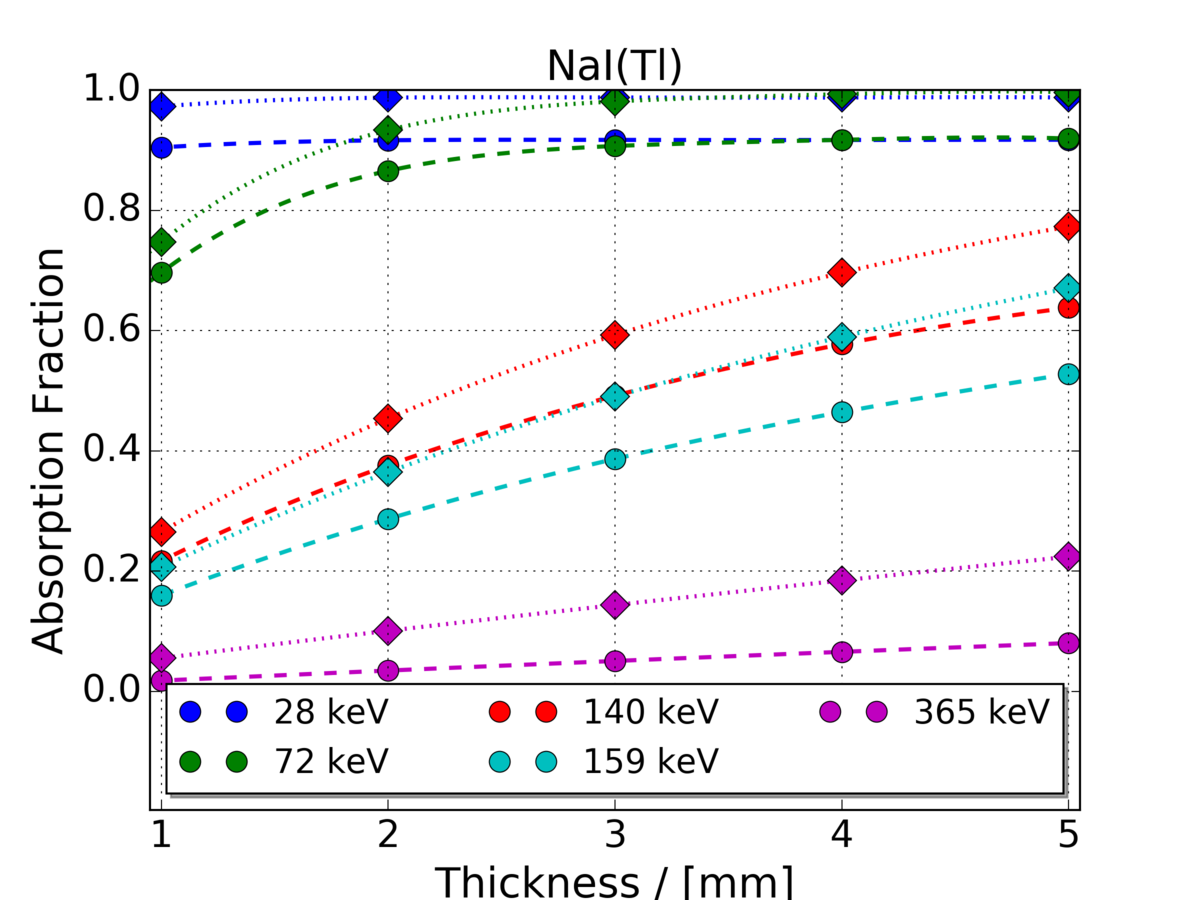}
        \label{fig:r1a}
    \end{subfigure}
    \begin{subfigure}
    \centering
        \includegraphics[width=0.425\textwidth, trim = {0 0 25 10},clip]{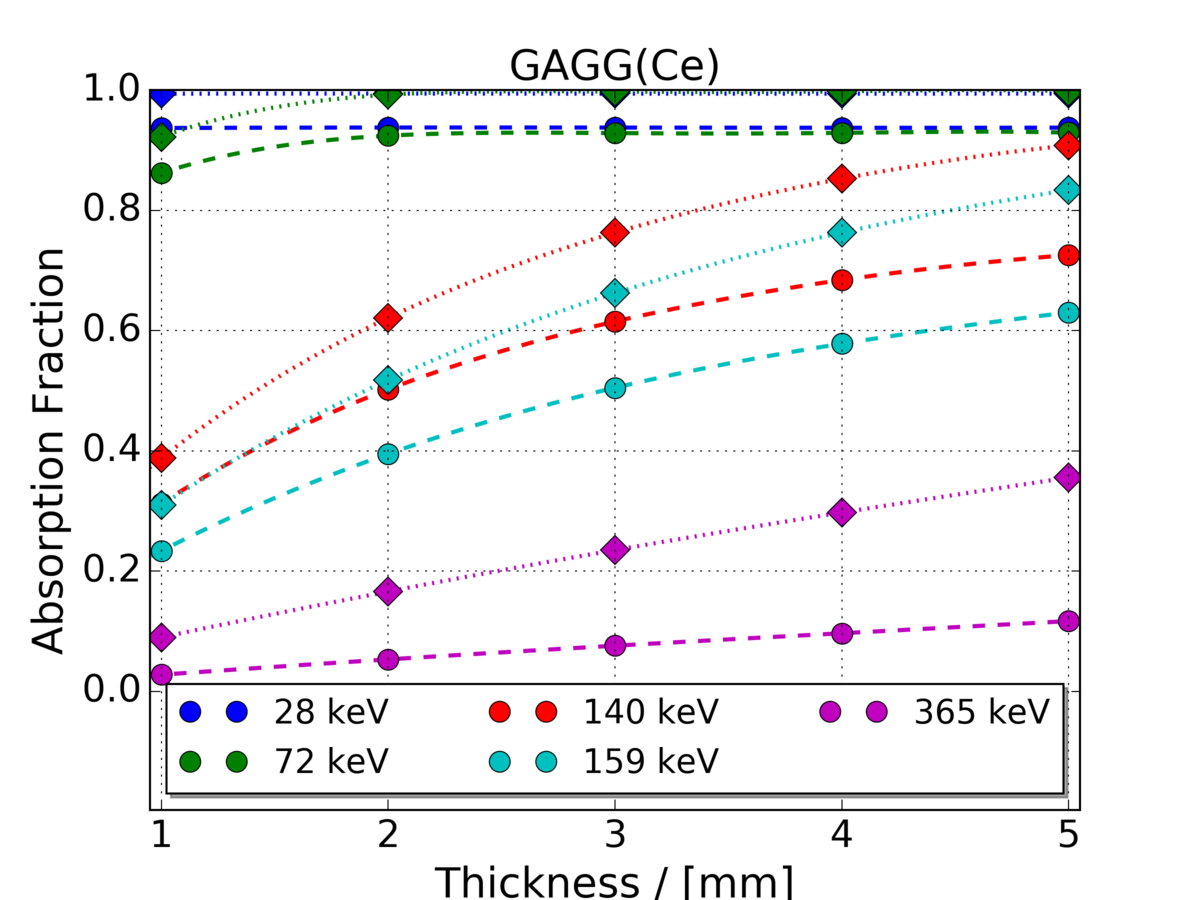}
        \label{fig:r1b}
    \end{subfigure}

    \begin{subfigure}
    \centering 
        \includegraphics[width=0.425\textwidth, trim = {0 0 25 10},clip]{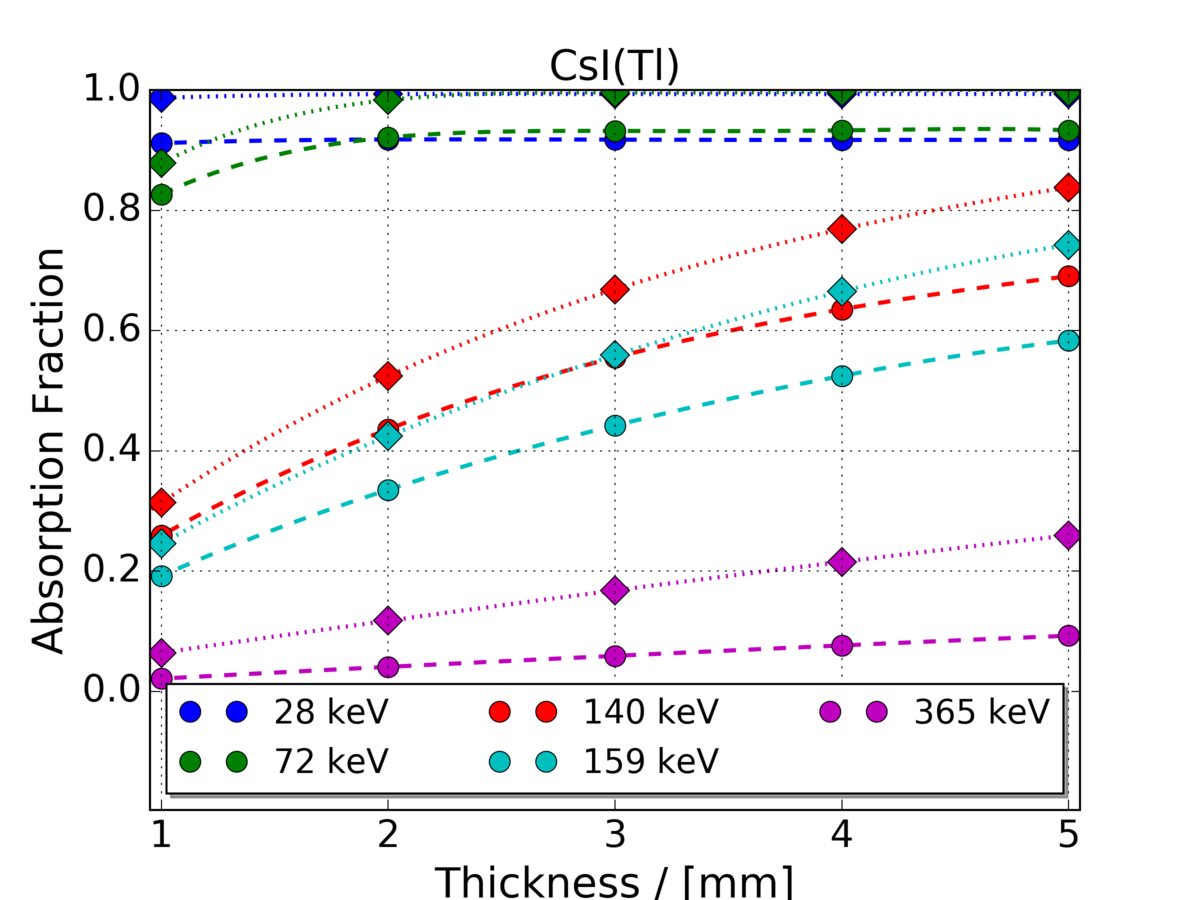}
        \label{fig:r1c}
    \end{subfigure}
    \begin{subfigure}
    \centering
        \includegraphics[width=0.425\textwidth, trim = {0 0 25 10},clip]{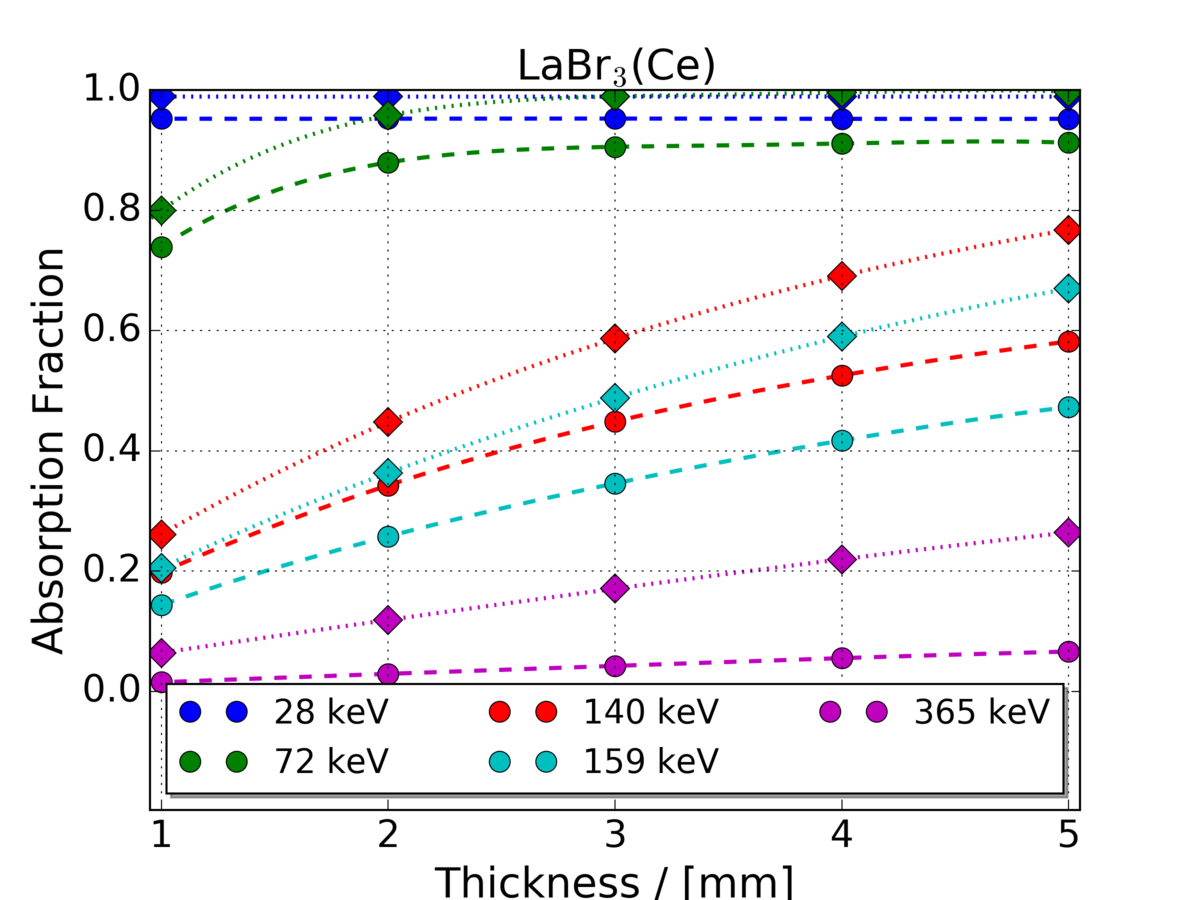}
        \label{fig:r1d}
    \end{subfigure}
\caption{Gamma/x-ray total absorption fraction (diamond marker) and photoelectric absorption fraction on the first interaction (circle marker) of the four different scintillator crystal materials, NaI(Tl), GAGG(Ce), CsI(Tl) and LaBr$_{3}$(Ce), as a function of incident gamma/x-ray energy and crystal thickness. The coloured dash lines, gamma/x-ray total absorption fraction (dotted) and photoelectric absorption fraction on the first interaction (dashed), correspond to a fitted polynomial surrogate function for each incident gamma/x-ray energy to illustrate the general trend as a function of crystal thickness.}
\label{fig:r1}
\end{figure}

The gamma/x-ray total absorption fraction, and photoelectric absorption fraction on the first interaction for each material as a function of incident energy and crystal thickness can be seen in Fig. \ref{fig:r1}. For all four materials the gamma/x-ray total absorption fraction for a given incident energy increases as a function of material thickness and, with the exception of the 28 keV profiles, the maximum value of each profile scales with incident energy. This lower than expected gamma/x-ray total absorption fraction at 28 keV in all four materials can be attributed to the non-negligible interaction cross-section of the ESR and glue layer at the front surface of each scintillator crystal at this energy. Similarly for each material, the photoelectric absorption fraction on the first interaction for a given incident energy increases as a function of material thickness and, with the exception of the 28 keV profile in CsI(Tl), the maximum value of each profile scales with its energy. In this case, the lower than expected photoelectric absorption fraction on the first interaction at 28 keV in CsI(Tl) can be attributed to the interplay between two factors: 1) the non-negligible interaction cross-section of 28 keV gamma/x-rays in the ESR and glue layer at the front surface of each scintillator crystal, and 2) an over 95\% contribution of photoelectric absorption towards the total interaction cross-section of CsI(Tl) at 72 keV for crystal thickness of greater than 3 mm \cite{xcom2019}. Finally, the ranking from maximum to minimum value of the four materials with both these FoMs corresponds directly to the total and relative photoelectric cross-section of each material (e.g. GAGG(Ce), CsI(Tl), NaI(Tl) and LaBr$_{3}$(Ce)).

\begin{figure}[tbh]    
    \centering
    \begin{subfigure}
    \centering 
        \includegraphics[width=0.425\textwidth, trim = {0 0 25 10},clip]{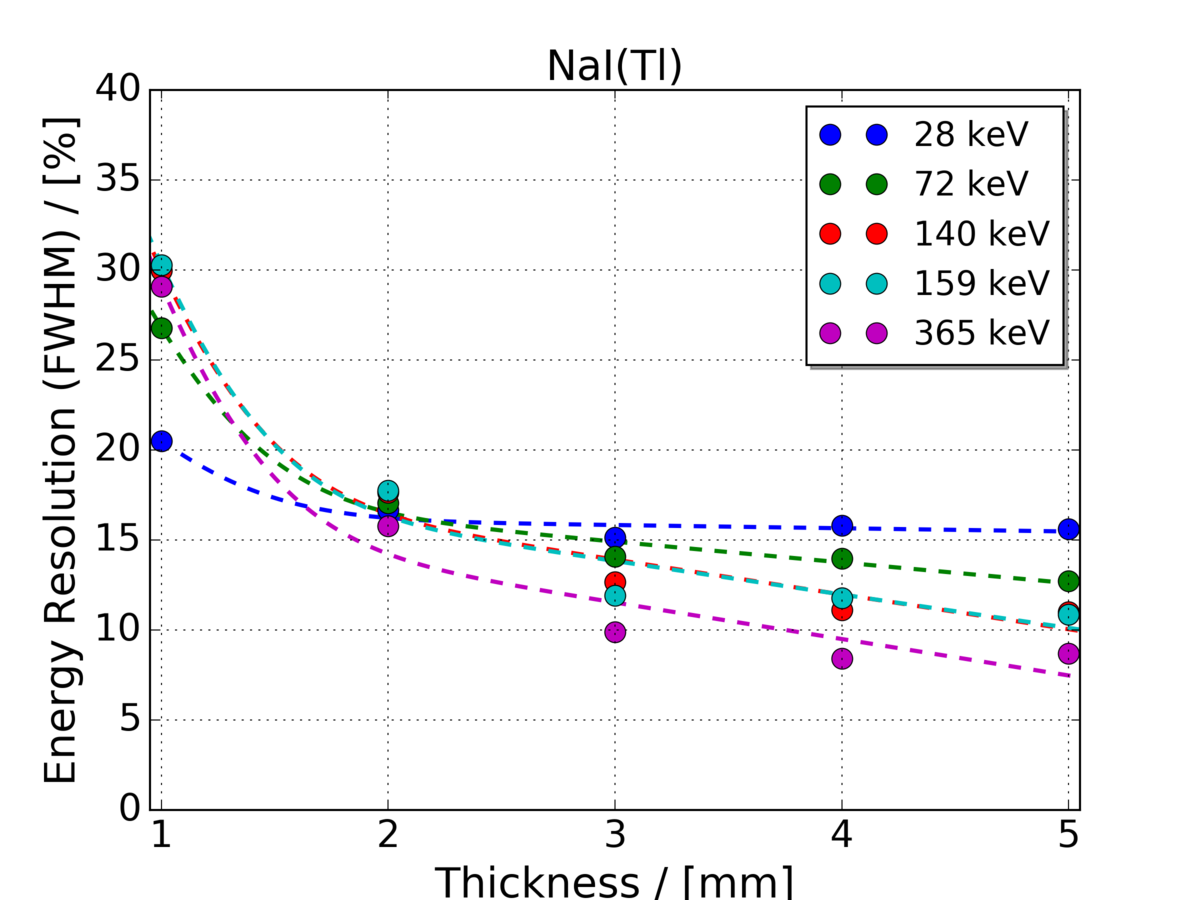}
        \label{fig:r2a}
    \end{subfigure}
    \begin{subfigure}
    \centering
        \includegraphics[width=0.425\textwidth, trim = {0 0 25 10},clip]{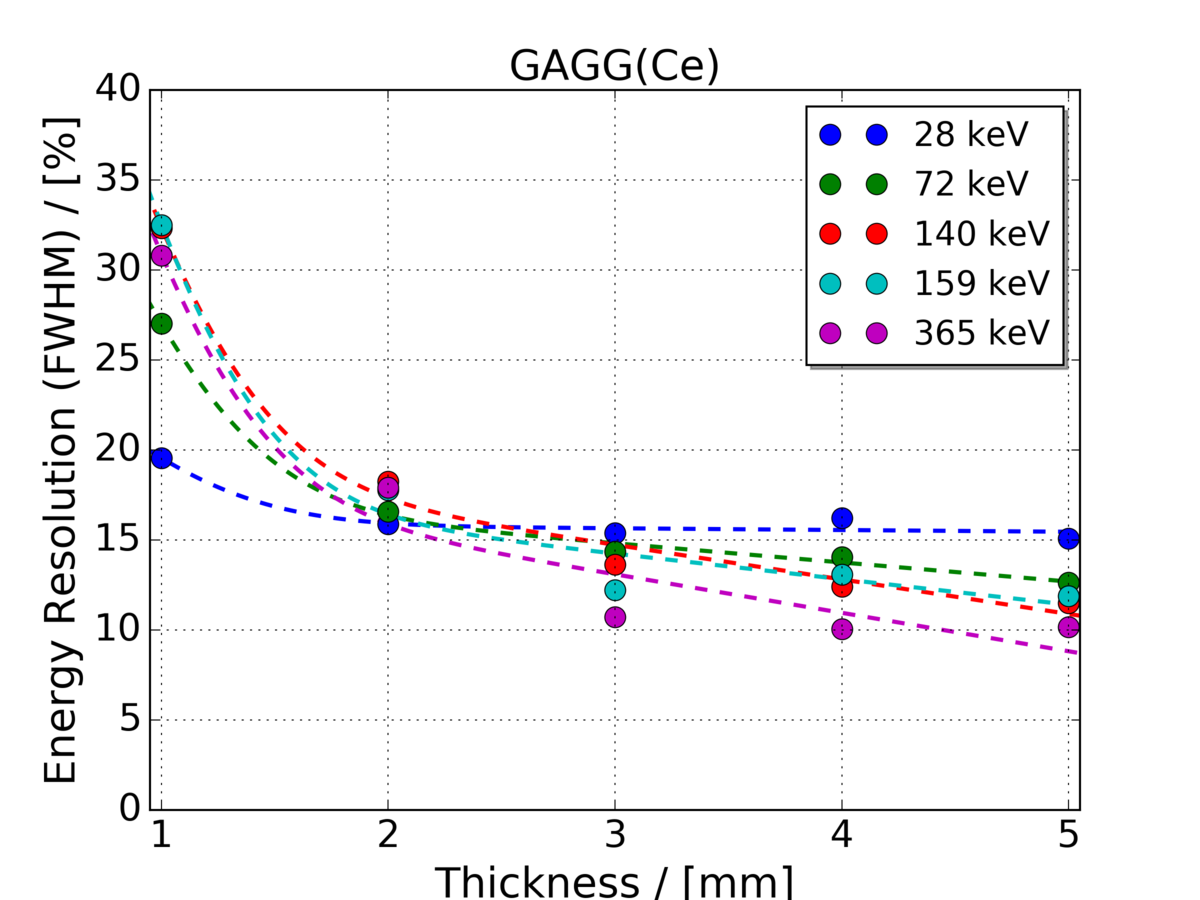}
        \label{fig:r2b}
    \end{subfigure}

    \begin{subfigure}
    \centering 
        \includegraphics[width=0.425\textwidth, trim = {0 0 25 10},clip]{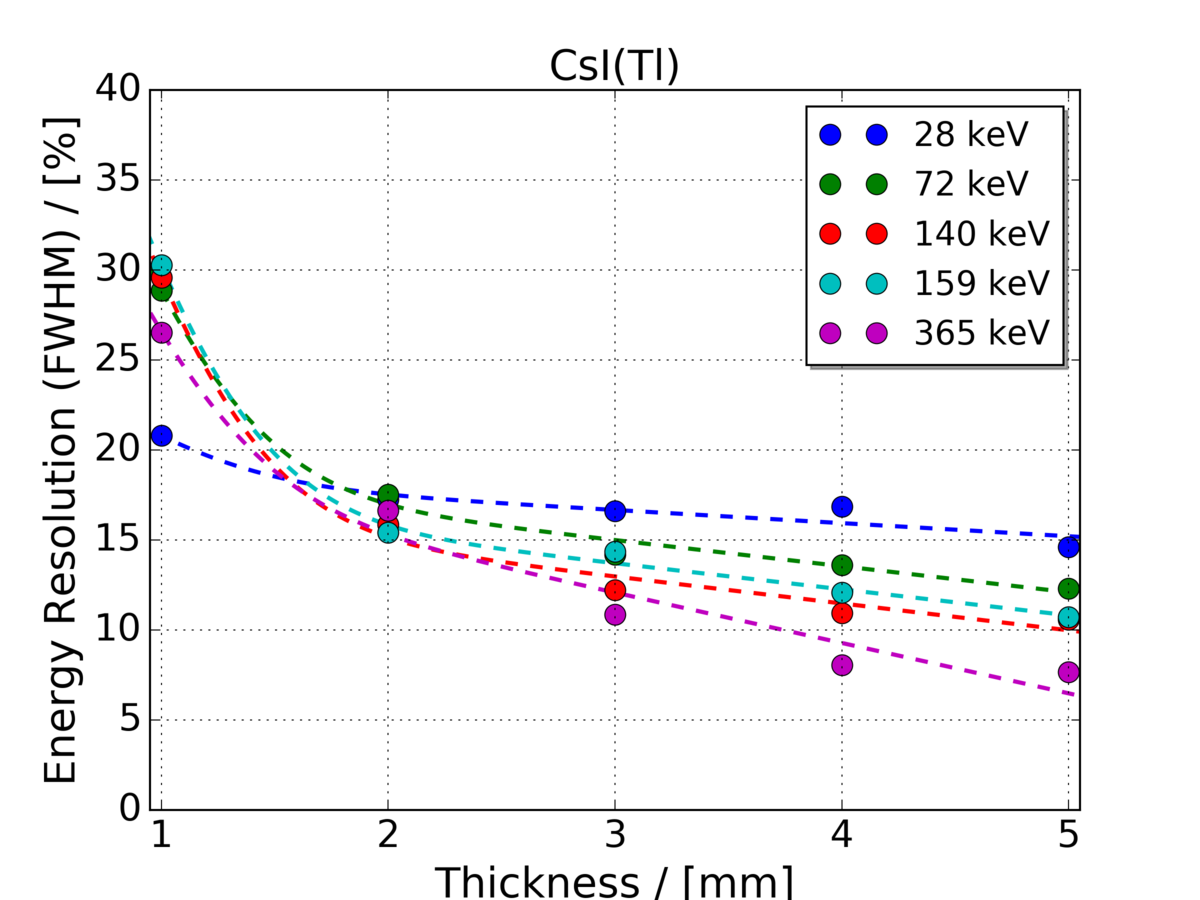}
        \label{fig:r2c}
    \end{subfigure}
    \begin{subfigure}
    \centering
        \includegraphics[width=0.425\textwidth, trim = {0 0 25 10},clip]{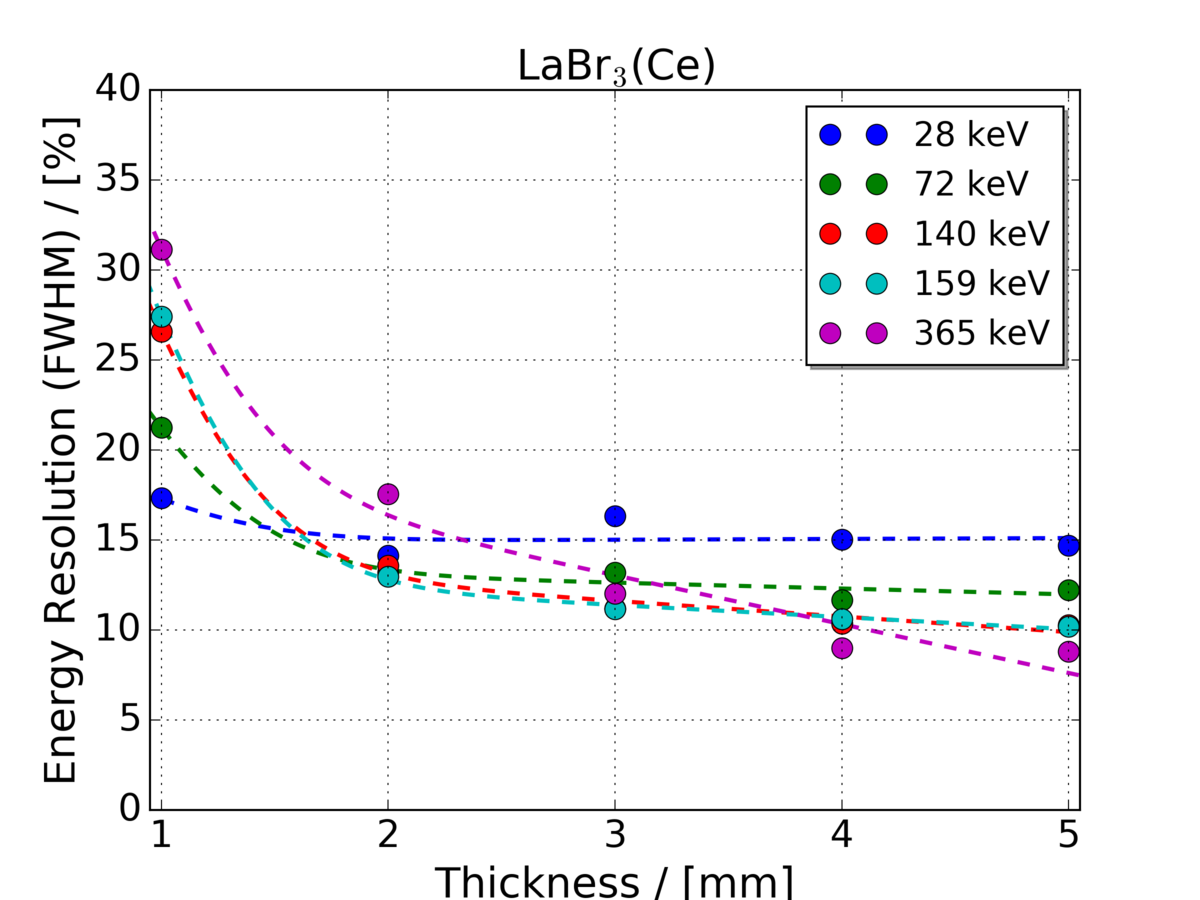}
        \label{fig:r2d}
    \end{subfigure}
    
\caption{Energy resolution (FWHM) of the four different scintillator crystal materials, NaI(Tl), GAGG(Ce), CsI(Tl) and LaBr$_{3}$(Ce), as a function of incident gamma/x-ray energy and crystal thickness. The coloured dash lines correspond to a fitted polynomial surrogate function for each incident gamma/x-ray energy to illustrate the general trend as a function of crystal thickness.}
\label{fig:r2}
\end{figure}

Figure \ref{fig:r2} presents the photopeak energy resolution (\% FWHM) of the four different scintillator crystal materials as a function of incident gamma/x-ray energy and material thickness. All four materials display a direct relationship between material thickness and energy resolution for the five explored gamma/x-ray energies. These material thickness profiles highlight that a minimum crystal thickness of 3 mm is required to ensure a photopeak energy resolution of approximately 15 \% for all four material regardless of incident gamma/x-ray energy. Below this 3 mm threshold there is a higher probability of material specific fluorescence x-ray escape after the photoelectric absorption of gamma/x-rays that distorts the shape and increases the width of photopeaks in measured energy spectra \cite{Gilmore2011}. Furthermore, this physical process of fluorescence x-ray escape is responsible for the distorted, and almost inverse, relationship between incident gamma/x-ray energy and photopeak energy resolution that can be observed for all four materials below 3 mm. Above this 3 mm threshold a direct relationship for all four materials between incident gamma/x-ray energy and photopeak energy resolution can be observed, with some interplay between the 140 and 159 keV profiles, which improves on the order of 0.5 to 1 \% FWHM with each additional mm increase in crystal thickness.

From these data shown in Fig. \ref{fig:r2} it can be seen that LaBr$_{3}$(Ce) possesses the best energy resolution performance on average for all tested gamma/x-ray energies, followed closely by CsI(Tl), NaI(Tl), and then GAGG(Ce). Here, the lower than expected performance of the GAGG(Ce) with respect to NaI(Tl), based on their optical photon yields per MeV (see Table \ref{tab:2}) and SiPM PDEs outlined in \cite{Frach2009}, can be attributed to the fact that GAGG(Ce) has a high level of self-absorption for its emitted optical scintillation photons resulting in a net loss in those which propagate the full distance from their emission site to the SiPM. Furthermore in the case of LaBr$_{3}$(Ce), its performance is also lower than expected based on its optical photon yield per MeV alone due to the Philips SiPM possessing a low effective PDE with respect to its optical emission spectrum \cite{Frach2009}.

\begin{figure}[tbh]   
   \centering
   \includegraphics[width=0.425\textwidth,trim = {0 0 25 20},clip]{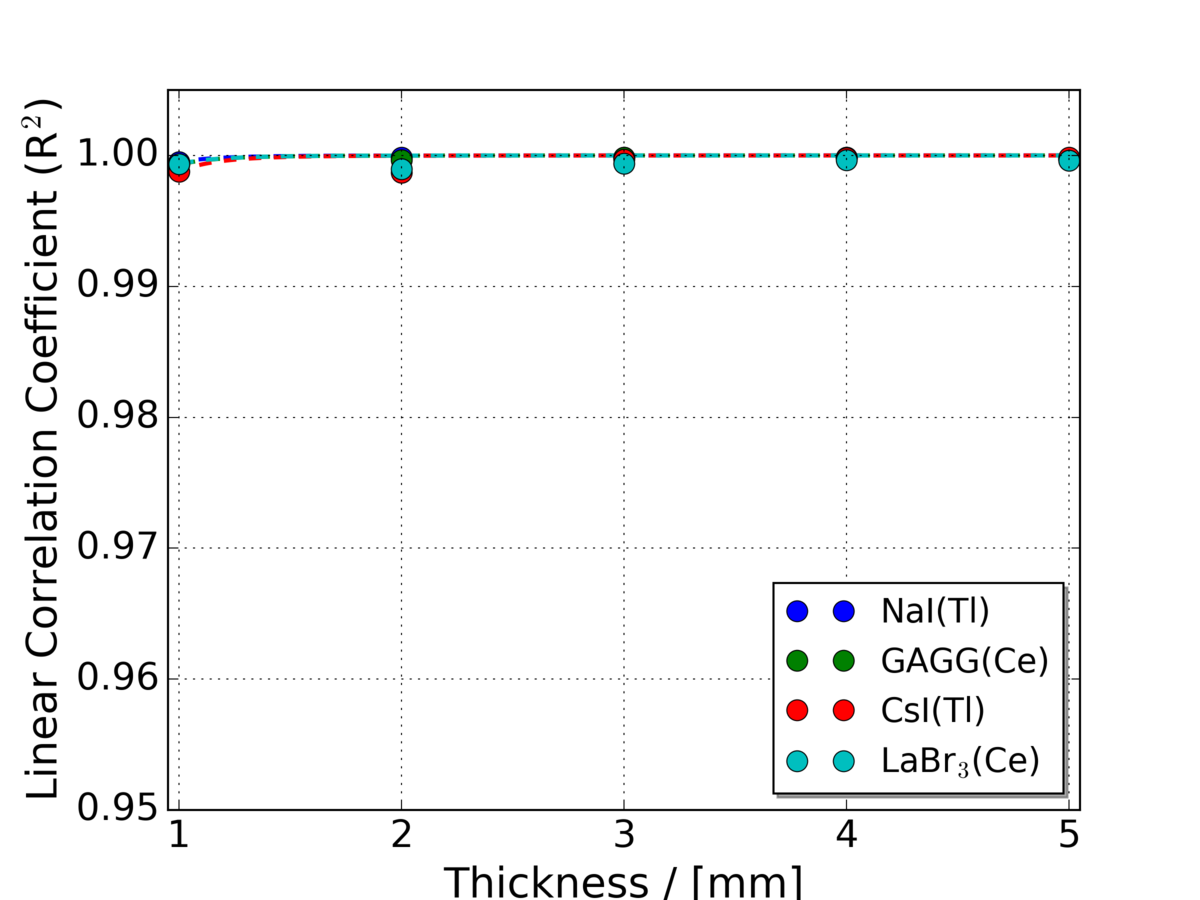}
\caption{Energy linearity of the four different scintillator crystal materials, NaI(Tl), GAGG(Ce), CsI(Tl) and LaBr$_{3}$(Ce), as a function of crystal thickness. The coloured dash lines correspond to a fitted polynomial surrogate function for each material to illustrate the general trend as a function of crystal thickness.}
\label{fig:r3}
\end{figure}

Energy linearity of each material for the 5 simulated gamma/x-rays as a function of thickness can be seen in Fig. \ref{fig:r3}. All four materials approach a near perfect energy linearity for material thicknesses above 4 mm, with CsI(Tl) and LaBr$_{3}$(Ce) performing worse than NaI(Tl) and GAGG(Ce) below 4 mm. This lower performance of CsI(Tl) and LaBr$_{3}$(Ce) for thinner crystal thicknesses can
be attributed to their higher probability of fluorescence x-ray escape after the photoelectric absorption of gamma/x-rays that distorts the shape and estimated centroid position of photopeaks in measured energy spectra \cite{Gilmore2011}. However it should be stated that all four materials across the range of explored material thicknesses possessed a $R^{2}$ of over 0.998, indicating a very high level of energy linearity.

\begin{figure}[tbh]    
    \centering
    \begin{subfigure}
    \centering 
        \includegraphics[width=0.425\textwidth, trim = {0 0 25 10},clip]{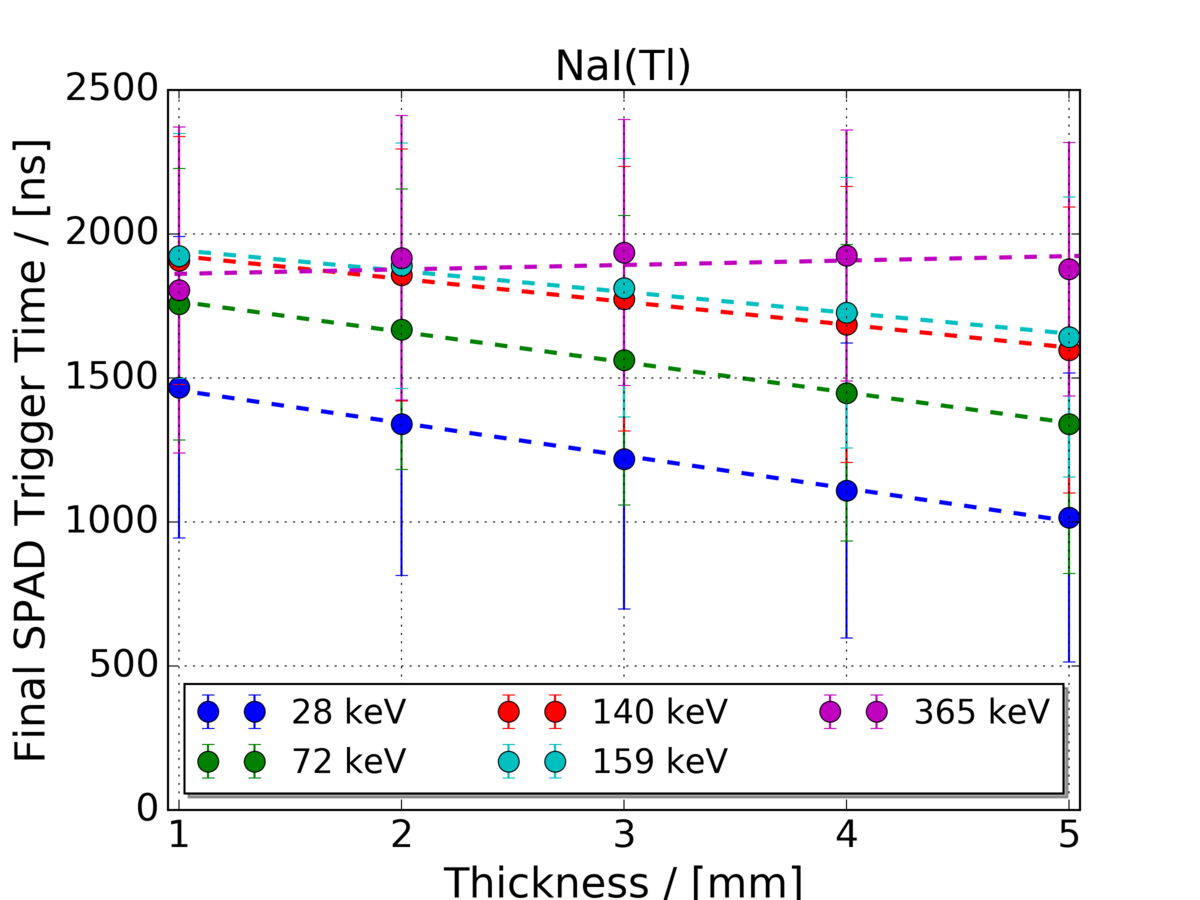}
        \label{fig:r4a}
    \end{subfigure}
    \begin{subfigure}
    \centering
        \includegraphics[width=0.425\textwidth, trim = {0 0 25 10},clip]{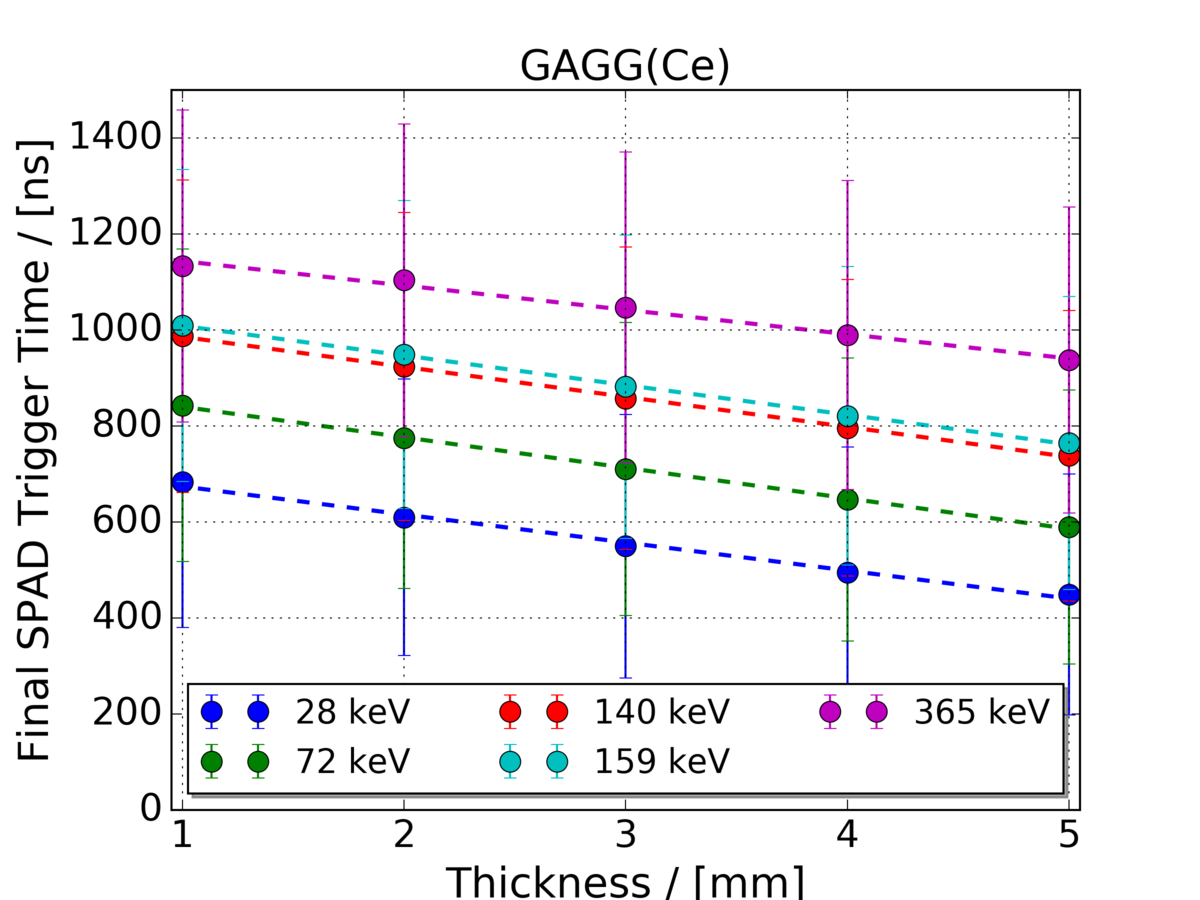}
        \label{fig:r4b}
    \end{subfigure}

    \begin{subfigure}
    \centering 
        \includegraphics[width=0.425\textwidth, trim = {0 0 25 10},clip]{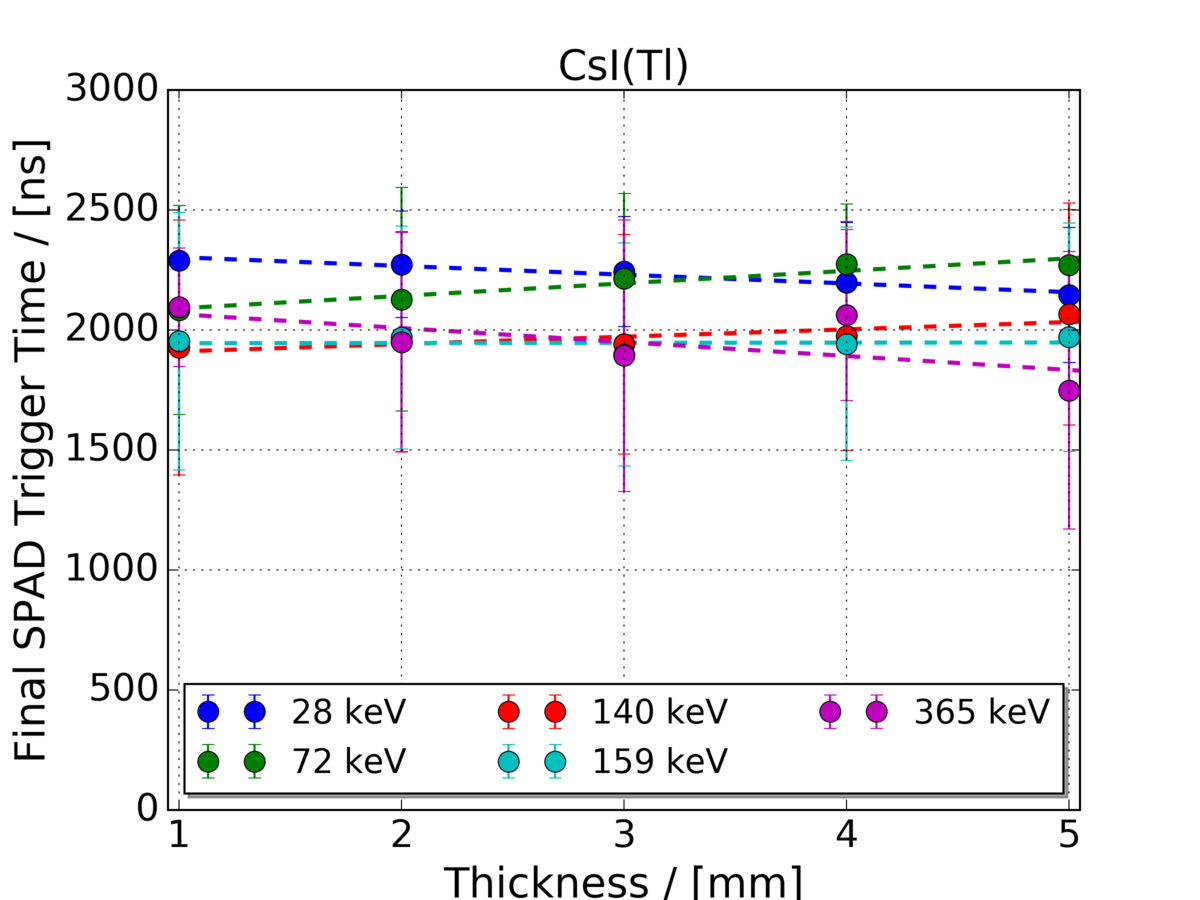}
        \label{fig:r4c}
    \end{subfigure}
    \begin{subfigure}
    \centering
        \includegraphics[width=0.425\textwidth, trim = {0 0 25 10},clip]{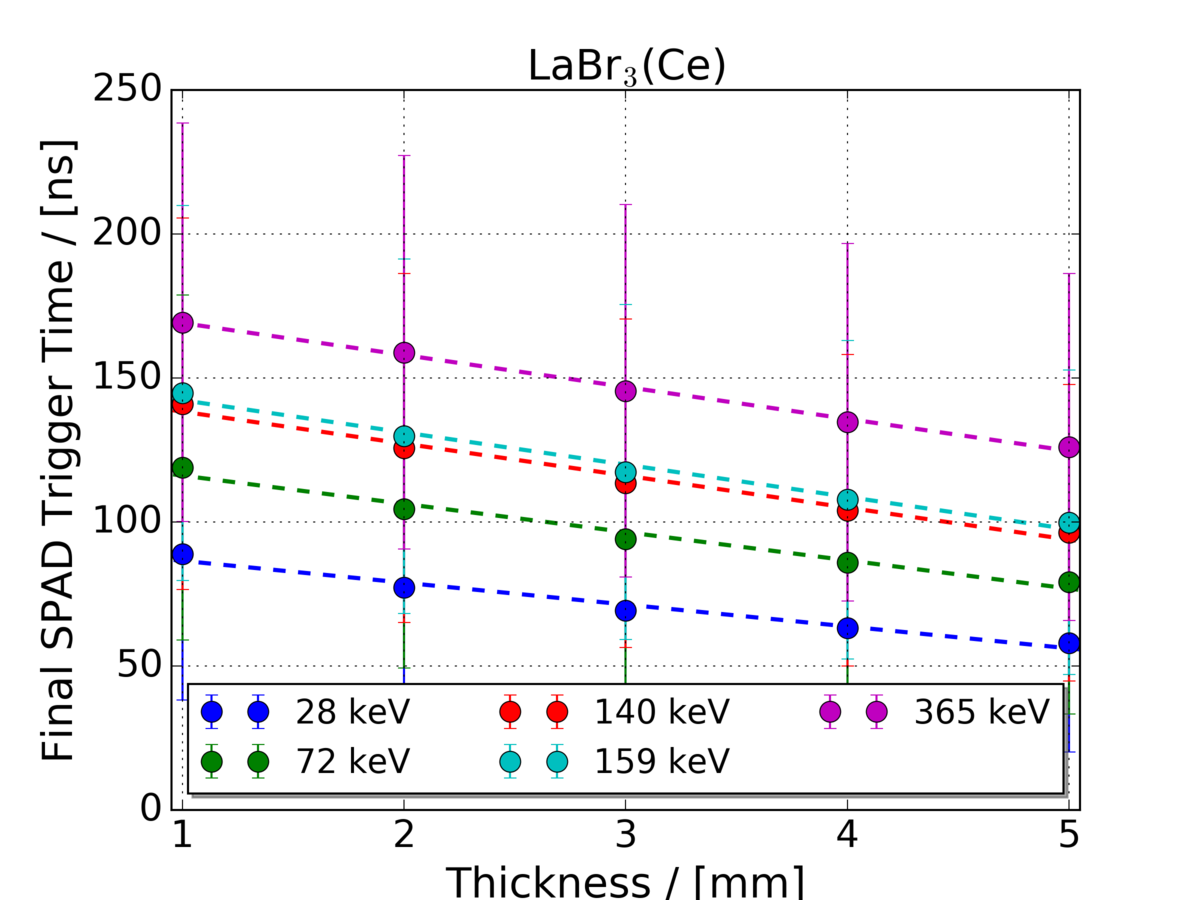}
        \label{fig:r4d}
    \end{subfigure}
\caption{Mean and standard deviations of the relative final SPAD trigger time per gamma/x-ray interaction for the four different scintillator crystal materials, NaI(Tl), GAGG(Ce), CsI(Tl) and LaBr$_{3}$(Ce), as a function of incident gamma/x-ray energy and crystal thickness. The coloured dash lines correspond to a fitted polynomial surrogate function for each incident gamma/x-ray energy to illustrate the general trend as a function of crystal thickness.}
\label{fig:r4}
\end{figure}

The mean and standard deviation of the relative final SPAD trigger times per first gamma/x-ray interaction for each material as a function of incident energy and crystal thickness can be seen in Fig. \ref{fig:r4}. With the exception of CsI(Tl), the three other materials exhibit inverse relationships between both these parameters (incident gamma/x-ray energy and crystal thickness) and the mean relative final SPAD trigger times per first gamma/x-ray interaction. Comparison of all four material data-sets enables for a clear ranking between the four materials from shortest to longest mean relative final SPAD trigger times per gamma/x-ray: LaBr$_{3}$(Ce), GAGG(Ce), NaI(Tl), and CsI(Tl). Finally if it is assumed that the maximum mean SPECT radiation detector count rate before pile up is dependent on scintillation crystal alone, then for all tested energies and thicknesses NaI(Tl), GAGG(Ce), CsI(Tl) and LaBr$_{3}$(Ce) would possess approximate maximum counts per second (cps) rates of 500,000 cps, 830,000 cps, 400,000 cps, and 5,000,000 cps respectively. 

\begin{figure}[p]    
    \centering
    \begin{subfigure}
    \centering 
        \includegraphics[width=0.425\textwidth, trim = {0 0 25 10},clip]{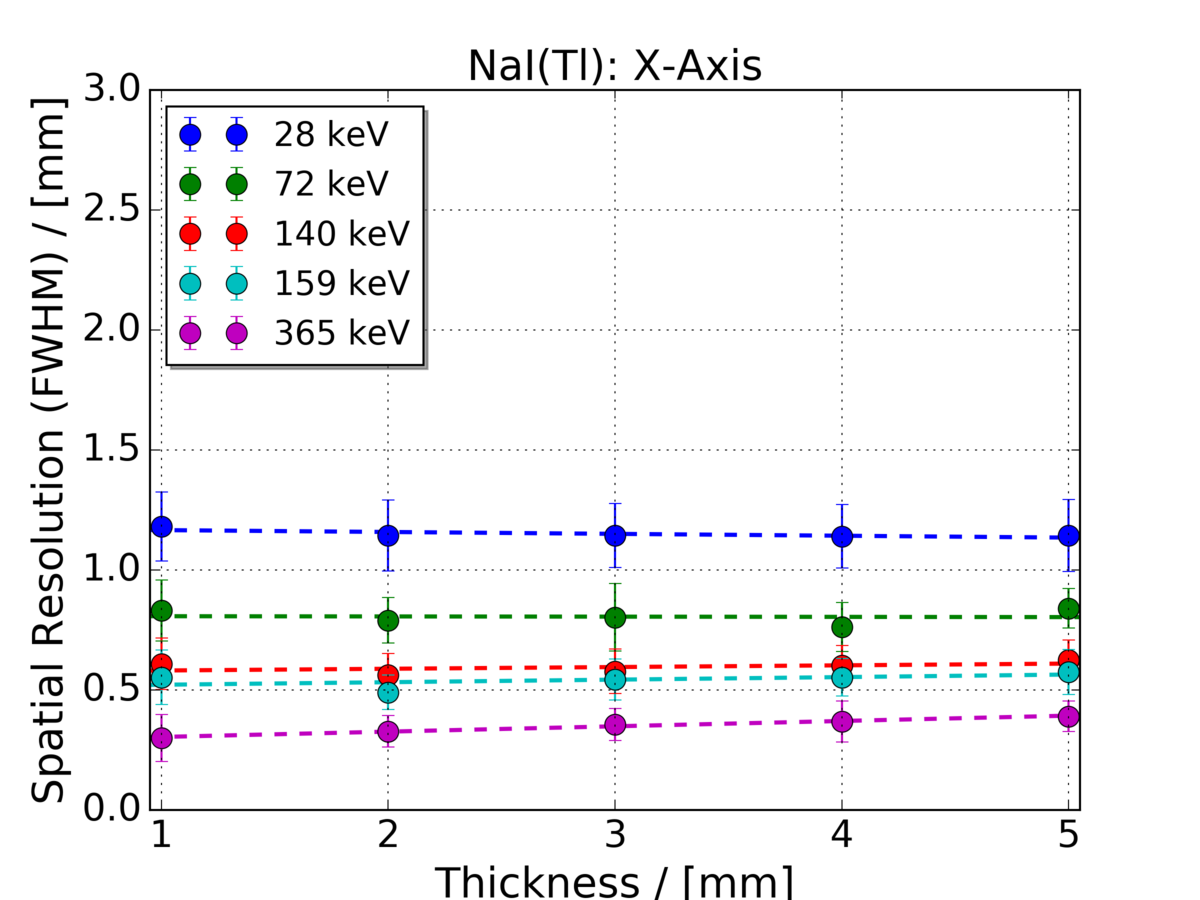}
        \label{fig:r5a}
    \end{subfigure}
    \begin{subfigure}
    \centering
        \includegraphics[width=0.425\textwidth, trim = {0 0 25 10},clip]{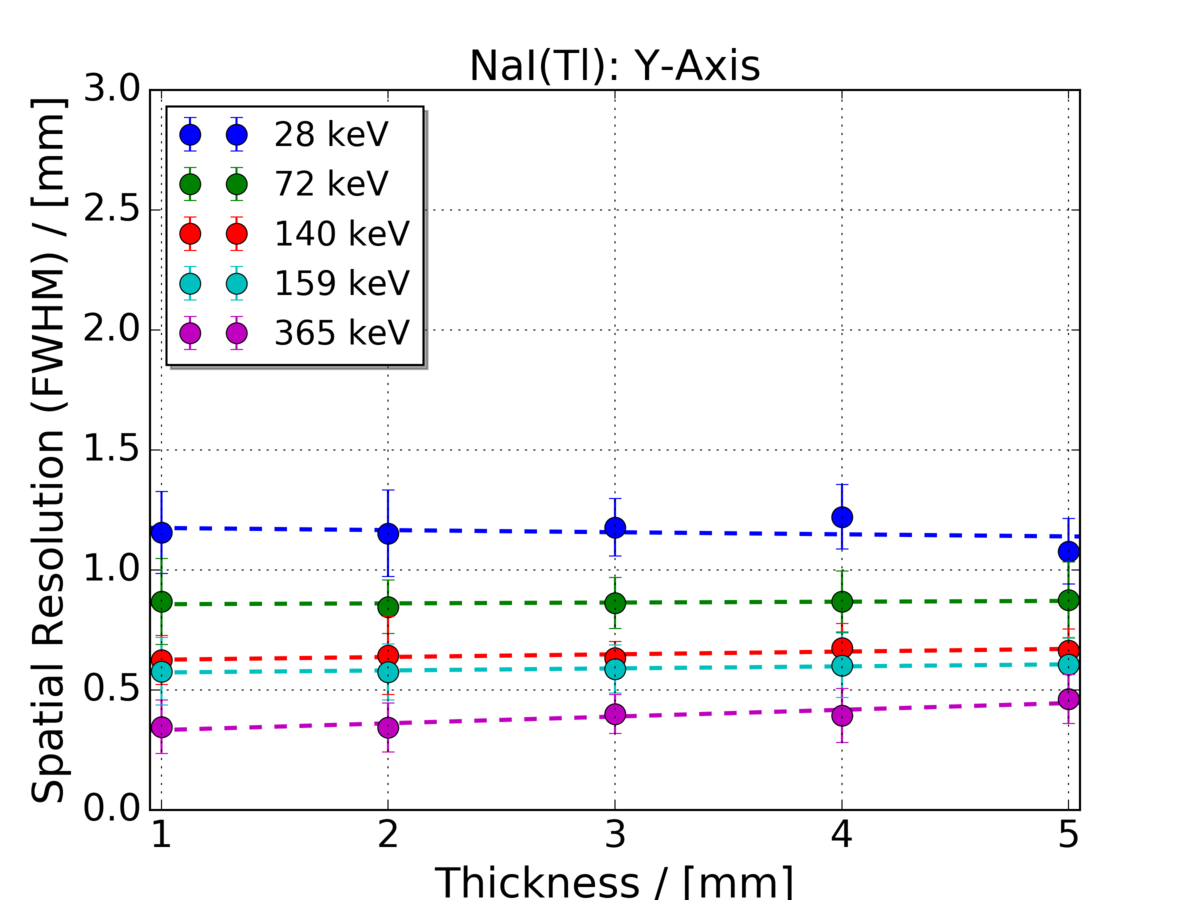}
        \label{fig:r5b}
    \end{subfigure}

    \begin{subfigure}
    \centering 
        \includegraphics[width=0.425\textwidth, trim = {0 0 25 10},clip]{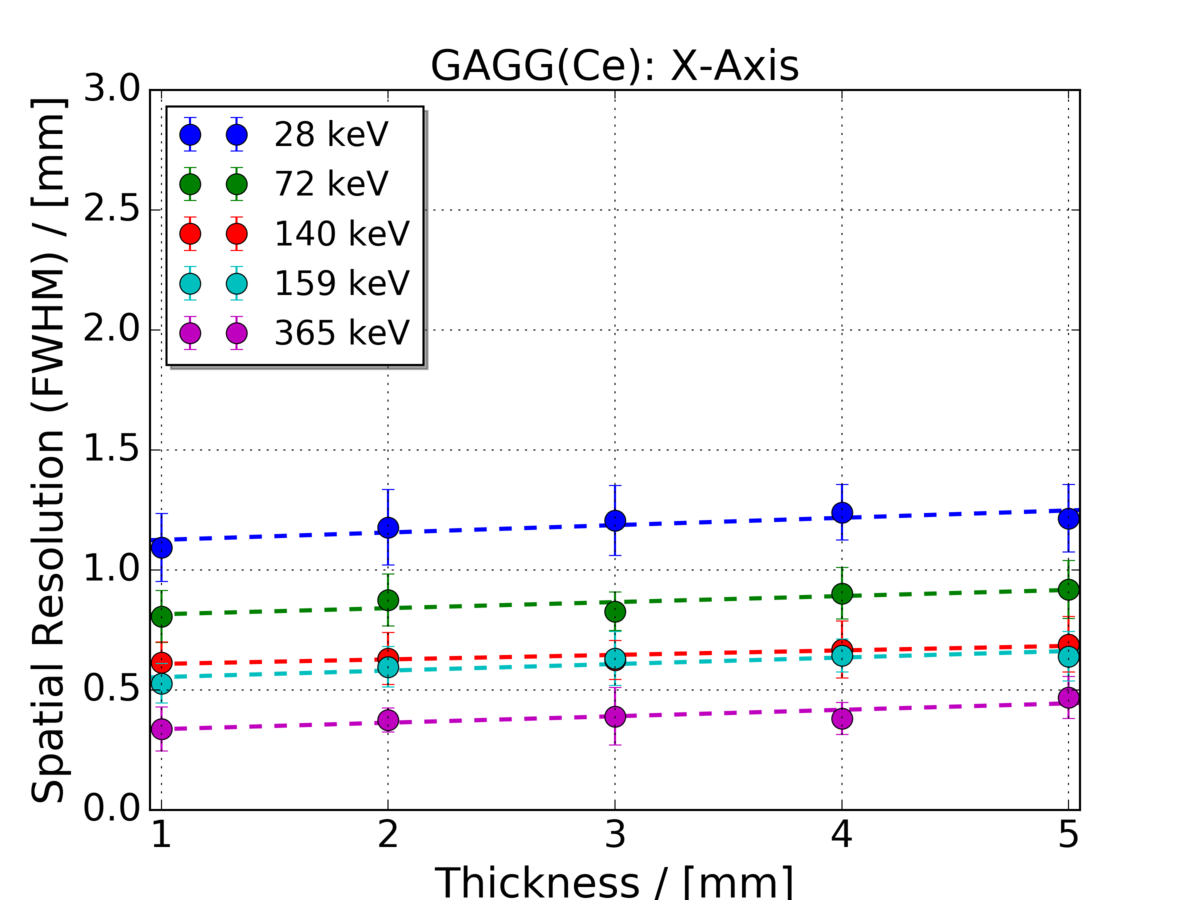}
        \label{fig:r5c}
    \end{subfigure}
    \begin{subfigure}
    \centering
        \includegraphics[width=0.425\textwidth, trim = {0 0 25 10},clip]{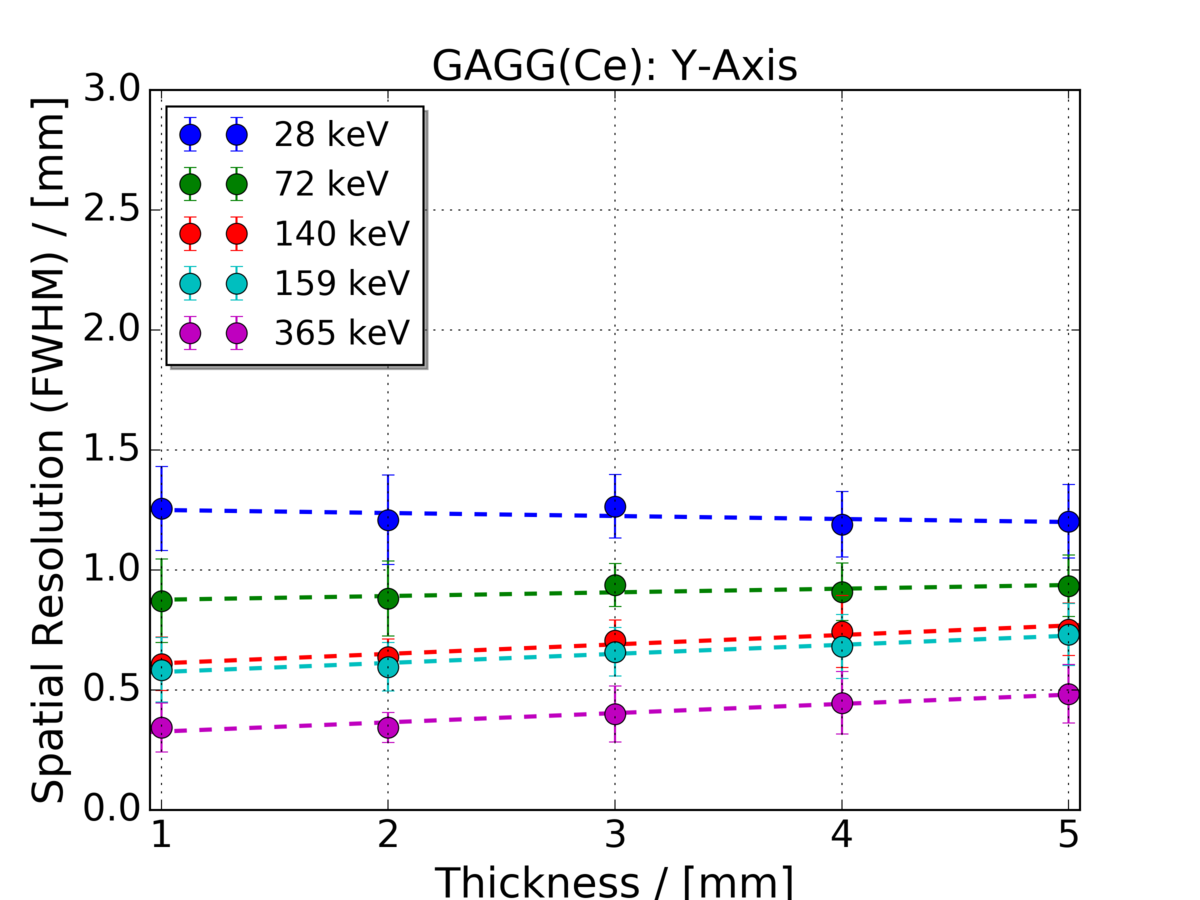}
        \label{fig:r5d}
    \end{subfigure}

    \begin{subfigure}
    \centering 
        \includegraphics[width=0.425\textwidth, trim = {0 0 25 10},clip]{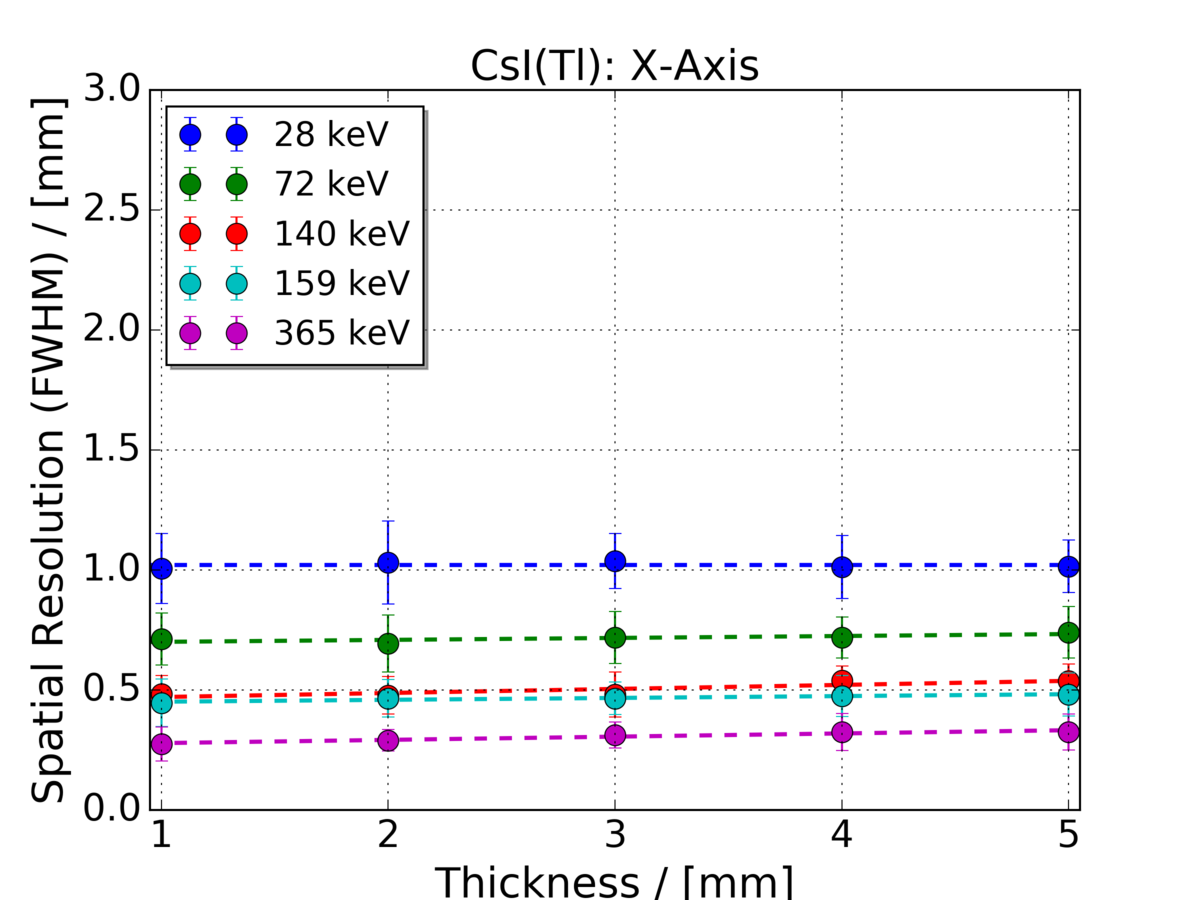}
        \label{fig:r5e}
    \end{subfigure}
    \begin{subfigure}
    \centering
        \includegraphics[width=0.425\textwidth, trim = {0 0 25 10},clip]{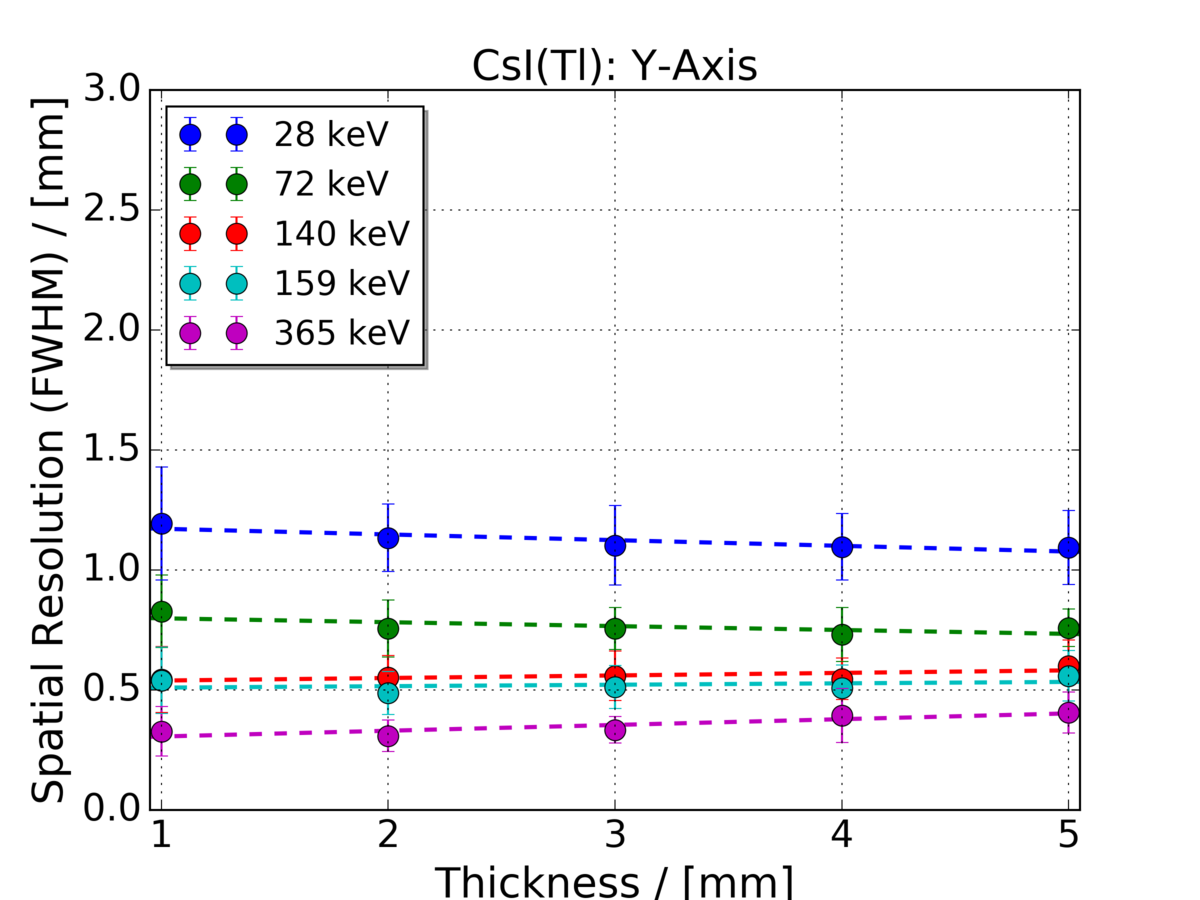}
        \label{fig:r5f}
    \end{subfigure}

    \begin{subfigure}
    \centering 
        \includegraphics[width=0.425\textwidth, trim = {0 0 25 10},clip]{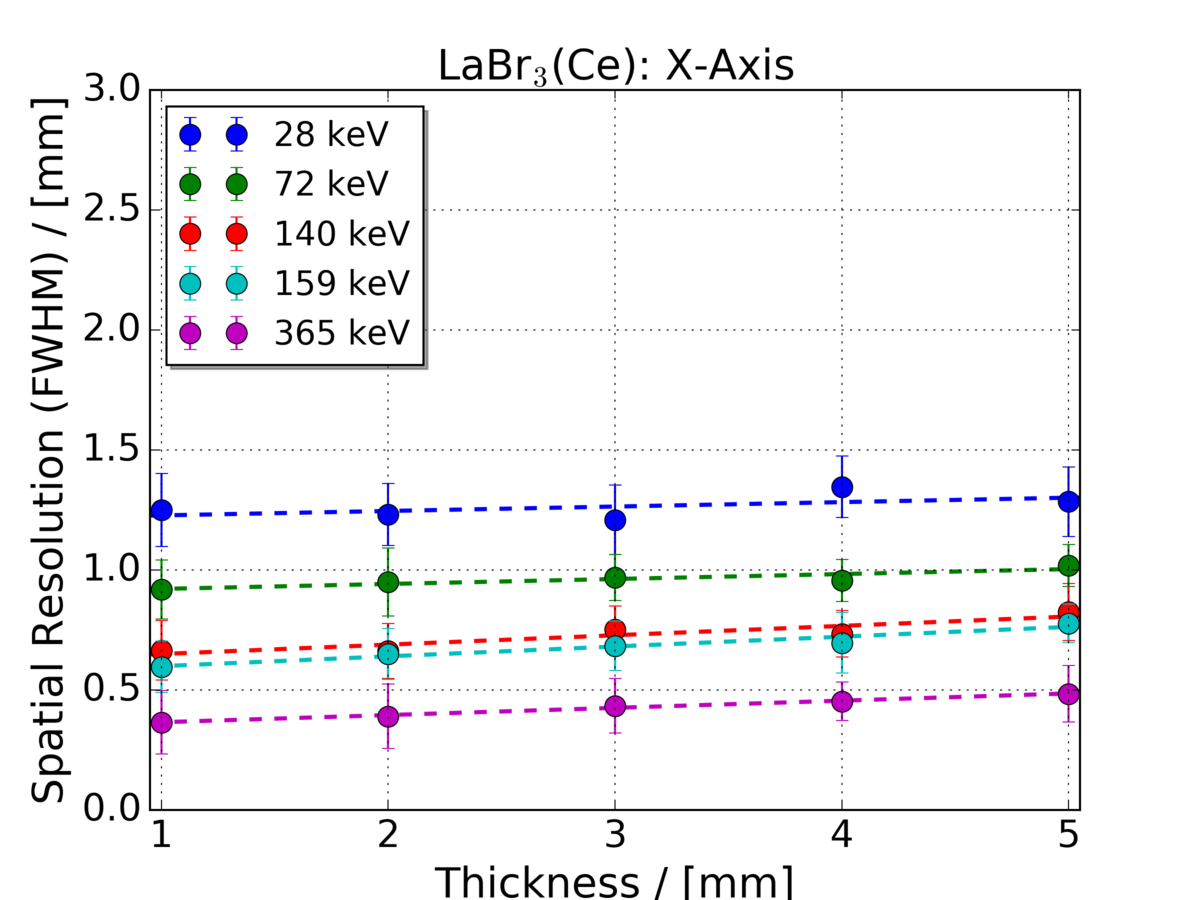}
        \label{fig:r5g}
    \end{subfigure}
    \begin{subfigure}
    \centering
        \includegraphics[width=0.425\textwidth, trim = {0 0 25 10},clip]{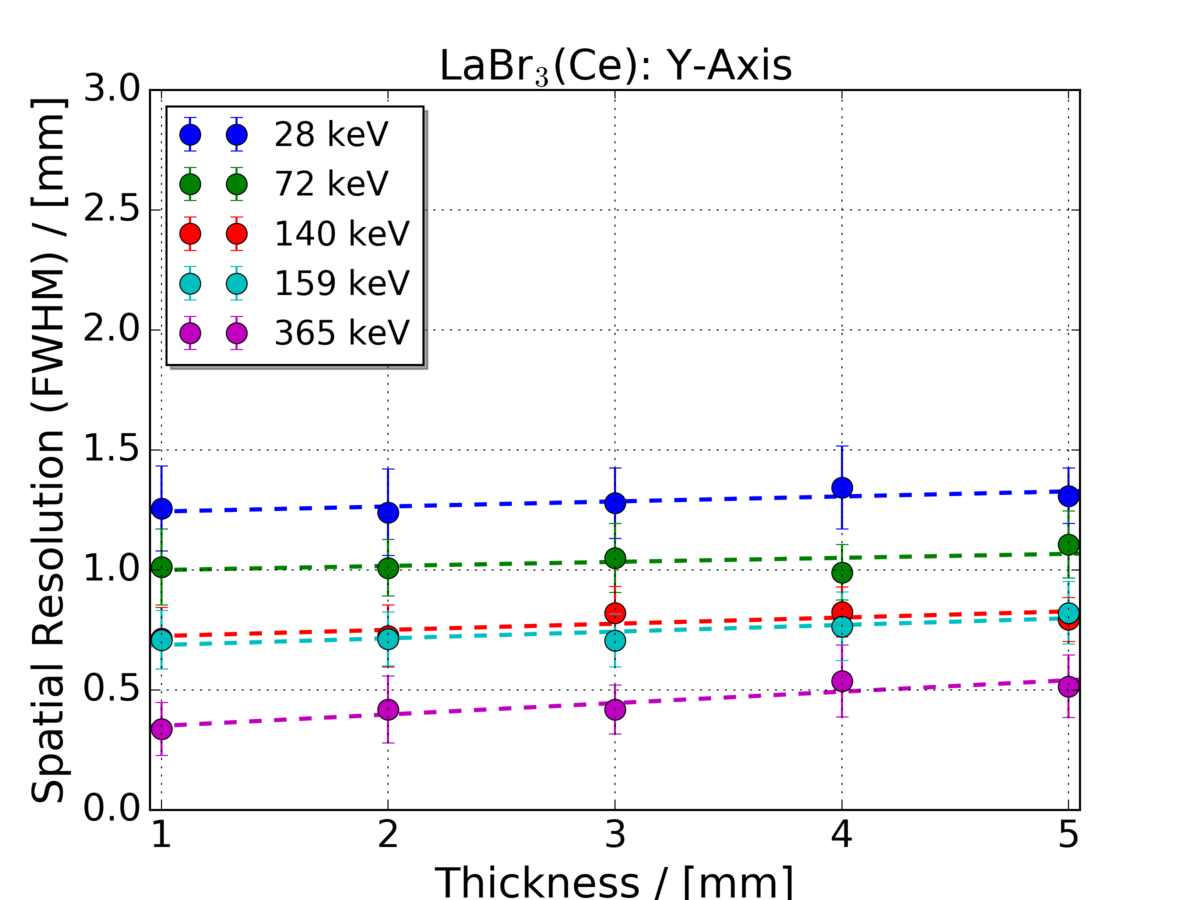}
        \label{fig:r5h}
    \end{subfigure}    
    
\caption{Mean and standard deviations of the irradiation spot $x$- and $y$-axial spatial resolution (FWHM) for the four different scintillator crystal materials, NaI(Tl), GAGG(Ce), CsI(Tl) and LaBr$_{3}$(Ce), as a function of incident gamma/x-ray energy, and crystal thickness for $\alpha=0$ corresponding to no truncation. The coloured dash lines correspond to a fitted polynomial surrogate function for each incident gamma/x-ray energy to illustrate the general trend as a function of crystal thickness.}
\label{fig:r5}
\end{figure}

\begin{figure}[p]    
    \centering
    \begin{subfigure}
    \centering 
        \includegraphics[width=0.425\textwidth, trim = {0 0 25 10},clip]{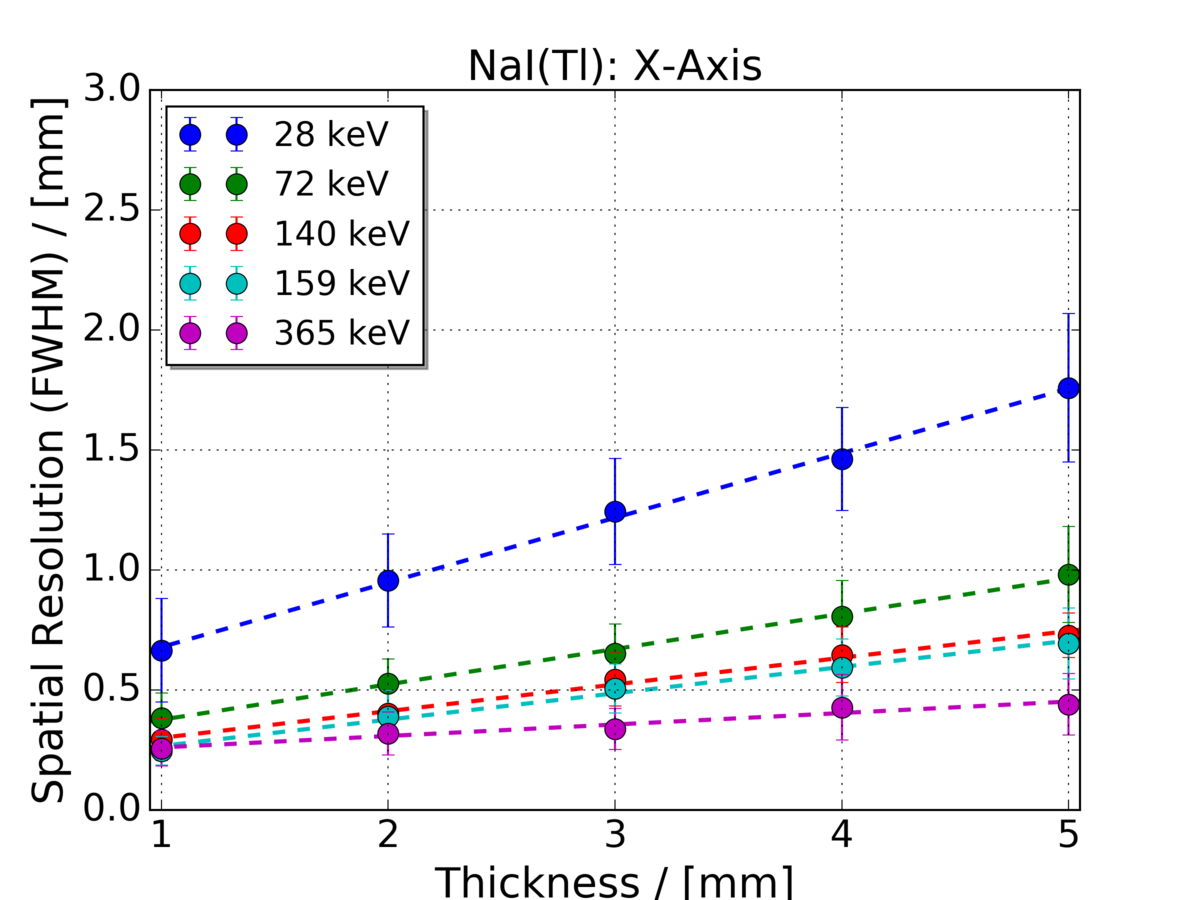}
        \label{fig:r6a}
    \end{subfigure}
    \begin{subfigure}
    \centering
        \includegraphics[width=0.425\textwidth, trim = {0 0 25 10},clip]{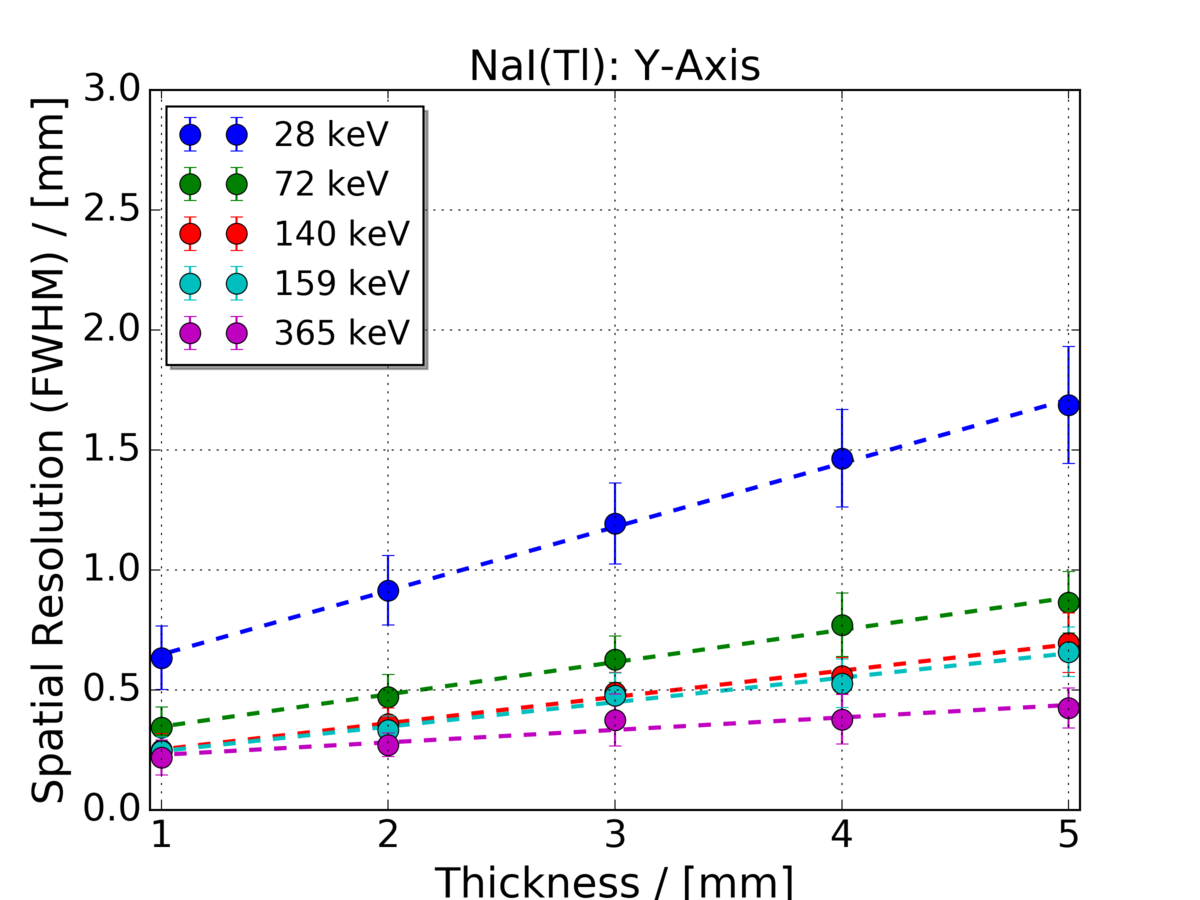}
        \label{fig:r6b}
    \end{subfigure}

    \begin{subfigure}
    \centering 
        \includegraphics[width=0.425\textwidth, trim = {0 0 25 10},clip]{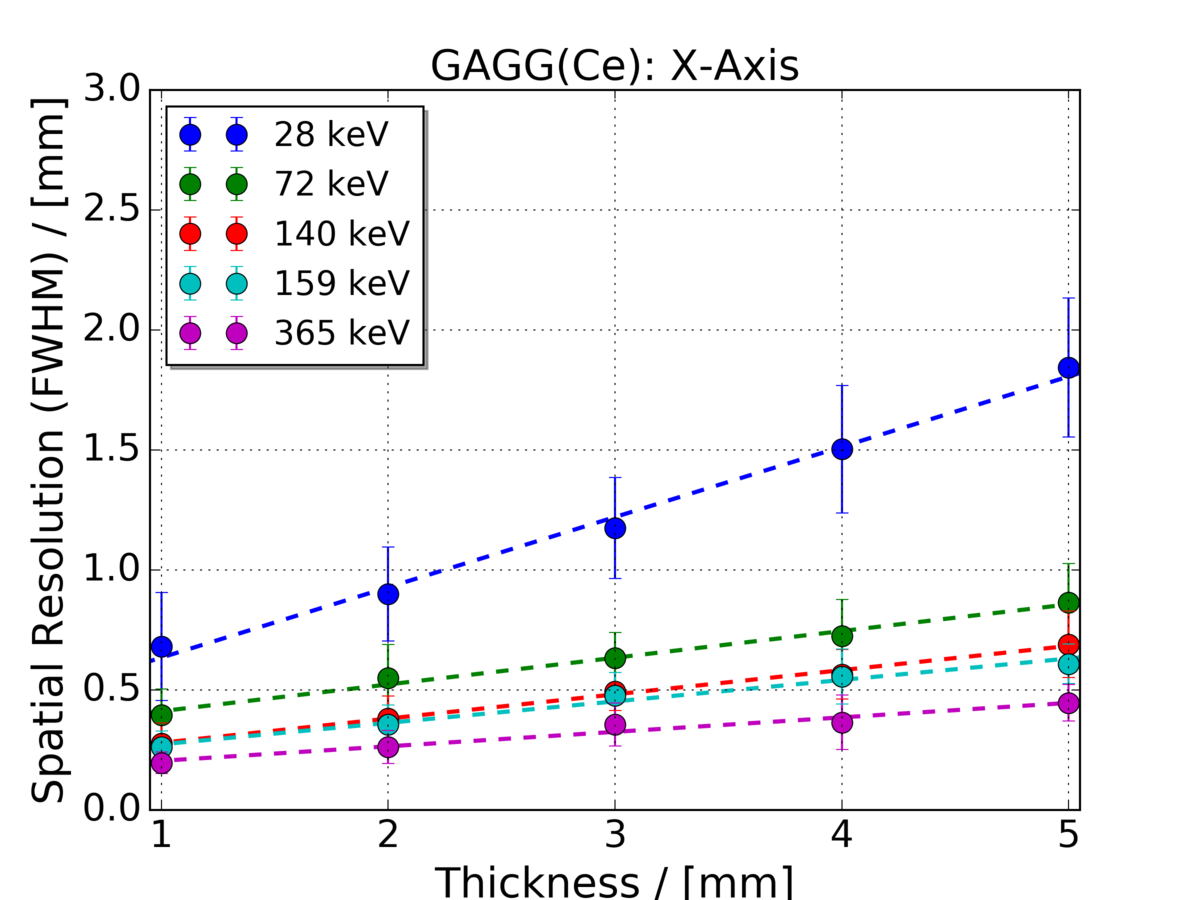}
        \label{fig:r6c}
    \end{subfigure}
    \begin{subfigure}
    \centering
        \includegraphics[width=0.425\textwidth, trim = {0 0 25 10},clip]{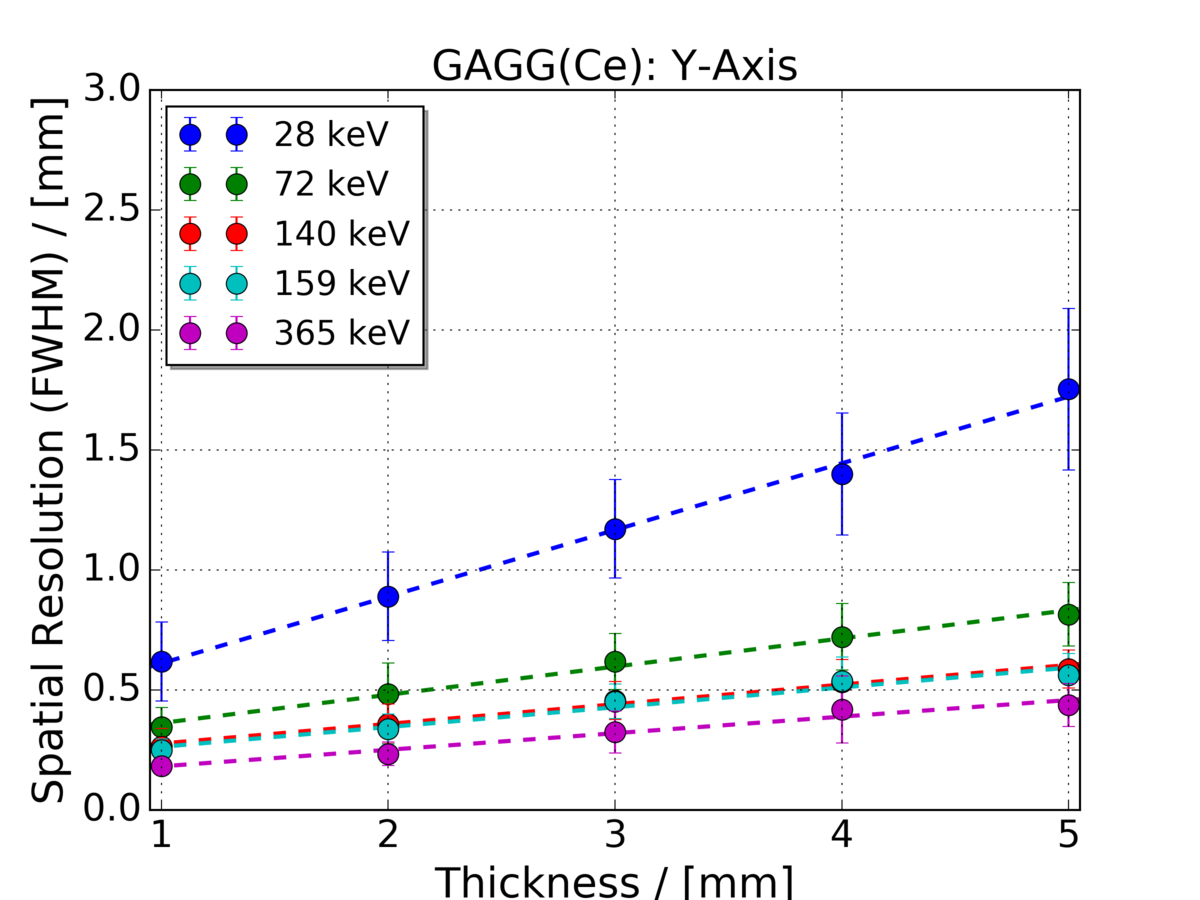}
        \label{fig:r6d}
    \end{subfigure}

    \begin{subfigure}
    \centering 
        \includegraphics[width=0.425\textwidth, trim = {0 0 25 10},clip]{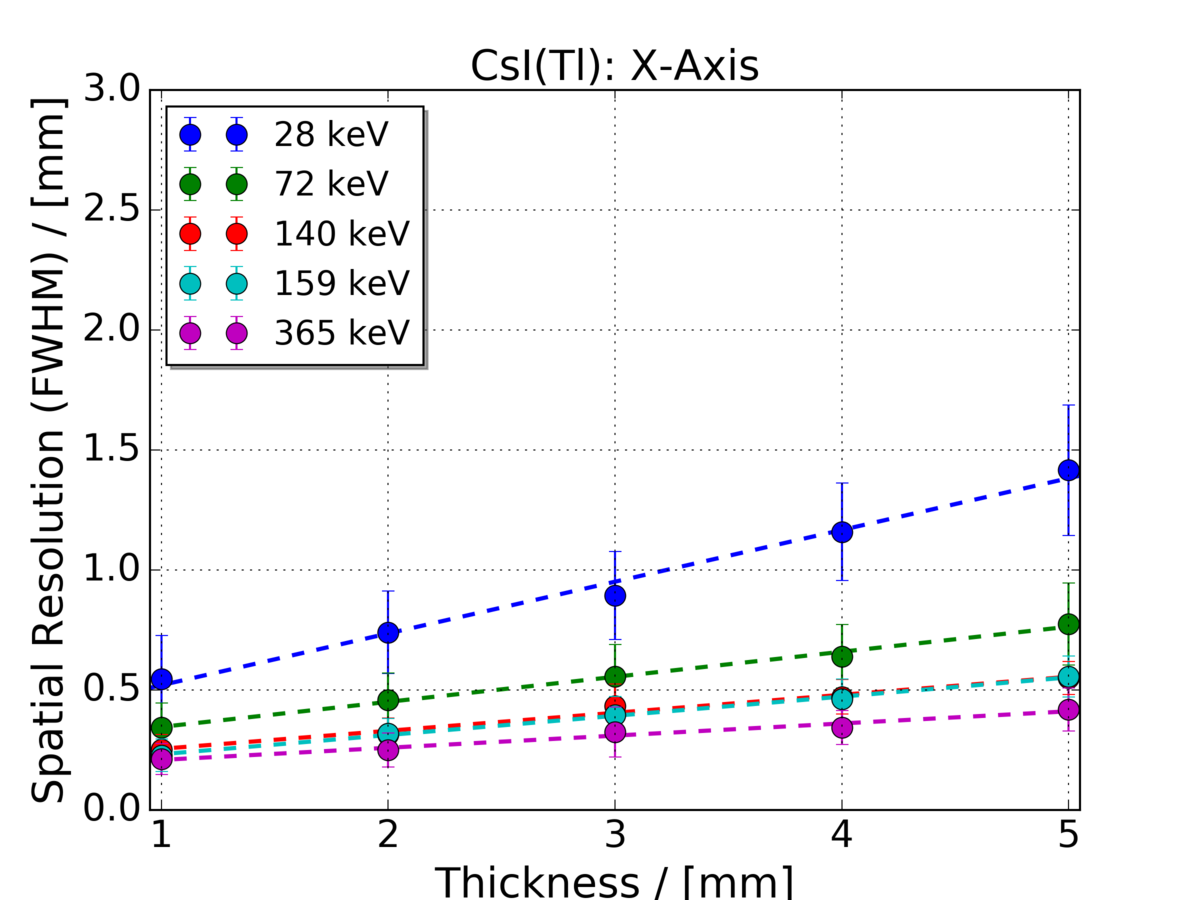}
        \label{fig:r6e}
    \end{subfigure}
    \begin{subfigure}
    \centering
        \includegraphics[width=0.425\textwidth, trim = {0 0 25 10},clip]{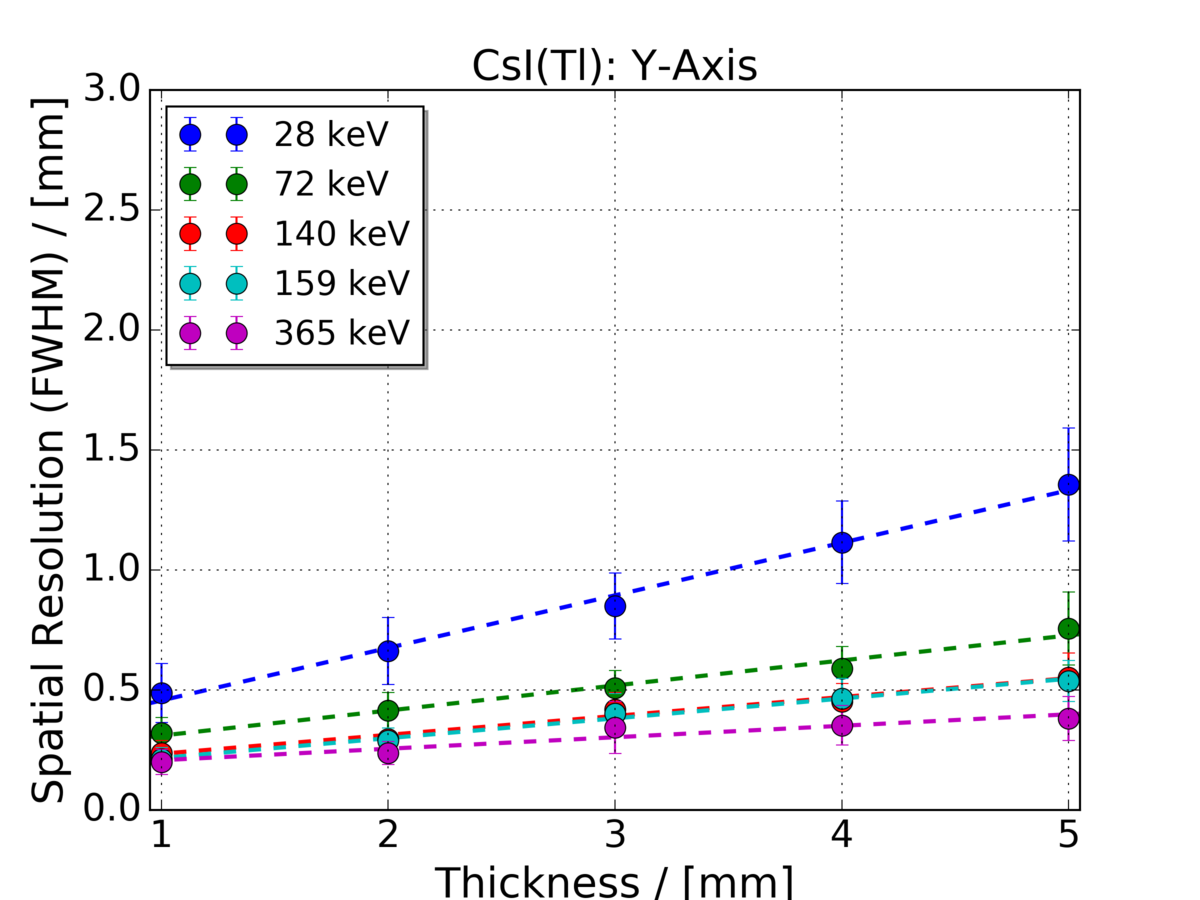}
        \label{fig:r6f}
    \end{subfigure}

    \begin{subfigure}
    \centering 
        \includegraphics[width=0.425\textwidth, trim = {0 0 25 10},clip]{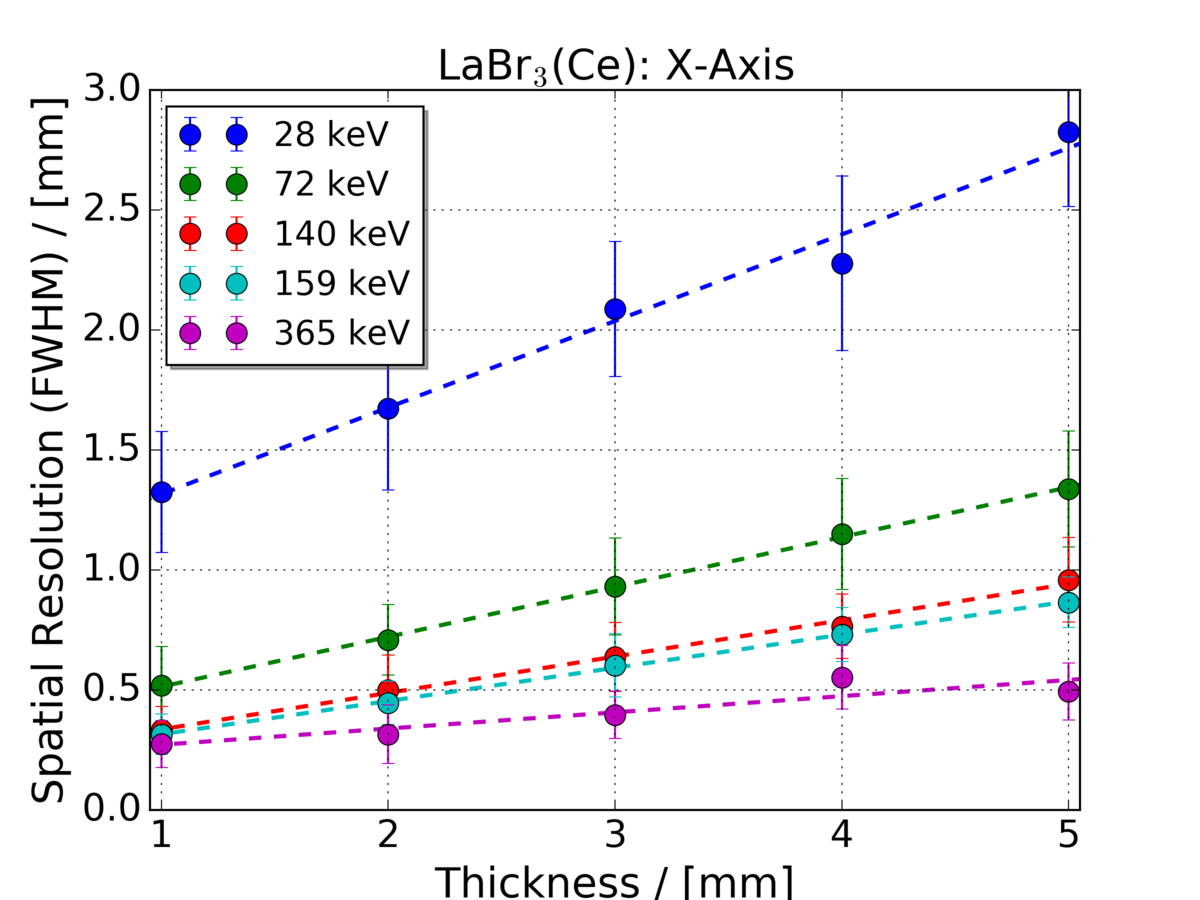}
        \label{fig:r6g}
    \end{subfigure}
    \begin{subfigure}
    \centering
        \includegraphics[width=0.425\textwidth, trim = {0 0 25 10},clip]{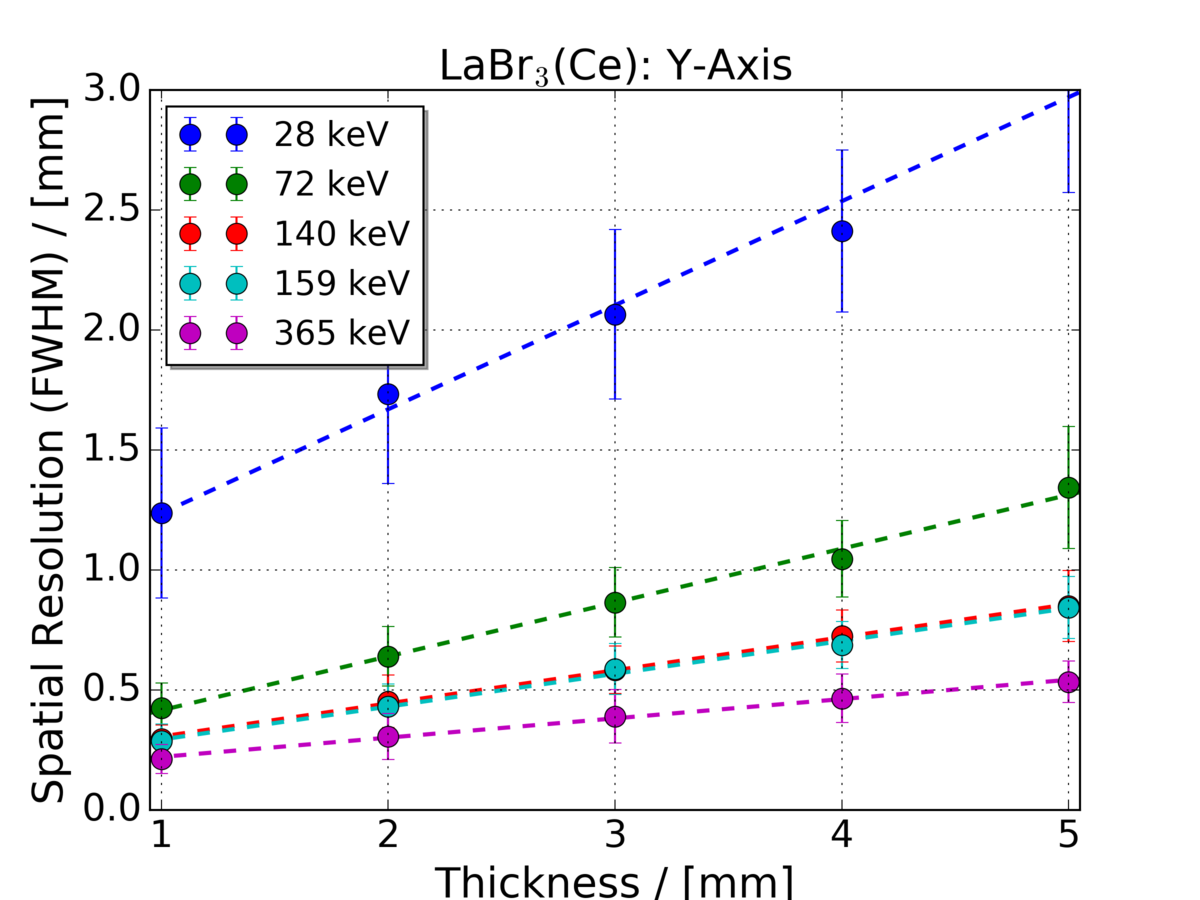}
        \label{fig:r6h}
    \end{subfigure}    
    
\caption{Mean and standard deviations of the irradiation spot $x$- and $y$-axial spatial resolution (FWHM) for the four different scintillator crystal materials, NaI(Tl), GAGG(Ce), CsI(Tl) and LaBr$_{3}$(Ce), as a function of incident gamma/x-ray energy, and crystal thickness for $\alpha = 0.02$ corresponding to the fraction uncertainty of when all active pixel SPADs are triggered. The coloured dash lines correspond to a fitted polynomial surrogate function for each incident gamma/x-ray energy to illustrate the general trend as a function of crystal thickness.}
\label{fig:r6}
\end{figure}

Figures \ref{fig:r5} and \ref{fig:r6} present the mean and standard deviations of the irradiation spot $x$-axial and $y$-axial spatial resolution (FWHM) for all four materials as a function of incident gamma/x-ray energy, crystal thickness, and CoG truncation factor. Here the observed asymmetry between the $x$-axial and $y$-axial data for each given material, incident gamma/x-ray energy, crystal thickness, and CoG truncation factor combination can be attributed to the non-symmetrical structure of the Philips DPC3200 SiPM \cite{DPCManual2016}. For all four materials an inverse relationship can be observed between the incident gamma/x-ray energy and improvement of spot spatial resolution along both axes (i.e. with increasing gamma/x-ray energy the FWHM of each spot decreases along both axes). A similar, but weaker, relationship can be observed for all four materials between the crystal thickness and spot spatial resolution along both axes. However, the impact of CoG truncation factor on spot spatial resolutions for all four materials is more complex, with all 28 keV profiles showing degraded performance with its application for crystal thickness greater than 3 mm. Similarly, all of the NaI(Tl) and LaBr$_{3}$(Ce) gamma/x-ray energy profiles also show degraded performance with the application of a non-zero CoG truncation factor above crystal thicknesses of 3 to 4 mm. Overall Figs. \ref{fig:r5} and \ref{fig:r6} illustrate that the four materials' performance as a factor of incident gamma/x-ray energy, crystal thickness, and CoG truncation factor exhibit similar trends, with CsI(Tl) coming out on top of GAGG(Ce), NaI(Tl), and finally LaBr$_{3}$(Ce) for all tested gamma/x-ray energies. This lower than expected performance for LaBr$_{3}$(Ce) can again be attributed to the effect of the Philips SiPM possessing a low effective PDE with respect to its optical emission spectrum \cite{Frach2009}.

\begin{figure}[p]    
    \centering
    \begin{subfigure}
    \centering 
        \includegraphics[width=0.425\textwidth, trim = {0 0 25 10},clip]{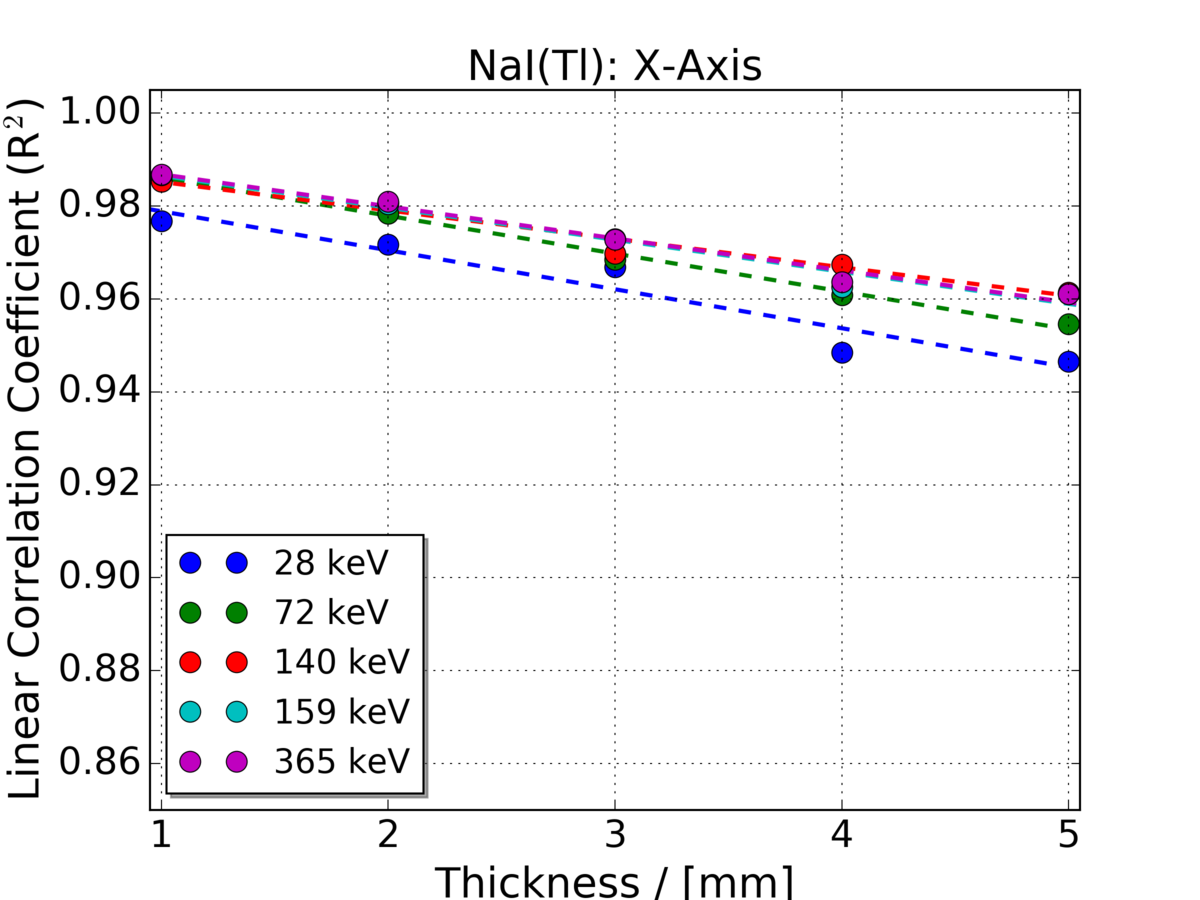}
        \label{fig:r7a}
    \end{subfigure}
    \begin{subfigure}
    \centering
        \includegraphics[width=0.425\textwidth, trim = {0 0 25 10},clip]{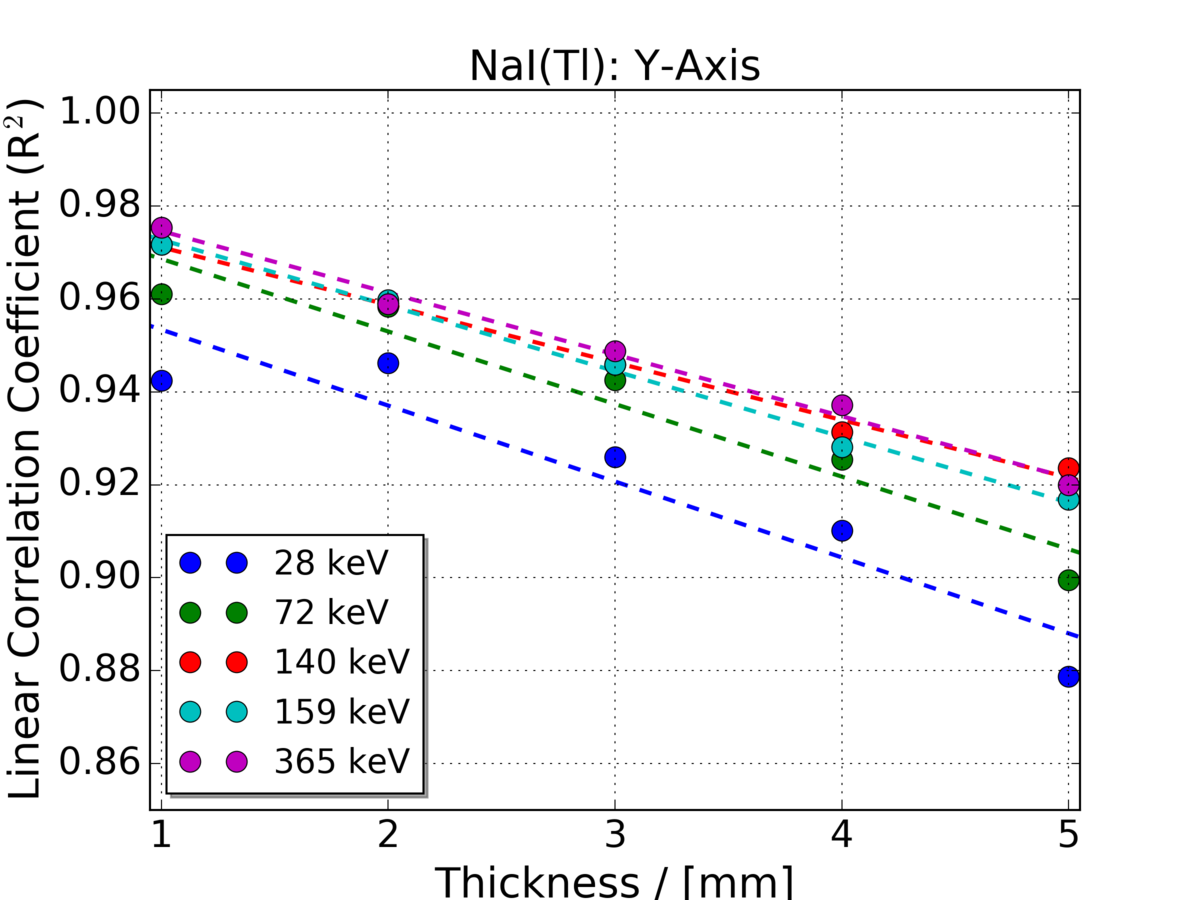}
        \label{fig:r7b}
    \end{subfigure}

    \begin{subfigure}
    \centering 
        \includegraphics[width=0.425\textwidth, trim = {0 0 25 10},clip]{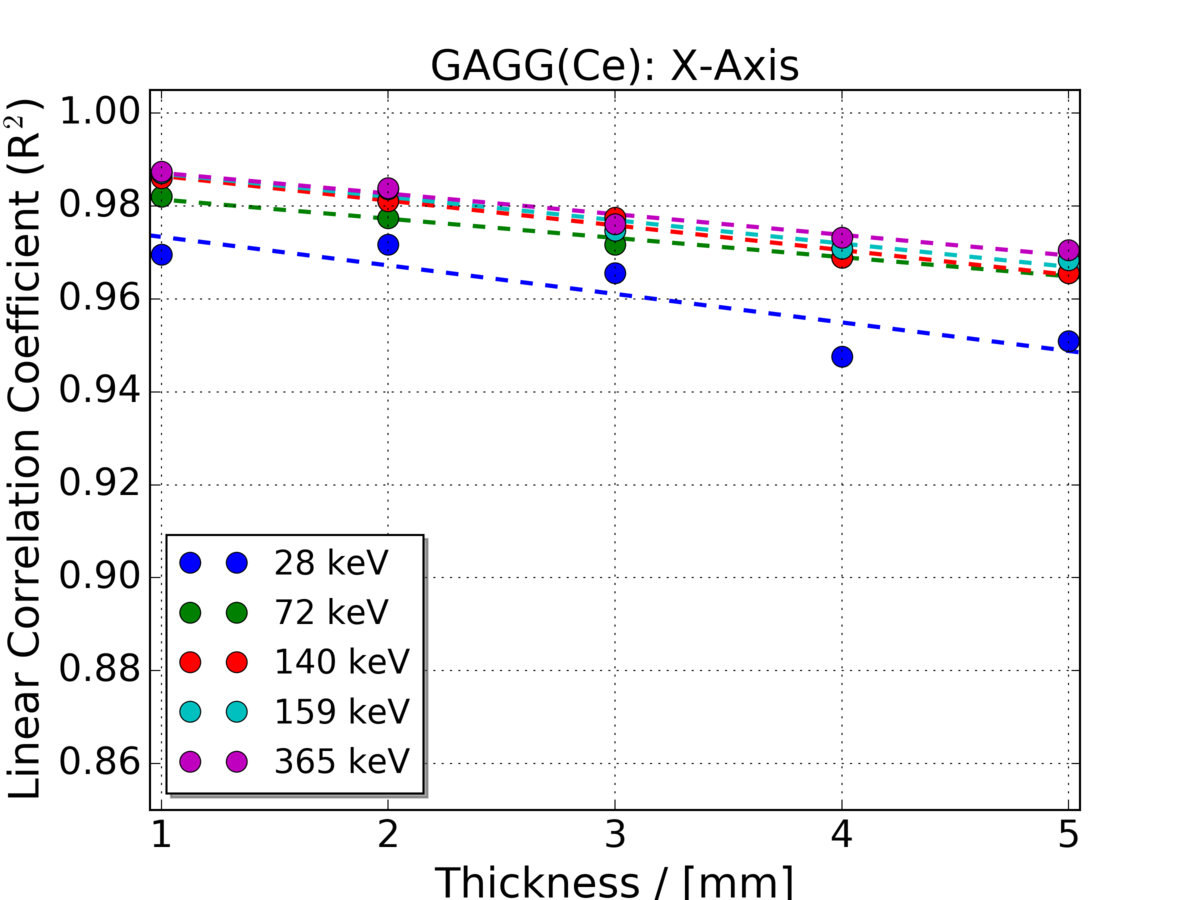}
        \label{fig:r7c}
    \end{subfigure}
    \begin{subfigure}
    \centering
        \includegraphics[width=0.425\textwidth, trim = {0 0 25 10},clip]{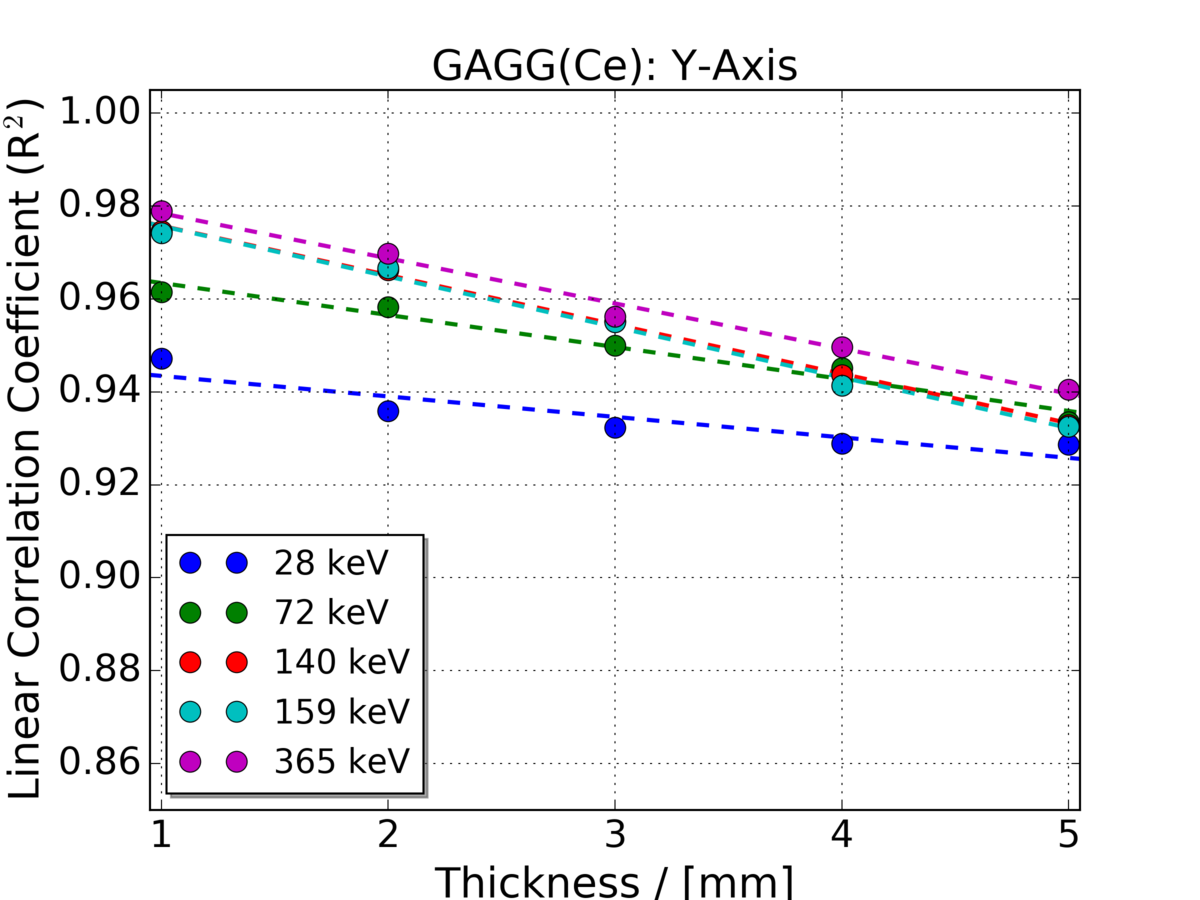}
        \label{fig:r7d}
    \end{subfigure}

    \begin{subfigure}
    \centering 
        \includegraphics[width=0.425\textwidth, trim = {0 0 25 10},clip]{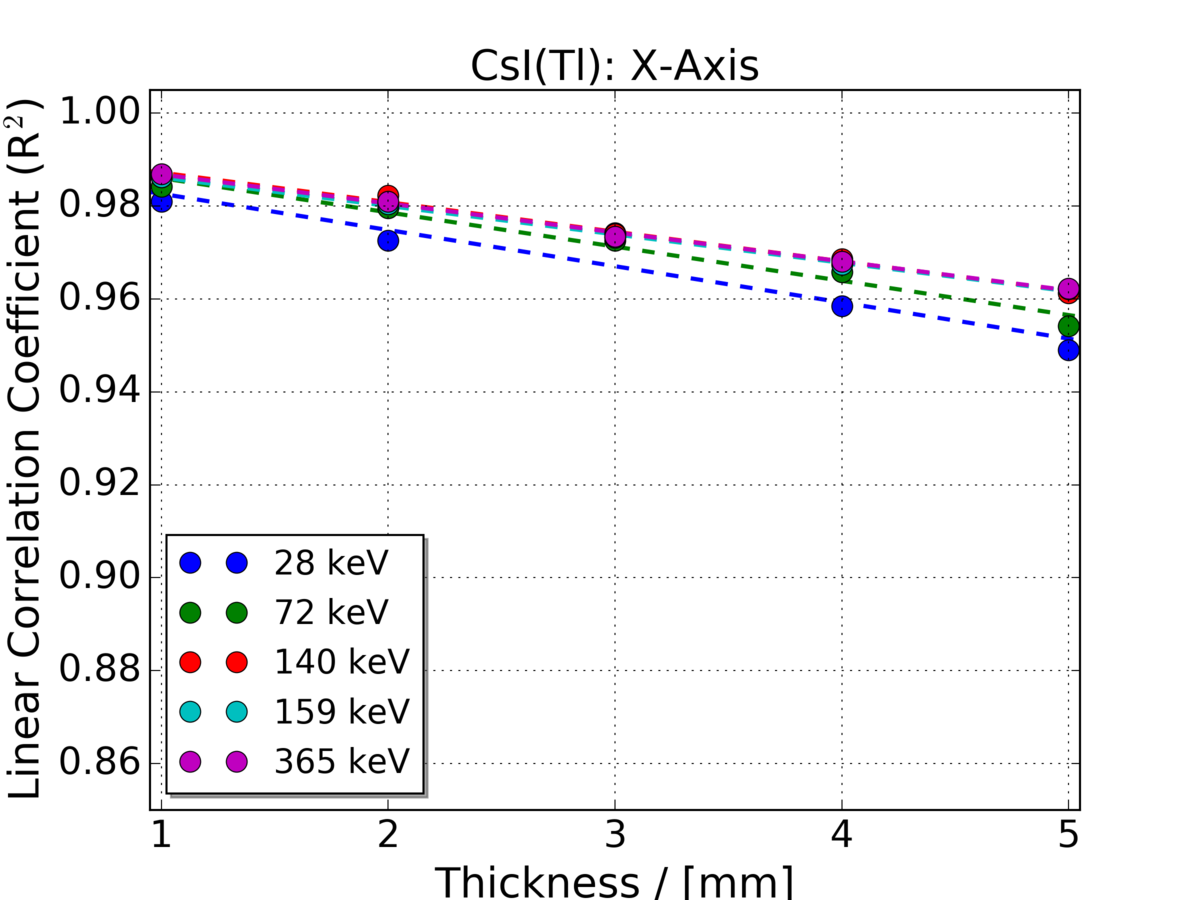}
        \label{fig:r7e}
    \end{subfigure}
    \begin{subfigure}
    \centering
        \includegraphics[width=0.425\textwidth, trim = {0 0 25 10},clip]{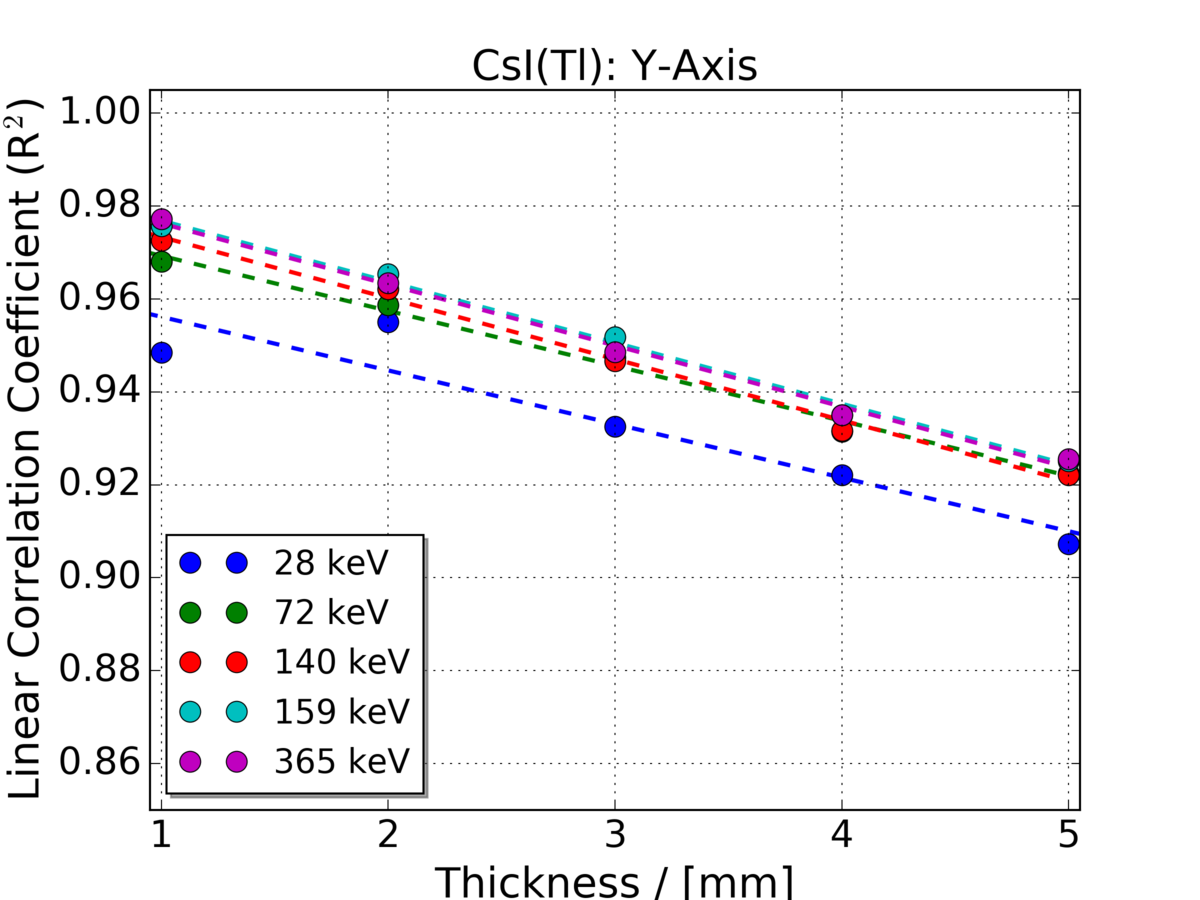}
        \label{fig:r7f}
    \end{subfigure}

    \begin{subfigure}
    \centering 
        \includegraphics[width=0.425\textwidth, trim = {0 0 25 10},clip]{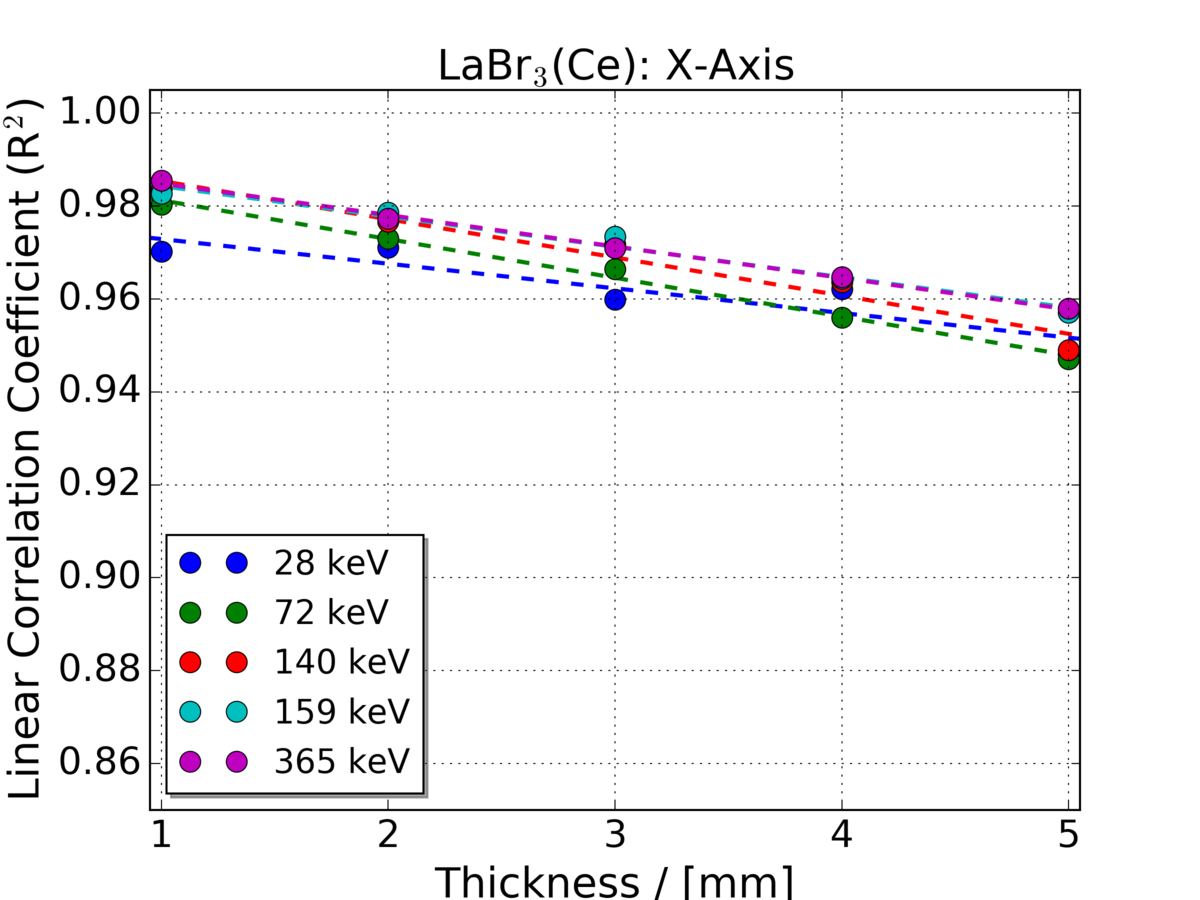}
        \label{fig:r7g}
    \end{subfigure}
    \begin{subfigure}
    \centering
        \includegraphics[width=0.425\textwidth, trim = {0 0 25 10},clip]{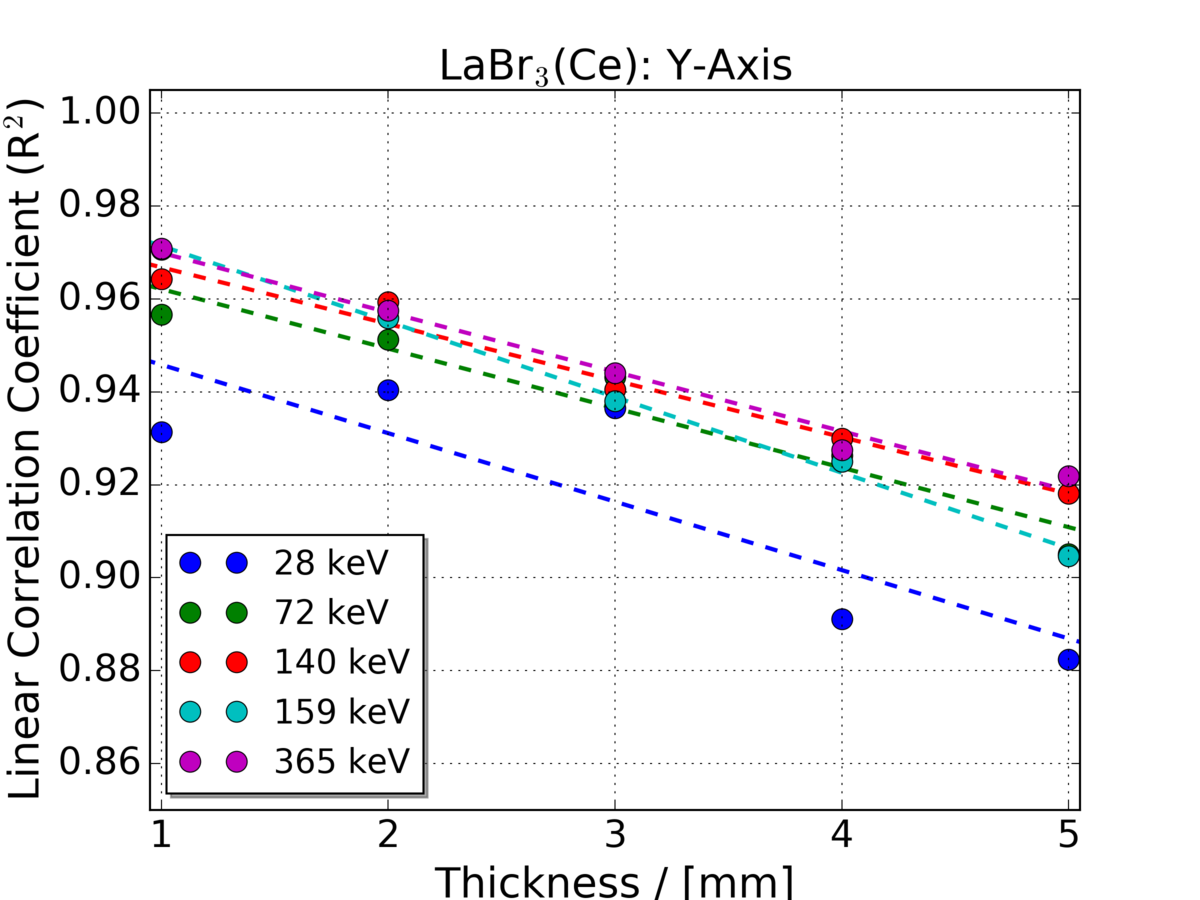}
        \label{fig:r7h}
    \end{subfigure}    
    
\caption{Axial spatial linearity of irradiation spot locations for the four different scintillator crystal materials, NaI(Tl), GAGG(Ce), CsI(Tl) and LaBr$_{3}$(Ce), as a function of incident gamma/x-ray energy, and crystal thickness for $\alpha=0$ corresponding to no truncation. The coloured dash lines correspond to a fitted polynomial surrogate function for each incident gamma/x-ray energy to illustrate the general trend as a function of crystal thickness.}
\label{fig:r7}
\end{figure}

\begin{figure}[p]    
    \centering
    \begin{subfigure}
    \centering 
        \includegraphics[width=0.425\textwidth, trim = {0 0 25 10},clip]{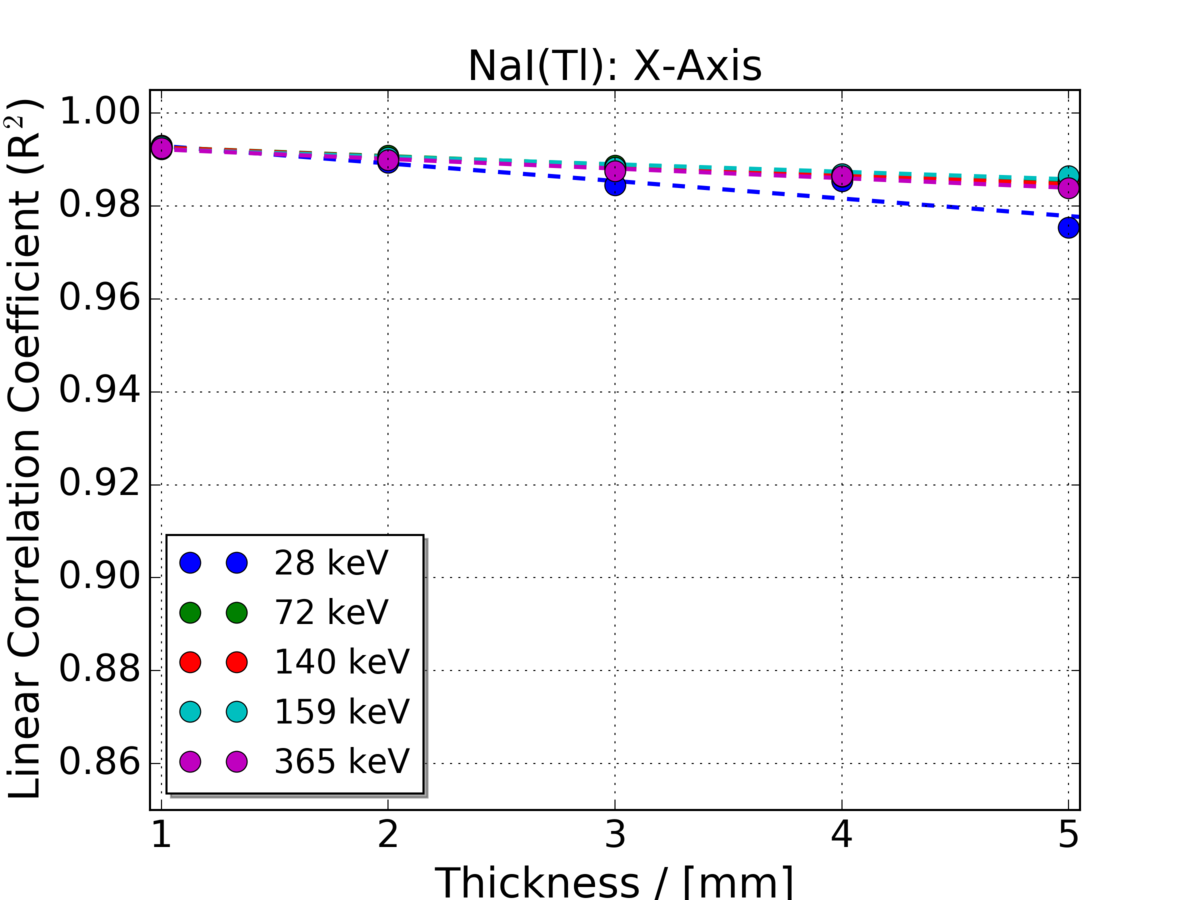}
        \label{fig:r8a}
    \end{subfigure}
    \begin{subfigure}
    \centering
        \includegraphics[width=0.425\textwidth, trim = {0 0 25 10},clip]{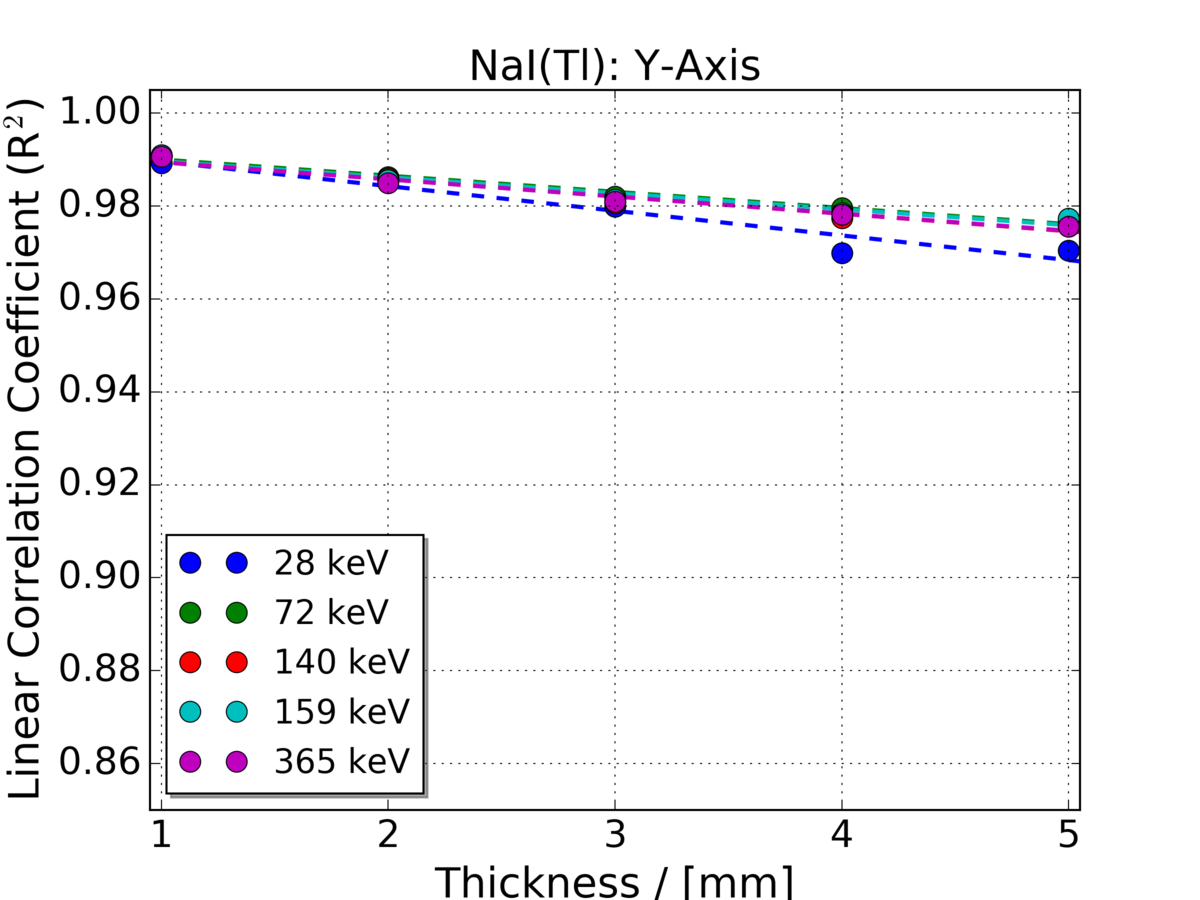}
        \label{fig:r8b}
    \end{subfigure}

    \begin{subfigure}
    \centering 
        \includegraphics[width=0.425\textwidth, trim = {0 0 25 10},clip]{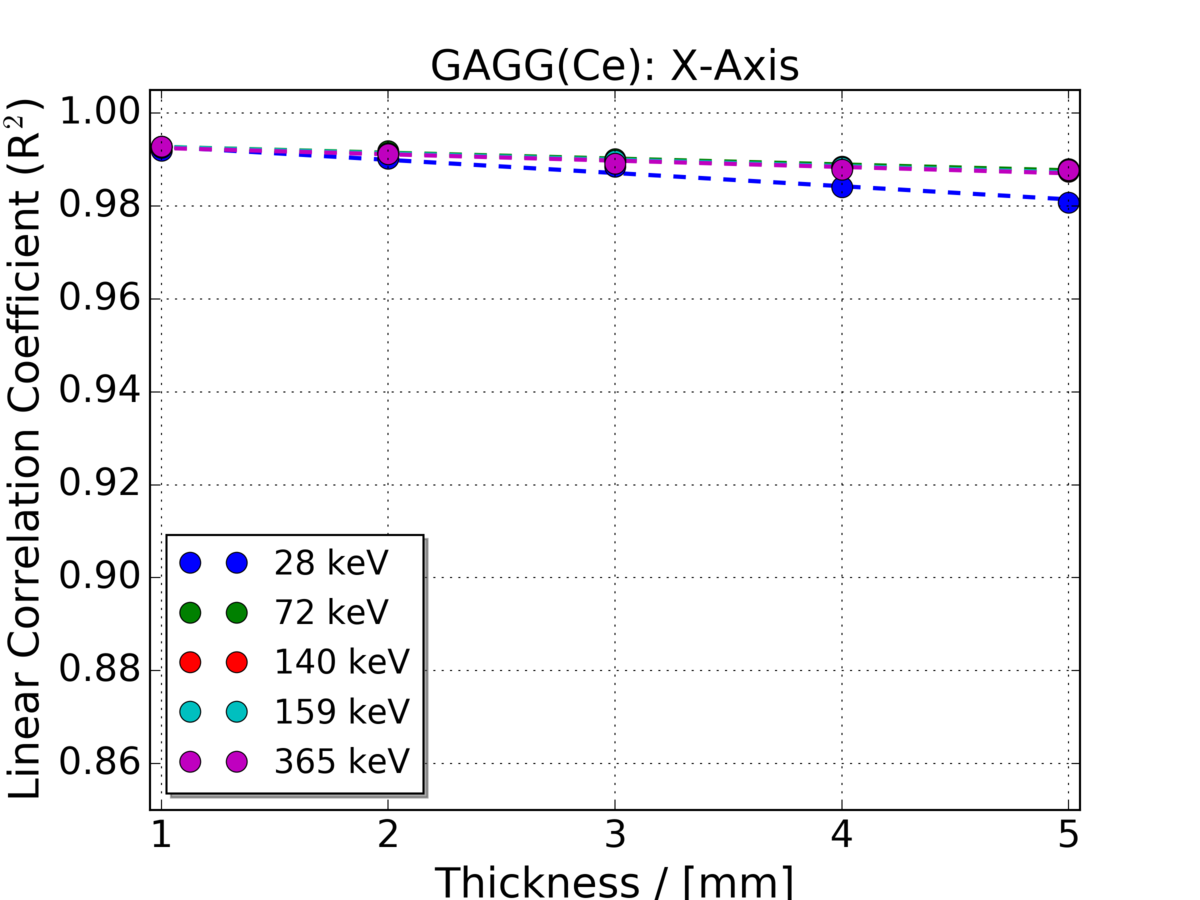}
        \label{fig:r8c}
    \end{subfigure}
    \begin{subfigure}
    \centering
        \includegraphics[width=0.425\textwidth, trim = {0 0 25 10},clip]{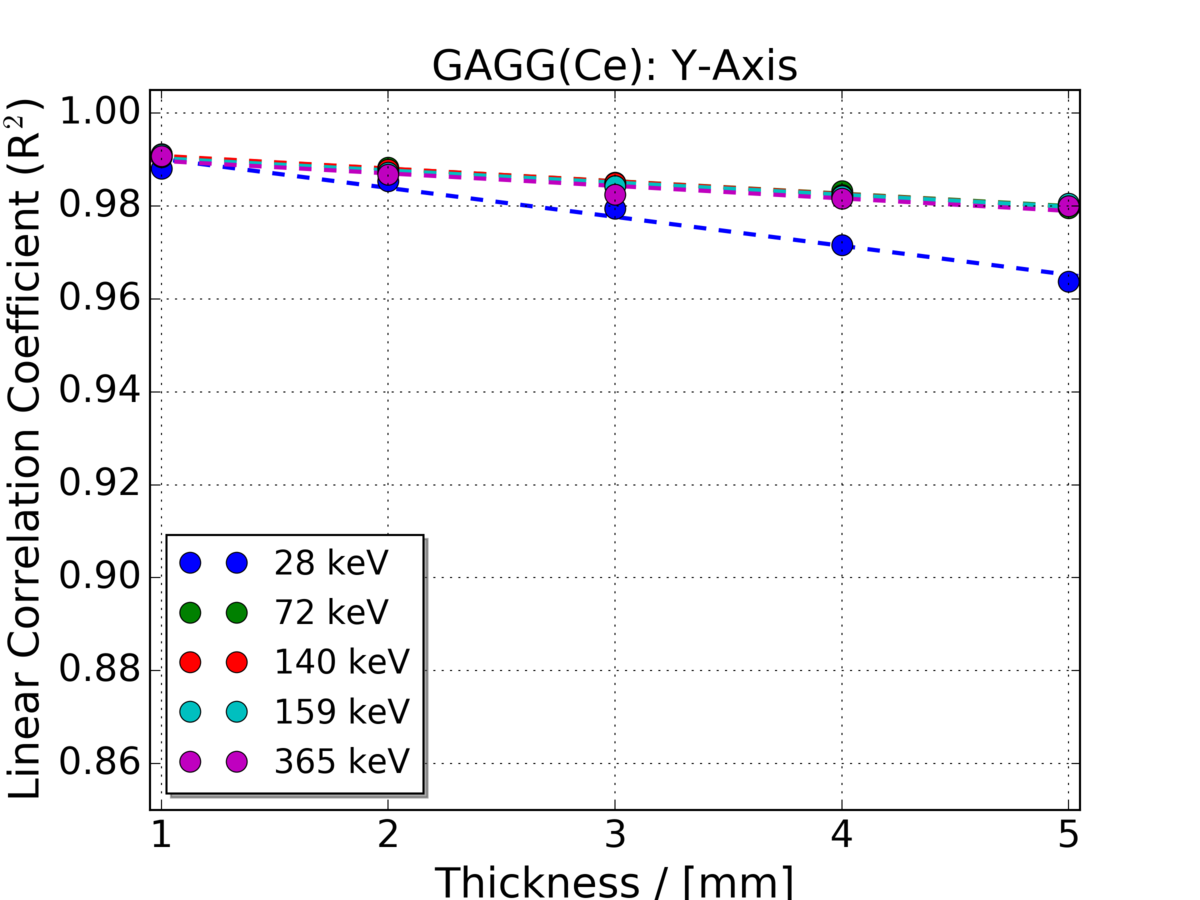}
        \label{fig:r8d}
    \end{subfigure}

    \begin{subfigure}
    \centering 
        \includegraphics[width=0.425\textwidth, trim = {0 0 25 10},clip]{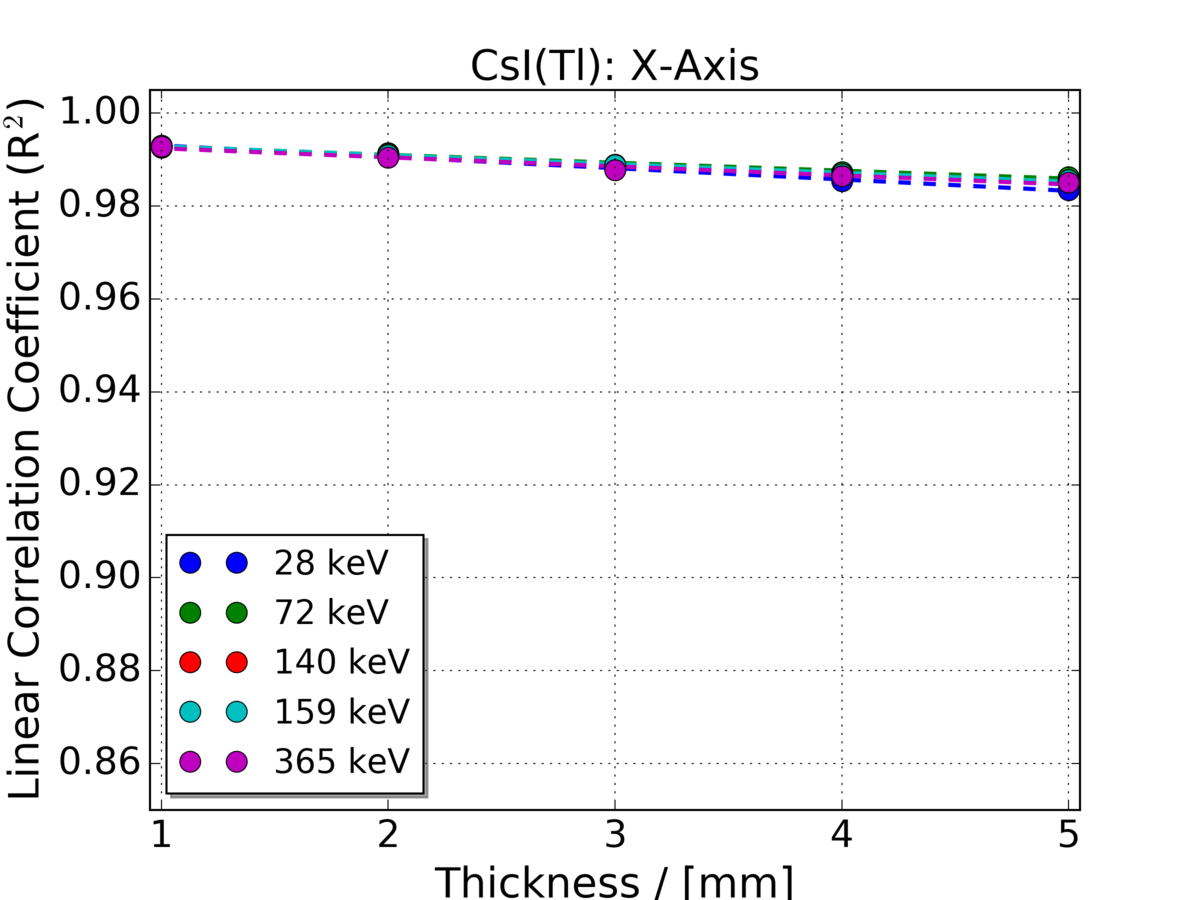}
        \label{fig:r8e}
    \end{subfigure}
    \begin{subfigure}
    \centering
        \includegraphics[width=0.425\textwidth, trim = {0 0 25 10},clip]{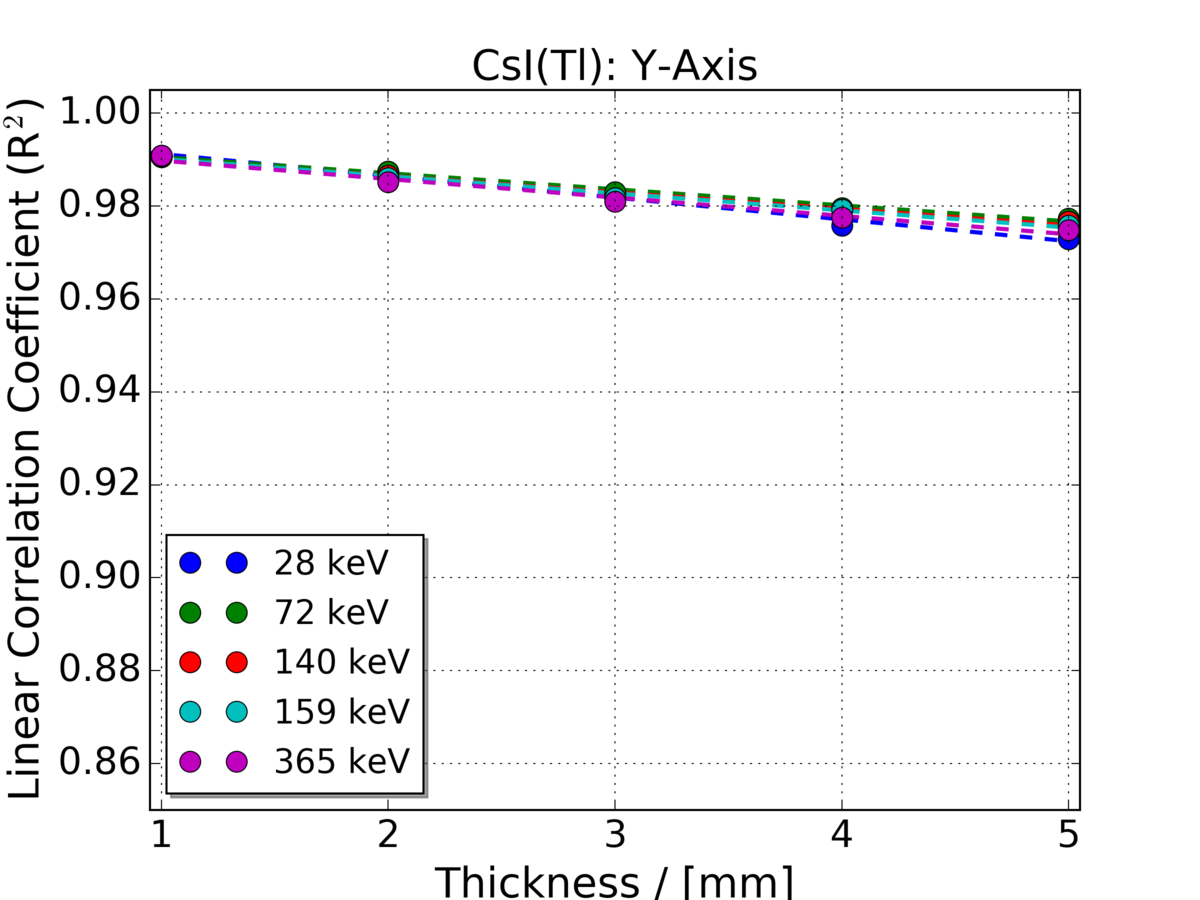}
        \label{fig:r8f}
    \end{subfigure}

    \begin{subfigure}
    \centering 
        \includegraphics[width=0.425\textwidth, trim = {0 0 25 10},clip]{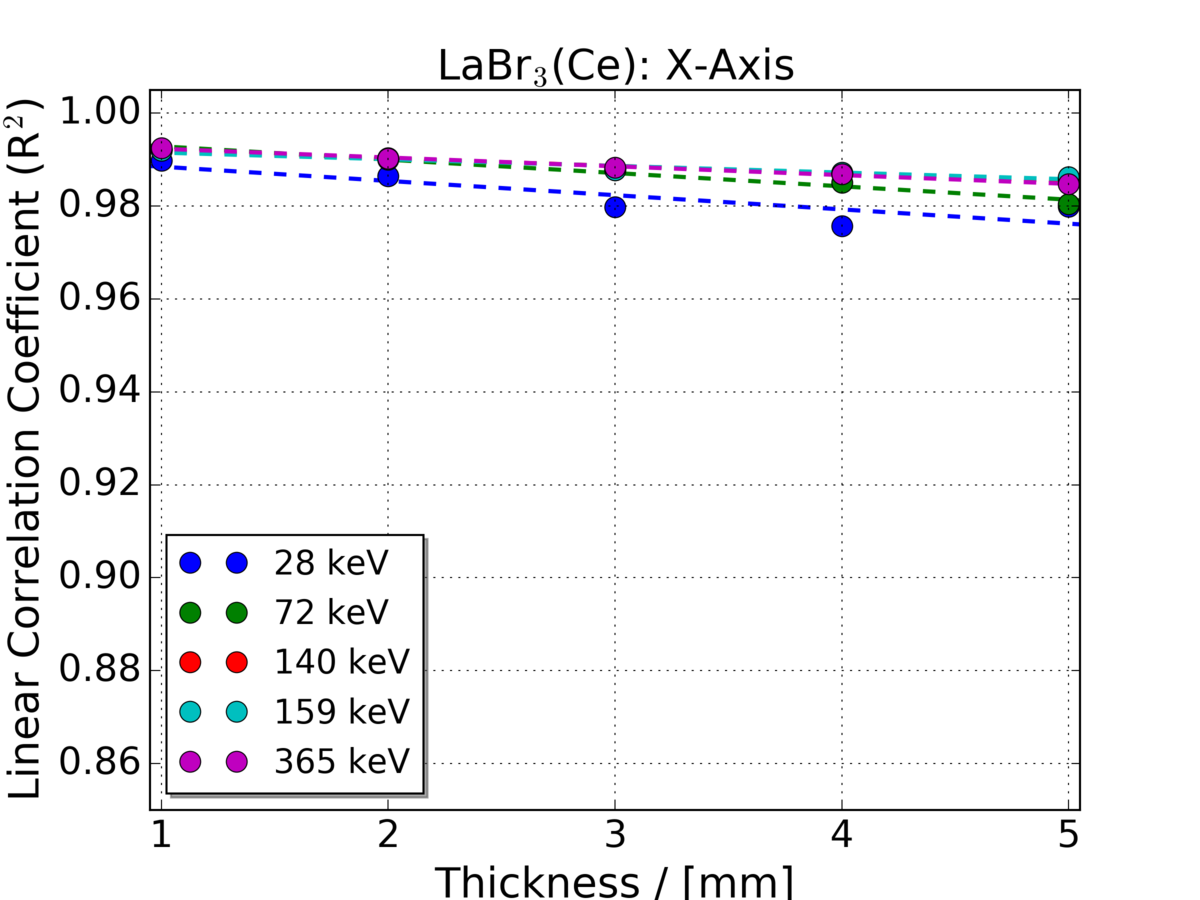}
        \label{fig:r8g}
    \end{subfigure}
    \begin{subfigure}
    \centering
        \includegraphics[width=0.425\textwidth, trim = {0 0 25 10},clip]{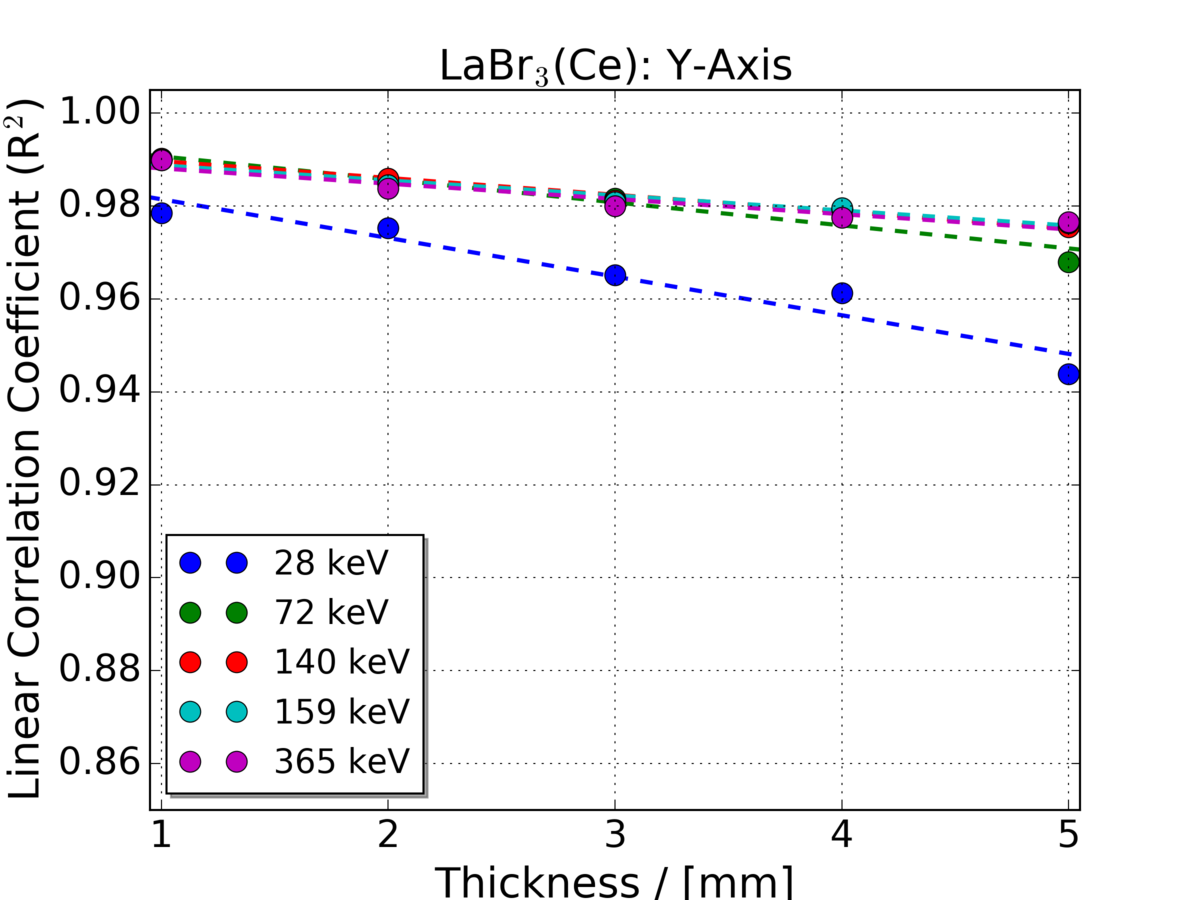}
        \label{fig:r8h}
    \end{subfigure}    
    
\caption{Axial spatial linearity of irradiation spot locations for the four different scintillator crystal materials, NaI(Tl), GAGG(Ce), CsI(Tl) and LaBr$_{3}$(Ce), as a function of incident gamma/x-ray energy, and crystal thickness for $\alpha = 0.02$ corresponding to the fraction uncertainty of when all active pixel SPADs are triggered. The coloured dash lines correspond to a fitted polynomial surrogate function for each incident gamma/x-ray energy to illustrate the general trend as a function of crystal thickness.}
\label{fig:r8}
\end{figure}

The axial spatial linearity of irradiation spot locations for the four scintillator crystal materials as a function of incident gamma/x-ray energy, crystal thickness, and CoG truncation factor can be seen in Figs. \ref{fig:r7} and \ref{fig:r8}. An inverse relationship can be observed between crystal thickness and the spatial linearity for all four materials along both axes (i.e. with increasing crystal thickness the linear correlation coefficient ($R^2$) decreases). In contrast a general direct relationship between incident gamma/x-ray energy and spatial linearity is present for all four materials. Further inspection of Figs. \ref{fig:r7} and \ref{fig:r8} illustrates that an axial asymmetry in spatial linearity as a function of incident gamma/x-ray energy and crystal thickness is present for all four materials, and that the extent of this axial asymmetry is suppressed when a CoG truncation factor of $\alpha$ = 0.02 is implemented. This observed asymmetry between the $x$-axial and $y$-axial data can again be attributed to the non-symmetrical structure of the Philips DPC3200 SiPM \cite{DPCManual2016}. Of the four tested materials GAGG(Ce) and CsI(Tl) display the best performance on average for the tested incident gamma/x-ray energies, followed by NaI(Tl) and then LaBr$_{3}$(Ce). 

\section{Discussion}
\label{sec:D}

The in-silico optimisation of a Philips DPC3200 SiPM photosensor-based thin monolithic scintillator detector for SPECT applications was undertaken through the use of seven FoMs. For the gamma/x-ray total absorption fraction and photoelectric absorption fraction on the first interaction FoMs, the ranking of the four tested materials corresponded directly to the total and relative photoelectric cross-section of each material (e.g. GAGG(Ce), CsI(Tl), NaI(Tl) and LaBr$_{3}$(Ce)). Assessment of the material types based on photopeak energy resolution (FWHM) illustrated that a minimum crystal thickness of 3 mm is required for all four materials to ensure an approximately 15 \% energy resolution regardless of incident gamma/x-ray energy. LaBr$_{3}$ and CsI(Tl)'s energy resolution was found to be superior over the other two tested materials for tested gamma/x-ray energies, with all four materials showing a high level of energy linearity regardless of crystal thickness. Of the four materials LaBr$_{3}$ was determined to possess the highest maximum count rate before saturation regardless of incident gamma/x-ray energy or material thickness, followed by GAGG(Ce), NaI(Tl), and CsI(Tl). Assessment of spatial resolution saw CsI(Tl) obtaining the best result over that of GAGG(Ce), NaI(Tl) and finally LaBr$_{3}$(Ce) across the tester range of gamma/x-ray energies. Moreover, it was shown that for all materials the application of a CoG truncation factor of $\alpha$=0.02 degraded the spatial resolution performance for crystal thicknesses of less than 3 mm at 28 keV. This trend was also observed at high incident gamma/x-ray energies for NaI(Tl) and LaBr$_{3}$(Ce) crystal thickness of greater than 3 mm. Furthermore, it should be noted that for each detector configuration the spatial resolution was nonuniform and degraded with increasing distance of each irradiation location from the crystal centre. Finally, the outcome of the assessment of spatial linearity for the four materials as a function of incident gamma/x-ray energy, crystal thickness, and CoG truncation factor ($\alpha$) illustrated that setting $\alpha$ to 0.02 always resulted in an notable improvement in performance. For this FoM GAGG(Ce) and CsI(Tl) displayed the best performance on average for the tested incident gamma/x-ray energies, followed by NaI(Tl) and then LaBr$_{3}$(Ce). 

Based on these FoM results, one clear finding with respect to crystal thickness for all four materials can be obtained: a minimal crystal thickness of 3 mm is required to ensure an acceptable level of performance for all FoMs with the Philips DPC3200 SiPM. As crystal thickness increases above 3 mm, the photoeletric absorption fraction, energy resolution, energy linearity and relative final SPAD trigger time improves noticeably for all four materials. On the other hand, spatial resolution and spatial linearity decreases slightly with increasing crystal thickness for all for materials, with this trend for spatial linearity suppressed when the CoG truncation factor was set to $\alpha$=0.02. Therefore of the simulated crystal thicknesses, 4 to 5 mm appears to a viable thickness range for Philips digital SiPM based thin monolithic scintillator detectors composed of any of the four materials intended for SPECT applications. This 4 to 5 mm crystal thickness range achieves an acceptable trade-off between energy resolution, sensitivity and spatial resolution for the four materials whilst minimising the effect of oblique gamma/x-rays spatial resolution degradation and DoI \cite{Kupinski2005,Cherry2003,Bushberg2011,Garcia2011}. 

Of the investigated gamma/x-ray energies, the FoM data from the 140 keV gamma-ray simulations is of particular importance as it corresponds to the primary emission line from the most commonly used radionuclide in SPECT imaging (i.e. $^{99m}$Tc). Table \ref{tab:D1} outlines a summary of the seven FoMs for 5 mm thick crystals at 140 keV and a CoG truncation factor of $\alpha=0.02$. From these results, GAGG(Ce) and CsI(Tl) are on average the two best performing scintillator crystal materials of the four at 140 keV. In addition Table \ref{tab:D1} also outlines three other key considerations in the selection of scintillator crystal materials to construct a Philips digital SiPM based thin monolithic scintillator detector: MR compatibility, hygroscopy, and cost. Out of the four materials CsI(Tl) and LaBr$_{3}$(Ce) possess a high level of MR compatibility, followed by NaI(Tl), and then GAGG(Ce) which has effectively zero MR compatibility due to the presence of Gadolinium (a common MRI contrast agent). This consideration is particularly important when considering a radiation detector design for SPECT/MR applications. Of the four materials only GAGG(Ce) has been continuously found to possess zero hygroscopy, the phenomenon of absorbing and trapping water from surrounding environments, with CsI(Tl) following closely behind with very slight hygroscopy that can lead to degraded performance over time in high humidity environments. Conversely, NaI(Tl) and LaBr$_{3}$(Ce) are extremely hyroscopic \cite{Lecoq2016} and require full encapsulation to operate in standard environments making them more difficult to work with. Finally, CsI(Tl) and NaI(Tl) are on the order of 10 to 25 times more cost effective than LaBr$_{3}$(Ce) and GAGG(Ce). Based on these factors CsI(Tl) represents the most promising, and cost effective, material to construct tileable Philips digital SiPM based thin monolithic scintillator detectors for SPECT applications.

\begin{table}
\begin{tabular}{l||c|c|c|c}

  & NaI(Tl) & GAGG(Ce) & CsI(Tl) & LaBr$_3$(Ce) \\
 \hline
 \hline
T.A. Fraction     &  0.740 &  0.880 &  0.809 &  0.719 \\
 P.A. Fraction      &  0.638 & 0.726 & 0.690 & 0.582 \\
 Energy Resolution (\%)  &  11.0 & 11.5 & 10.6 & 10.3 \\
 Energy Linearity          &  0.9997 & 0.9997 & 0.9999 & 0.9999 \\
 Final SPAD Trigger (ns)  &  1597 & 738 & 2067 & 96 \\
 Spatial Resolution (mm) & 0.713 & 0.639 & 0.551 & 0.905 \\
 Spatial Linearity  &  0.9806 & 0.9838 &  0.9810 & 0.9806 \\
 MR Compatibility        &  Low & None & High & High \\
 Hygroscopy                &  High & None & Slight & High \\
 Cost                      &  Low & Medium & Low & High \\
\end{tabular}
 \caption[]{ A summary of the seven FoMs for 5 mm thick crystals at 140 keV and a CoG truncation factor of $\alpha=0.02$, and other key considerations of the four different scintillator crystal materials (NaI(Tl), GAGG(Ce), CsI(Tl) and LaBr$_{3}$(Ce)).  Data on these material properties, i.e. hydroscopy and MR compatibility, were taken from the literature and further information can be found in \cite{David2015,Lecoq2016,Avdeichikov1994,Yamamoto2003}.}
\label{tab:D1}
\end{table}

This investigation is part of a larger research program to develop a novel multiple radiomolecular tracer imaging platform for small animals within the Department of Radiation Science and Technology at the Delft University of Technology (The Netherlands). As the first phase of this new imaging platform will use a parallel hole collimator, two experimental Philips DPC3200 SiPM photosensor-based monolithic scintillator detector prototypes using 5 mm thick CsI(Tl) crystals have begun construction. The performance of these units will be explored not only as a function of incident gamma/x-ray energy, but also as a function of unit temperature, Philips DPC3200 SiPM photosensor trigger setting, and readout algorithm.

\section{Conclusion}
\label{sec:C}

An in-silico investigation into the optimal design of a Philips DPC3200 SiPM photosensor-based thin monolithic scintillator detector for SPECT applications was undertaken using the Monte Carlo radiation transport modelling toolkit Geant4 version 10.5. The performance of the 20 different SPECT radiation detector configurations, 4 scintillator materials (NaI(Tl), GAGG(Ce), CsI(Tl) and LaBr$_{3}$(Ce)) and 5 thicknesses (1 to 5 mm), were determined through the use of seven FoMs. Based on these FoMs, it was found that a crystal thickness range of 4 to 5 mm was required for all four materials to ensure acceptable energy resolution, sensitivity and spatial resolution performance with the Philips DPC3200 SiPM. Any thinner than this and the performance of all materials was found to degrade rapidly due to a high probability of material specific fluorescence x-ray escape after incident gamma/x-ray photoelectric absorption. When these findings were weighted in combination with each material's MR compatibility, hygroscopy, and cost, it was found that CsI(Tl) represents the most promising material to construct tileable Philips digital SiPM based thin monolithic scintillator detectors for SPECT applications. Further work is underway to construct a pair of 5 mm thick CsI(Tl) prototype units for a novel parallel hole mechanical collimator based multiple radiomolecular tracer imaging platform for small animals within the Department of Radiation Science and Technology at the Delft University of Technology (The Netherlands).

\section*{Acknowledgements}

J.~M.~C.~Brown would like to acknowledge both F.~G.~A.~Quarati from the Department of Radiation Science and Technology, Delft University of Technology (The Netherlands) and D.~M.~Paganin of the School of Physics and Astronomy, Monash University (Australia) for their helpful comments and suggestions. J.~M.~C.~Brown is supported by a Veni fellowship from the Dutch Organization for Scientific Research (NWO Domain AES Veni 16808 (2018)). This work was carried out on the Dutch national e-infrastructure with the support of SURF Cooperative (Grant No: 36673 (2019)).

\appendix
\section{Geant4 Simulation Platform Material Properties}
\label{appendix1}

\begin{figure}[tbh]    
    \centering
     \includegraphics[width=0.45\textwidth]{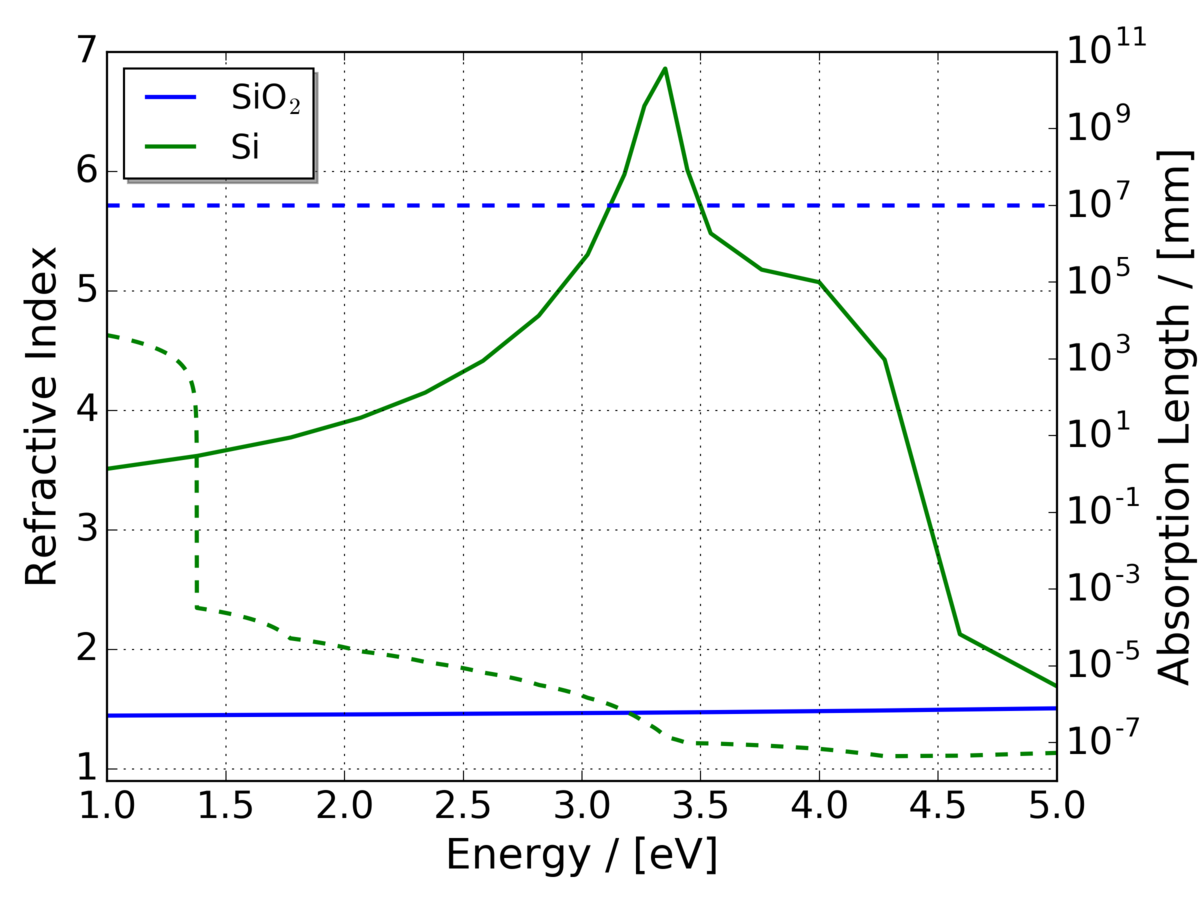}
\caption{DPC3200 quartz glass (SiO$_2$) and pixel (Si) material refractive index (solid line) and attenuation length (dashed line) data sets implemented in the Geant4 simulation platform.}
\label{fig:a1}
\end{figure}

\begin{figure}[tbh]    
    \begin{subfigure}
    \centering 
        \includegraphics[width=0.45\textwidth]{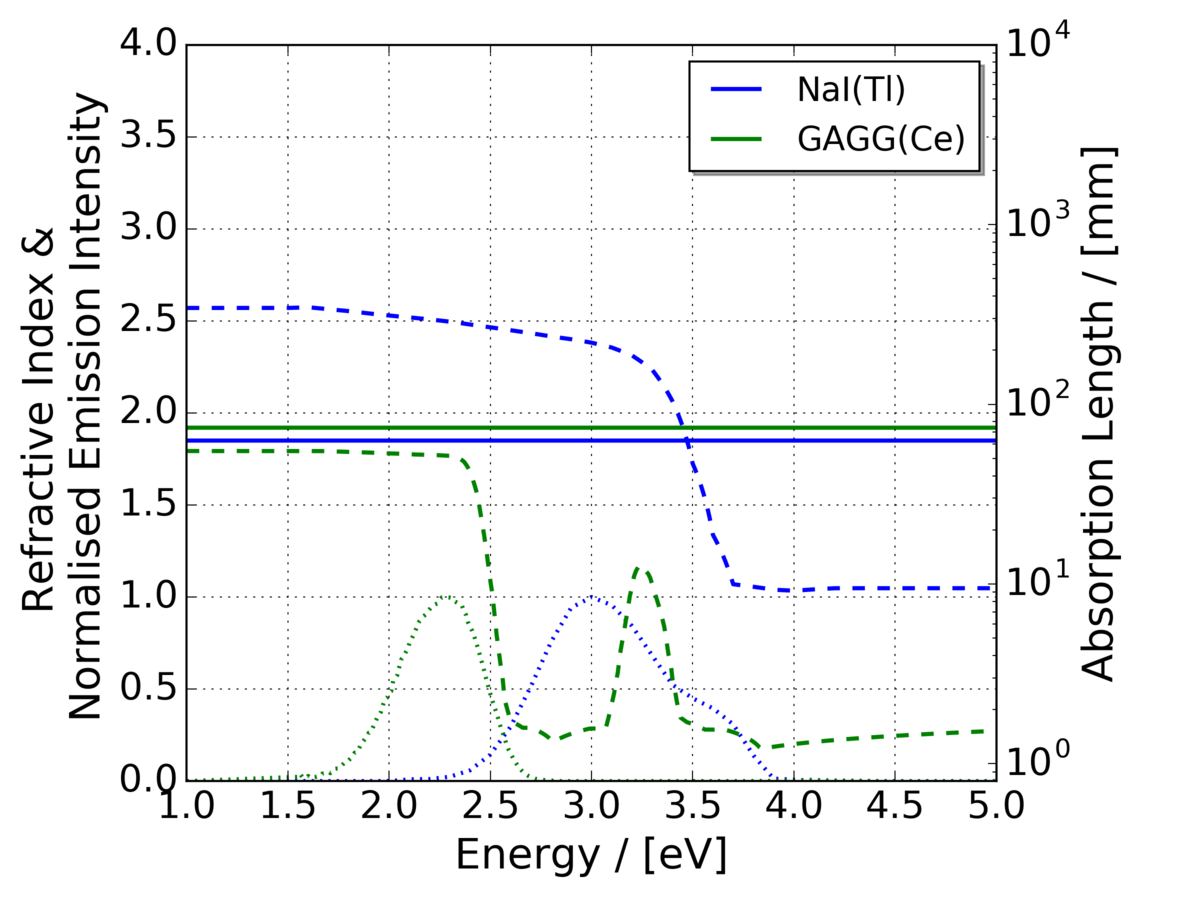}
        \label{fig:a2a}
    \end{subfigure}
    \begin{subfigure}
    \centering
        \includegraphics[width=0.45\textwidth]{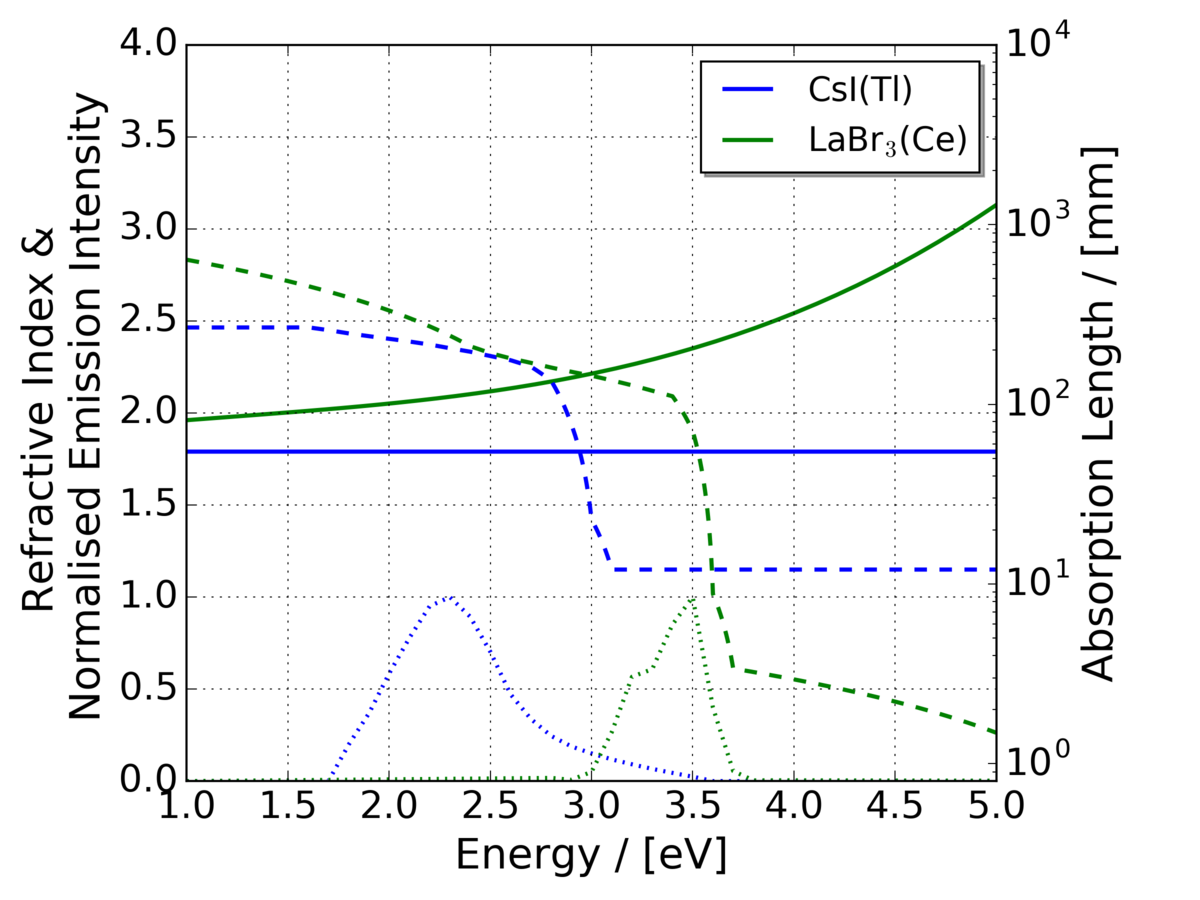}
        \label{fig:a2b}
    \end{subfigure}
\caption{NaI(Tl), GAGG(Ce), CsI(Tl) and LaBr$_{3}$(Ce) scintillator crystal material refractive indexes (solid line), attenuation lengths (dashed line) and normalised scintillation photon emission intensities (dotted line) data sets implemented in the Geant4 simulation platform.}
\label{fig:a2}
\end{figure}

The following appendix contains the density, elemental composition, and optical/scintillation properties of all materials utilised in the developed Geant4 simulation platform. Material data relating to the world volume, bonding glue, Vikuiti ESR foil, and implemented Philips DPC3200 SiPM is outlined in Table \ref{tab:1} and Fig. \ref{fig:a1}. Material data relating the four explored scintillator types, NaI(Tl), GAGG(Ce), CsI(Tl) and LaBr$_{3}$(Ce), can be seen in Table \ref{tab:2} and Fig. \ref{fig:a2}.

\begin{sidewaystable}[h]
\centering
\begin{tabular}{||c|c|c|c|c|c||}
\hline
\hline
             &              &           &            & Optical        &          \\
 Material    & Density      & Elemental & Refractive & Reflectivity / & Reference \\
             & (g/cm$^{3}$) & Composition & Index      & Absorption   & \\
 \hline
 \hline             
 Air         & 1.29$\times$10$^-3$ & C (0.01\%), N (75.52\%), & 1 & - & Geant4 Material  \\
             &       &  O (23.19\%), Ar (1.28\%) &  &  &   Database \cite{G42016} \\
 \hline                
 DELO glue   & 1.0   & H$_8$C$_5$O$_2$ & 1.5   & -   & \cite{Brown2019,Dachs2016} \\
  \hline   
 Vikuiti ESR & 1.29  & H$_8$C$_{10}$O$_4$& -     & 98\% / 2\%  & \cite{3M2019} \\
  \hline   
 DPC3200 PCB & 1.86  & SiO$_2$ (52.8\%), H$_1$C$_1$O$_1$ (47.2\%) & - & 0\% / 100\% & \cite{Brown2019,Dachs2016} \\
  \hline   
  DPC3200 Glass & 2.203 & SiO$_2$ & See Fig. \ref{fig:a1} & See Fig. \ref{fig:a1} & \cite{Brown2019,Dachs2016} \\
  \hline   
 DPC3200 Pixel &  2.33 & Si & See Fig. \ref{fig:a1} & See Fig. \ref{fig:a1} & \cite{Philipp1960} \\

 \hline   
 \hline
\end{tabular}
\\
\caption[]{Density, elemental composition, and optical material properties of the world volume, bonding glue, Vikuiti ESR foil and Philips DPC3200 SiPM implemented in the Geant4 simulation platform.}
\label{tab:1}
\end{sidewaystable}

\begin{sidewaystable}[h]
\centering
\begin{tabular}{||c|c|c|c|c|c|c|c||}
\hline
\hline
&             &           &             & Optical Yield,           &   Optical Decay   &                   &           \\
Material & Density      & Elemental & Refractive & Emission Spectrum,        &  Time Constants  & Resolution Scale  & Reference \\
& (g/cm$^{3}$) & Composition & Index      & Absorption Length         &       (ns)      &  (at 511 keV)     & \\

 \hline 
 \hline    
NaI(Tl)  & 3.67       & NaI &  See Fig. \ref{fig:a2}  & 41 Photons per eV, & Fast: 220 (96\%) & 3.50  & \cite{Moa2008} \\
  &        & (6.5\% Tl doping)           &                          & See Fig. \ref{fig:a2}  & Slow: 1500 (4\%) &   & \\ 
 \hline    
GAGG(Ce)  & 6.63       & Gd$_3$Al$_2$Ga$_3$O$_{12}$ &  See Fig. \ref{fig:a2} & 50 Photons per eV, & Fast: 87 (90\%) & 3.08  & \cite{Kobayashi2012} \\
  &        & (1\% Ce doping)           &                          & See Fig. \ref{fig:a2} & Slow: 255 (10\%) &   & \\  
 \hline    
CsI(Tl)  & 4.51       & CsI &  See Fig. \ref{fig:a2} & 54 Photons per eV, & 1000 (100\%) & 3.50  & \cite{Moa2008} \\
  &        & (0.08\% Tl doping)           &                          & See Fig. \ref{fig:a2} &  &   & \\ 

 \hline    
LaBr$_{3}$(Ce)  & 5.08       & LaBr$_{3}$ &  See Fig. \ref{fig:a2} & 63 Photons per eV, & Fast: 15 (97\%) & 1.5  & \cite{Glodo2005,vanDam2012} \\
  &        & (5\% Ce doping)           &                          & See Fig. \ref{fig:a2} & Slow: 55 (3\%) &   & \\
 \hline   
 \hline
\end{tabular} 
 \caption[]{Density, elemental composition, and optical properties of the four scintillator materials, NaI(Tl), GAGG(Ce), CsI(Tl) and LaBr$_{3}$(Ce), implemented in the Geant4 simulation platform.}
\label{tab:2}
\end{sidewaystable}

\newpage
\begin{thebibliography}{00}

\bibitem[]{Kupinski2005}
  Kupinski~M.~A. and Barrett~H.~H. \textit{Small-Animal SPECT Imaging}, Springer Science 2005.

\bibitem{Brown2013}
  Brown~J.~M.~C., Gillam~J.~E., Paganin~D.~M. and Dimmock~M.~R. \textit{Laplacian erosion: an image deblurring technique for multi-plane gamma-cameras}, IEEE Transactions on Nuclear Science 60(5): 3333-3342 (2013).
  
\bibitem{Cherry2003}
  Cherry~S.~R., Sorenson~J.~A. and Phelps~M.~E. \textit{Physics in Nuclear Medicine}, Elsevier Science 2003.

\bibitem{Bushberg2011}
  Bushberg~J.~T. and Boone J.~M. \textit{The essential physics of medical imaging}, Lippincott, Williams \& Wilkins (2011).

\bibitem{Lewellen2008}
  Lewellen TK. \textit{Recent developments in PET detector technology}, Physics in Medicine \& Biology 53(17): R287 (2008).

\bibitem{Schaart2016}
  Schaart~D.~R., Charbon~E., Frach~T. and Schulz~V. \textit{Advances in digital SiPMs and their application in biomedical imaging}, Nuclear Instruments and Methods in Physics Research Section A 809: 31-52 (2016).

\bibitem{Bisogni2018}
  Bisogni~M.~G., Del~Guerra~A. and Belcari.~N. \textit{Medical applications of silicon photomultipliers}, Nuclear Instruments and Methods in Physics Research Section A 926: 118-128 (2018). 
 
\bibitem{Judenhofer2007}
  Judenhofer~M.~S., Catana~C., Swann~B.~K., Siegel~S.~B., Jung~W.~I., Nutt~R.~E., Cherry~S.~R., Claussen~C.~D. and Pichler~B.~J. \textit{PET/MR images acquired with a compact MR-compatible PET detector in a 7-T magnet}, Radiology 244(3): 807-814 (2007).

\bibitem{Schlemmer2008}
  Schlemmer~H.~P.~W., Pichler~B.~J., Schmand~M., Burbar~Z., Michel~C., Ladebeck~R., Jattke~K., Townsend~D., Nahmias~C., Jacob~P.~K. and Heiss~W.~D. \textit{Simultaneous MR/PET imaging of the human brain: feasibility study}, Radiology 248(3): 1028-1035 (2008).

\bibitem{Wehrl2009}
  Wehrl~H.~F., Judenhofer~M.~S., Wiehr~S. and Pichler~B.~J. \textit{Pre-clinical PET/MR: technological advances and new perspectives in biomedical research}, European Journal of Nuclear Medicine and Molecular Imaging 36: 56-68 (2009).

\bibitem{Delso2011}
  Delso~G., F\:{u}rst~S., Jakoby~B., Ladebeck~R., Ganter~C., Nekolla~S.~G., Schwaiger~M. and Ziegler~S.~I. \textit{Performance measurements of the Siemens mMR integrated whole-body PET/MR scanner}, Journal of Nuclear Medicine 52(12): 1914-1922 (2011).

\bibitem{Weissler2015}
  Weissler~B. et al. \textit{A digital preclinical PET/MRI insert and initial results}, IEEE Transactions on Medical Imaging 34(11): 2258-2270 (2015).

\bibitem{Gonzalez2016}
  Gonz\'{a}lez~A.~J. et al. \textit{The MINDView brain PET detector, feasibility study based on SiPM arrays}, Nuclear Instruments and Methods in Physics Research Section A 818: 82-90 (2016).
  
\bibitem{Benlloch2018}
  Benlloch~J.~M. et al. \textit{The MINDVIEW project: First results} European Psychiatry 50:21-27 (2018).
  
\bibitem{Gonzalez2018}
  Gonz\'{a}lez~A.~J., S\'{a}nchez~F., Benlloch~J.~M. \textit{Organ-dedicated molecular imaging systems}, IEEE Transactions on Radiation and Plasma Medical Sciences 2(5): 388-403 (2018).

\bibitem{Hypmed2019}
  HYPMED Consortium, http://www.hypmed.eu/ (2019).

\bibitem{Frach2009}
  Frach~T., Prescher~G., Degenhardt~C., de~Gruyter~R., Schmitz~A. and Ballizany~R. \textit{The digital silicon photomultiplier: Principle of operation and intrinsic detector performance}, 2009 IEEE Nuclear Science Symposium Conference Record (NSS/MIC): 1959-1965 (2009).

\bibitem{Frach2010}
  Frach~T., Prescher~G., Degenhardt~C. and Zwaans~B. \textit{The digital silicon photomultiplier—System architecture and performance evaluation}, 2010 IEEE Nuclear Science Symposium Conference Record (NSS/MIC): 1722-1727 (2010).
  
\bibitem{Heemskerk2010}
 Heemskerk~J.~W., Korevaar~M.~A., Huizenga~J., Kreuger~R., Schaart~D.~R., Goorden~M.~C. and Beekman~F.~J. \textit{An enhanced high-resolution EMCCD-based gamma camera using SiPM side detection}, Physics in Medicine \& Biology 55(22): 6773-6784 (2010).

\bibitem{Georgiou2014}
  Georgiou~M., Borghi~G., Spirou~S.~V., Loudos~G. and Schaart~D.~R. \textit{First performance tests of a digital photon counter (DPC) array coupled to a CsI (Tl) crystal matrix for potential use in SPECT}, Physics in Medicine \& Biology 59(10): 2415-2430 (2014).

\bibitem{Busca2015}
  Busca~P., Occhipinti~M., Trigilio~P., Cozzi~G., Fiorini~C., Piemonte~C., Ferri~A., Gola~A., Nagy~K., B\:{u}kki~T. and Rieger~J. \textit{Experimental evaluation of a SiPM-based scintillation detector for MR-compatible SPECT systems}, IEEE Transactions on Nuclear Science 62(5): 2122-2128 (2015).

\bibitem{David2015}
  David~S., Georgiou~M., Fysikopoulos~E. and Loudos~G. \textit{Evaluation of a SiPM array coupled to a Gd$_3$Al$_2$Ga$_3$O$_12$:Ce (GAGG:Ce) discrete scintillator}, Physica Medica 31(7): 763-766 (2015).

\bibitem{Busca2014}
  Busca~P. et~al. \textit{Simulation of the expected performance of INSERT: A new multi-modality SPECT/MRI system for preclinical and clinical imaging}, Nuclear Instruments and Methods in Physics Research Section A 734: 141-146 (2014).

\bibitem{Hutton2018}
  Hutton~B.~F. et.~al \textit{Development of clinical simultaneous SPECT/MRI}, The British Journal of Radiology 91: 20160690 (2018).

\bibitem{Carminati2019}
 Carminati~M., D’Adda~I., Morahan~A.~J., Erlandsson~K., Nagy~K., Czeller~M., T\"{o}lgyesi~B., Nyitrai~Z., Savi~A., van~Mullekom,~P. and Hutton~B.~F. \textit{Clinical SiPM-Based MRI-Compatible SPECT: Preliminary Characterization}, IEEE Transactions on Radiation and Plasma Medical Sciences: In Press (2019).
 
\bibitem{Lecoq2016}
  Lecoq~P., Gektin~A. and Korzhik~M. \textit{Inorganic scintillators for detector systems: physical principles and crystal engineering}, Springer (2016).

\bibitem{G42003}
  Agostinelli~S. et al. \textit{Geant4-a simulation toolkit}, Nuclear Instruments and Methods in Physics Research Section A 506(3): 250-303 (2003).

\bibitem{G42006}
  Allison~J. et al. \textit{Geant4 developments and applications}, IEEE Transactions on Nuclear Science 53(1): 270-278 (2006).

\bibitem{G42016}
  Allison~J. et al. \textit{Recent developments in Geant4}, Nuclear Instruments and Methods in Physics Research Section A 835: 186-225 (2016).
 
\bibitem{DPCManual2016}
  Module-TEK User Manual, Philips Digital Photon Counting (2016). 
 
\bibitem{Brown2019}
  Brown~J.~M.~C., Brunner~S.~E. and Schaart~D.~R. \textit{A High Count-Rate and Depth-of-Interaction Resolving Single Layered One-Side Readout Pixelated Scintillator Crystal Array for PET Applications}, IEEE Transactions on Radiation and Plasma Medical Sciences 4(3): 361-370 (2020).

\bibitem{Levin1996}
  Levin~A. and Moisan~C. \textit{A more physical approach to model the surface treatment of scintillation counters and its implementation into DETECT}, 1996 IEEE Nuclear Science Symposium Conference Record 2: 702-706 (1996).

\bibitem{G4Phys2019}
  GEANT4 Collaboration, \textit{GEANT4 Physics Reference Manual-Version 10.5},  https://geant4.web.cern.ch (2019).

\bibitem{VanderLaan2010}
  Van~der~Laan~D.~J., Schaart~D.~R., Maas~M.~C., Beekman~F.~J., Bruyndonckx~P. and van~Eijk~C.W. \textit{Optical simulation of monolithic scintillator detectors using GATE/GEANT4}, Physics in Medicine \& Biology 55(6): 1659-1675 (2010).

\bibitem{Nilsson2015}
  Nilsson~J., Cuplov~V. and Isaksson~M. \textit{Identifying key surface parameters for optical photon transport in GEANT4/GATE simulations}, Applied Radiation and Isotopes 103: 15-24 (2015).
  
\bibitem{Steinbach1996}
  Steinbach~D., Majewski~S., Williams~M., Kross~B., Weisenberger~A.~G. and Wojcik~R. \textit{Development of a small field of view scintimammography camera based on a YAP crystal array and a position sensitive PMT}. 1996 IEEE Nuclear Science Symposium Conference Record 2: 1251-1256) (1996).

\bibitem{Wojcik2001}
  Wojcik~R., Majewski~S., Kross~B., Popov~V. and Weisenberger~A.~G. \textit{Optimized readout of small gamma cameras for high resolution single gamma and positron emission imaging}, In 2001 IEEE Nuclear Science Symposium Conference Record : 1821-1825 (2001).

\bibitem{Garcia2011}
  Garcia~E.~V., Faber~T.~L. and Esteves~F.~P. \textit{Cardiac dedicated ultrafast SPECT cameras: new designs and clinical implications}, Journal of Nuclear Medicine 52(2): 210-217 (2011).

\bibitem{xcom2019}
  Berger~M.~J., Hubbell~J.~H., Seltzer~S.~M., Chang~J., Coursey~J.~S., Sukumar~R., Zucker~D.~S., and Olsen~K. \textit{XCOM: Photon Cross Section Database (version 1.5)}, [Online] Available: http://physics.nist.gov/xcom (2019).

\bibitem{Gilmore2011}
  Gilmore~G. \textit{Practical gamma-ray spectroscopy}, John Wiley \& Sons (2011).

\bibitem{Avdeichikov1994}
  Avdeichikov~V.~V., Bergholt~L., Guttormsen~M., Taylor~J.~E., Westerberg~L., Jakobsson~B., Klamra~W. and Murin, Y.A. \textit{Light output and energy resolution of CsI, YAG, GSO, BGO and LSO scintillators for light ions}, Nuclear Instruments and Methods in Physics Research Section A 349(1): 216-224 (1994).

\bibitem{Yamamoto2003}
  Yamamoto~S., Kuroda~K. and Senda~M. \textit{Scintillator selection for MR-compatible gamma detectors}. IEEE Transactions on Nuclear Science 50(5): 1683-1685 (2003).

\bibitem{Dachs2016}
  Dachs F. 2016, \textit{Monte-Carlo simulation of a new ultra-fast gamma detector design in Geant4}. Masters' Dissertation, Vienna University of Technology (Vienna, Austria).

\bibitem{3M2019}
  3M Vikuiti Enhanced Spectral Reflector Datasheet, http://www.3m.com (2019).

\bibitem{Philipp1960}
  Philipp~H.~R. and Taft~E.~A. \textit{Optical constants of silicon in the region 1 to 10 eV}, Physical Review 120(1): 37-38 (1960).

\bibitem{Moa2008}
  Mao~R., Zhang~L. and Zhu~R.~Y. \textit{Optical and scintillation properties of inorganic scintillators in high energy physics}, IEEE Transactions on Nuclear Science 55(4): 2425-2431 (2008).

\bibitem{Kobayashi2012}
  Kobayashi~M., Tamagawa~Y., Tomita~S., Yamamoto~A., Ogawa~I. and Usuki~Y. \textit{Significantly different pulse shapes for $\gamma$-and $\alpha$-rays in Gd$_3$Al$_2$Ga3$_3$O$_{12}$:Ce$^{3+}$ scintillating crystals}, Nuclear Instruments and Methods in Physics Research Section A 694: 91-94 (2012).
 
\bibitem{Glodo2005}
  Glodo~J., Moses~W.~W., Higgins~W.~M., Van~Loef~E.~V.~D., Wong~P., Derenzo~S.~E., Weber~M.~J. and Shah~K.~S. \textit{Effects of Ce concentration on scintillation properties of LaBr$_{3}$(Ce)}, IEEE Transactions on Nuclear Science 52(5): 1805-1808 (2005).
 
\bibitem{vanDam2012} 
  van~Dam~H~T., Seifert~S., Drozdowski~W., Dorenbos~P. and Schaart~D.~R. \textit{Optical Absorption Length, Scattering Length, and Refractive Index of LaBr$_{3}$:Ce$^{3+}$}, IEEE Transactions on Nuclear Science 59(3): 656-664 (2012).

\end{thebibliography}
\end{document}